\documentclass[a4paper,11pt]{article}
\usepackage{jheppub}
\usepackage[english]{babel}
\usepackage{subcaption}
\usepackage{placeins}
\usepackage{xspace}
\usepackage{fancyvrb}
\usepackage{relsize}
\usepackage[utf8]{inputenc}
\usepackage{lineno}
\usepackage{stackrel}
\usepackage{amsmath}
\usepackage{mathtools}
\usepackage{amsmath,amssymb,enumerate,alltt,xspace,multirow}
\usepackage{graphicx}
\usepackage{xcolor}

\newcommand{\rrad}{\mathrm{rad}}
\newcommand{\Phirad}[1]{\Phi_{\rrad}}

\newcommand{\Sherpa}{{\rmfamily\scshape Sherpa}\xspace}

\newcommand{\OpenLoops}{{\rmfamily\scshape OpenLoops}\xspace}

\newcommand{\Pythia}{{\tt Pythia\,8.307}\xspace}

\newcommand{\RivetThree}{{\rmfamily\scshape Rivet\,3}\xspace}

\newcommand\POWHEGBOX{{\tt POWHEG\,BOX}\xspace}
\newcommand\POWHEG{{POWHEG}\xspace}
\newcommand\POWHEGRES{{POWHEG--RES}\xspace}

\newcommand\MCatNLO{MC@NLO}
\newcommand\PythiaEight{{\tt Pythia8}}

\newcommand\RES{{\tt POWHEG\,BOX\,RES}\xspace}

\newcommand\bbfourl{{\tt bb4l}\xspace}
\newcommand\bbfourlH{{\tt bb4lH}\xspace}

\newcommand\allrad{{\tt allrad}\xspace}
\newcommand\ttH{{\tt ttH}\xspace}
\newcommand\tWH{{\tt tWH}\xspace}

\newcommand{\GeV}{\text{GeV}\xspace}
\newcommand{\TeV}{\text{TeV}\xspace}

\newcommand{\ttbar}{\ensuremath{t \bar t}\xspace}

\newcommand{\Wt}{\ensuremath{tW}\xspace}
\newcommand{\tW}{\Wt}

\def\beq{\begin{equation}}
\def\beqn{\begin{eqnarray}}
\def\eeq{\end{equation}}
\def\eeqn{\end{eqnarray}}

\def\({\left(} 
\def\){\right)} 
 
\newcommand     \MSB            {\ifmmode {\overline{\rm MS}} \else
                                 $\overline{\rm MS}$\fi}

\newcount\minutes 
\newcount\scratch 
\def\timestamp{%
\scratch=\time 
\divide\scratch by 60 
\edef\hours{\the\scratch} 
\multiply\scratch by 60 
\minutes=\time 
\advance\minutes by -\scratch 
---$\,$\hours:\null 
\ifnum\minutes< 10 0\fi 
\the\minutes}

\def\refeq#1{\mbox{(\ref{#1})}}

\def\reffi#1{\mbox{Fig.~\ref{#1}}}
\def\reffis#1#2{\mbox{Figs.~\ref{#1}--\ref{#2}}}
\def\refta#1{\mbox{Tab.~\ref{#1}}}

\def\refse#1{\mbox{Sect.~\ref{#1}}}
\def\refses#1#2{\mbox{Sects.~\ref{#1}--\ref{#2}}}

\def\citere#1{\mbox{Ref.~\cite{#1}}}

\newcommand{\bit}{\begin{itemize}}
\newcommand{\eit}{\end{itemize}}

\RequirePackage[T1]{fontenc}

\usepackage{newtxtext}
\RequirePackage{flushend}
\RequirePackage[numbers,sort&compress]{natbib}

\begin{document}
\setlength{\abovedisplayskip}{6pt}
\setlength{\belowdisplayskip}{6pt}
\setlength{\abovedisplayshortskip}{6pt}
\setlength{\belowdisplayshortskip}{6pt}

\title{Resonance-aware NLOPS matching for off-shell \boldmath$t\bar t+tW$ production in association with a Higgs boson}

\preprint{MS-TP-26-30}

\author[a]{Patrick E. Hurley,}
\author[b]{Tom\'a\v{s} Je\v{z}o,} 
\author[a]{Jonas M. Lindert} 

\emailAdd{p.hurley@sussex.ac.uk}
\emailAdd{tomas.jezo@uni-muenster.de}
\emailAdd{j.lindert@sussex.ac.uk}

\affiliation[a] {Department of Physics and Astronomy, University of Sussex, Brighton BN1 9QH, UK}
\affiliation[b]{Institut f{\"u}r Theoretische Physik, Westf{\"a}lische Wilhelms-Universit{\"a}t M{\"u}nster, \\Wilhelm-Klemm-Stra{\ss}e 9, D-48149 M{\"u}nster, Germany}


\abstract{
We present a new NLO QCD Monte Carlo generator for the complete off-shell final state encompassing $t\bar tH$ and $tWH$ production with dileptonic decays, matched to parton showers in the
resonance-aware \POWHEGRES{} framework. At Born level the calculation
corresponds to the process
$pp\to b\bar b\,e^+\nu_e\,\mu^-\bar\nu_\mu\,H$ in the four-flavour scheme
with massive $b$ quarks.
The complete matrix elements include 
all double-resonant, single-resonant and non-resonant
topologies, all interferences and spin correlations, with both top quarks and $W$
bosons treated in the complex-mass scheme. This approach provides a 
unified and consistent treatment of $\ttbar H$ and $tWH$ production modes without the need for a diagram-removal or diagram-subtraction prescription.
 The generator, dubbed \bbfourlH{}, is implemented in the
\RES{} framework with \OpenLoops{} amplitudes, and takes advantage of the multiple-radiation scheme (\texttt{allrad}) and of 
matrix-element--based resonance projectors. 
Semileptonic events are also supported and can be obtained from dileptonic ones, including QCD
radiation in the hadronic $W$-boson decay. 
In addition, we present two new on-shell \POWHEG{} generators for
$\ttbar H$ and $tWH$ production, the latter being the first of its kind.
%
We validate all generators against known fixed-order results and in
the narrow-width limit, and present NLO and NLOPS predictions for the LHC
in the dileptonic and semileptonic channels. 
Tuned comparisons with the corresponding on-shell descriptions show good agreement for inclusive observables, while sizeable differences arise in observables sensitive to the reconstructed top-quark line shape and in $tWH$-enriched regions.
These findings are of direct relevance for the LHC analysis of $\ttbar H$ events
and for the separation of $\ttbar H$ and $tWH$
production.
}

\maketitle

\section{Introduction}
\label{se:intro}

The associated production of a Higgs boson with a top-quark pair,
$\ttbar H$, provides direct experimental access to the top-quark Yukawa
coupling, which is the largest Yukawa coupling in the Standard Model (SM)
and one of the key parameters that determine the stability of the
electroweak vacuum.
Moreover, $\ttbar H$ production is a key probe of physics beyond the SM.

Since its observation by the ATLAS and CMS
collaborations~\cite{ATLAS:2018mme,CMS:2018uxb}, $\ttbar H$ production
has been measured in an increasingly wide range of Higgs-decay channels,
including $H\to b\bar b$, $H\to \gamma\gamma$ and multi-lepton final
states~\cite{ATLAS:2024gth,CMS:2024fdo,ATLAS:2022vkf,CMS:2022dwd}, and
differential measurements as well as studies of the CP structure of the
top--Higgs interaction~\cite{ATLAS:2020ior,CMS:2020cga} are becoming
possible.
The $\ttbar H$ signal strength has reached an experimental uncertainty of
about
$\pm 20\%$~\cite{CMS:2020mpn,CMS:2024fdo,ATLAS:2024gth,ATLAS:2025eua}.
With the full Run~3 data set and, in particular, at the HL-LHC, the
precision on the $\ttbar H$ signal strength is expected to reach the
level of a few percent, which translates into an uncertainty of about
$3$--$4\%$ on the top-Yukawa coupling~\cite{Cepeda:2019klc}.
At this point, the accuracy of the theoretical modelling of signal and
background processes at the level of the fully decayed final state
becomes a limiting factor.
At the same time, the associated production of a single top quark with a
Higgs boson is directly sensitive to the relative sign and magnitude of
the top-Yukawa and $HW^+W^-$ couplings, through the interference between
amplitudes where the Higgs boson is radiated off the top-quark line
($\propto \kappa_t$) or off the $W$ line ($\propto \kappa_V$).
In the SM this interference is destructive and nearly complete, such that
the $pp \to tWH$ cross section is anomalously small, while it is enhanced
by a factor of about $8$ for $\kappa_t = -1$ and
$\kappa_V = +1$~\cite{Demartin:2016axk}.
For corresponding effects in $tHq$ production see
Refs.~\cite{Farina:2012xp,Biswas:2012bd,Demartin:2015uha}.
Experimentally, $tWH$ production has so far only been probed as part of a
combined single-top plus Higgs signal, together with the dominant $tHq$
channel, in dedicated $tH$ searches~\cite{CMS:2018jeh,ATLAS:2025irr} and
in $\ttbar H$ multi-lepton analyses.

Theoretical predictions for on-shell $\ttbar H$ production are well
advanced.
The NLO QCD corrections have been known for more than two
decades~\cite{Beenakker:2001rj,Beenakker:2002nc,Reina:2001sf,Dawson:2002tg,Dawson:2003zu},
and have been supplemented by NLO electroweak (EW)
corrections~\cite{Yu:2014cka,Zhang:2014gcy,Frixione:2014qaa,Frixione:2015zaa},
by the resummation of soft-gluon
effects~\cite{Broggio:2015lya,Kulesza:2015vda,Kulesza:2017ukk,Broggio:2016lfj,Ju:2019lwp},
and recently by NNLO QCD corrections~\cite{Catani:2022mfv,Devoto:2024nhl}.
Monte Carlo generators that match NLO QCD calculations for on-shell
$\ttbar H$ production to parton showers (NLOPS) are available in the
\POWHEG{}~\cite{Garzelli:2011vp,Hartanto:2015uka},
\MCatNLO{}~\cite{Alwall:2014hca} and \textsc{Sherpa}~\cite{Sherpa:2019gpd}
frameworks, and are used extensively in experimental analyses.

For $tWH$ production, an NLO QCD Monte Carlo generator was presented in
\citere{Demartin:2016axk}, where the overlap with $\ttbar H$ production,
which arises at NLO from real-emission topologies of type
$gg\to tW^-\bar b H$ with a resonant intermediate anti-top quark, is
addressed by means of diagram-removal and diagram-subtraction
prescriptions~\cite{Frixione:2008yi}.
As is well known from the analogous case of $\tW$
production~\cite{Frixione:2008yi,Re:2010bp,Zhu:2001hw,Campbell:2005bb,White:2009yt}
and from supersymmetric squark
production~\cite{Beenakker:1996ch,Hollik:2012rc}, such prescriptions are
always subject to a certain degree of arbitrariness, which results from
ad-hoc projections, violations of gauge invariance, or the treatment of
interferences and off-shell effects.
So far, no \POWHEG{} generator for $tWH$ production has been available.

In all of the above tools and predictions, top quarks are treated as
stable particles at the level of the hard scattering process, and their
decays are modelled a posteriori, typically in terms of spin-correlated
LO matrix elements in the narrow-width approximation
(NWA)~\cite{Frixione:2007zp,Artoisenet:2012st} with a naive smearing of
the top-quark virtualities according to Breit--Wigner distributions.
In the following, such tools will be referred to as on-shell generators.

A fully consistent description of the physical final state requires
instead, e.g.\ in the dileptonic channel, the computation of the complete
$2\to 7$ process
$pp\to b\bar b\,\ell^+\nu_\ell\,\ell'^-\bar\nu_{\ell'}\,H$. The computation includes the 
exact matrix elements with all resonant and non-resonant
topologies, interferences and spin correlations, and treats unstable top quarks
and $W$ bosons in the complex-mass scheme~\cite{Denner:2005fg}.
Such off-shell calculations are available at NLO QCD~\cite{Denner:2015yca}
and NLO EW~\cite{Denner:2016wet} for the dileptonic channel in the
five-flavour number scheme (5FNS) with massless $b$ quarks, and at NLO
QCD also including Higgs-boson decays in the NWA~\cite{Stremmer:2021bnk}.
When performed in the four-flavour number scheme (4FNS), i.e.\ treating
$b$ quarks as massive partons that do not enter the initial state,
off-shell calculations provide a unified NLO description of $\ttbar H$
and $tWH$ production in the entire phase space, with a fully consistent
treatment of $\ttbar H$--$tWH$ interference effects and without any need
for diagram removal or subtraction. This is in complete analogy with off-shell
$\ttbar+\tW$ production~\cite{Cascioli:2013wga,Frederix:2013gra,Jezo:2016ujg}.
In this paper, such a 4FNS calculation is performed for the first time,
and is matched to parton showers.

Off-shell effects are known to be small for sufficiently inclusive
observables, but they can reach tens of percent in the tails of
distributions that are sensitive to the top-quark line shape, such as the
invariant mass of the reconstructed top quark, or in phase-space regions
where single-top contributions are enhanced.
Such regions are relevant for the reconstruction of the $\ttbar H$ system, and 
for the extraction of the top-Yukawa coupling in kinematic fits.
So far, however, no NLOPS generator for off-shell $\ttbar H+tWH$
production has been available.

The matching of off-shell NLO calculations to parton showers is
non-trivial, since standard matching techniques rely on kinematic
mappings between Born and real-emission configurations that do not
preserve the virtualities of intermediate resonances.
This leads to a violation of the collinear factorisation properties that
underlie the \POWHEG{}method, and can result in
unphysical distortions of resonance line shapes and in spurious
enhancements of the radiation pattern~\cite{Jezo:2015aia}.
A solution to this problem is provided by the resonance-aware \POWHEG{}
method of \citere{Jezo:2015aia}, referred to as \POWHEGRES{} in the
following.
This method is based on a probabilistic decomposition of events into
resonance histories, which correspond to the different combinations of
production and decay subprocesses that contribute to a given off-shell
process.
Within each resonance history, \POWHEG{} radiation is generated in a way
that preserves the virtualities of all resonances, such that resonances
have correct NLO shapes and, in the limit of small decay widths, NLOPS
predictions are consistent with the factorisation properties of the NWA.
Alternative resonance-aware matching approaches in the \MCatNLO{}
framework were proposed in
Refs.~\cite{Frederix:2016rdc,Denner:2026ztd,Hoche:2026dup}.
The \POWHEGRES{} generator \bbfourl{}~\cite{Jezo:2016ujg} describes
off-shell $\ttbar+\tW$ production with dileptonic decays in the 4FNS.
It has been widely used and scrutinised in experimental
analyses~\cite{ATLAS:2018ivx,ATLAS:2021pyq,ATLAS:2022jpn}, and was
recently extended to the semileptonic channel~\cite{Jezo:2023rht}, based
on dileptonic matrix elements supplemented by QCD radiation in the
hadronic $W$-boson decay via the \allrad{} scheme~\cite{Campbell:2014kua}.
In the same work, two important methodological improvements were
introduced, namely matrix-element--based resonance histories, which allow
for a more reliable separation of $\ttbar$ and $\tW$ contributions, and a
systematic treatment of spurious terms that arise from the perturbative
treatment of the NLO top-quark width in off-shell calculations.

In this paper we present a new \POWHEGRES{} generator for off-shell
$\ttbar H + tWH$ production in the 4FNS, in the following referred to as
\bbfourlH{}.
In the dileptonic channel, at LO this corresponds to the process
$pp\to b\bar b\,e^+\nu_e\,\mu^-\bar\nu_\mu\,H$ at
$\mathcal{O}(\alpha_s^2\alpha^5)$, including all doubly-resonant
$\ttbar H$, singly-resonant $tW^-H$ and $\bar tW^+H$, and non-resonant
topologies.
We compute the NLO QCD corrections to this process retaining all
off-shell effects of top quarks and $W$ bosons, while the Higgs boson is
treated as an on-shell particle that can be decayed by the parton shower.
The generator is implemented in the \RES{} framework, where all tree and
one-loop matrix elements are provided by
\OpenLoops{}~\cite{Cascioli:2011va,Buccioni:2017yxi,Buccioni:2019sur}, and
builds on the methodological developments of \citere{Jezo:2023rht}.
In particular, resonance histories are assigned by means of
matrix-element--based projectors, which guarantee a consistent treatment
of $\ttbar H$--$tWH$ interference effects and non-resonant contributions.
Spurious width effects are avoided through an inverse-width expansion
and \POWHEG{} radiation is generated from the production and each
top-quark decay subprocess in the \allrad{} approach.
Semileptonic events are obtained by re-processing dileptonic events
following \citere{Jezo:2023rht}.
Moreover, \bbfourlH{} supports modified top-Yukawa couplings, which makes
it possible to study the sensitivity of $tWH$ production, and of its
interference with $\ttbar H$ production, to the sign of the top-Yukawa
coupling.

In order to quantify the impact of off-shell and single-top effects, and
to provide a well-defined reference for experimental studies, we also
present two new on-shell generators for $\ttbar H$ and $tWH$ production
in the 5FNS referred to as $\texttt{ttH}$ and $\texttt{tWH}$, respectively. Both are implemented in the same \POWHEGRES{} framework and employ \OpenLoops{} amplitudes throughout.
In the $\texttt{tWH}$ generator, which is the first \POWHEG{} generator for this
process, the overlap with $\ttbar H$ production is removed locally in the
phase space by means of diagram
removal~\cite{Frixione:2008yi,Demartin:2016axk}.
In both on-shell generators, top quarks are decayed at the level of the
Les Houches events, including spin correlations and finite-width effects,
following the method of
Refs.~\cite{Frixione:2007zp,Alioli:2009je,Alioli:2011as}.
This makes it possible to carry out a tuned comparison of off-shell and
on-shell generators at the level of the fully decayed final state, with
identical fiducial cuts.

We validate the new generators against fixed-order results available in
the literature and, in the case of \bbfourlH{}, also against the NWA in the
limit of vanishing top-quark width.
Then we present NLO and NLOPS predictions in the dileptonic channel, as
well as a tuned comparison of off-shell and on-shell generators in the
dileptonic and semileptonic channels.
For Higgs-boson and leptonic observables, NLO, LHE and NLOPS predictions
of \bbfourlH{} agree at the few-percent level, and for $b$-jet observables
within about $10\%$, while large differences arise above the peak of the
reconstructed top-quark line shape, which is populated by soft QCD
radiation clustered into the $b$-jets.
In the comparison against on-shell generators, inclusive cross sections
agree within $2\%$, while in the fiducial phase space the on-shell
predictions lie $5$--$7\%$ below \bbfourlH{}, mainly as a result of QCD
radiation in top-quark decays, which in the on-shell generators is
entirely generated by the parton shower.
The shapes of Higgs-boson, leptonic and $b$-jet distributions agree at
the few-percent level, whereas differences of $20$--$40\%$ show up in the
reconstructed top-quark line shape and above the kinematic edge of the
lepton--$b$-jet invariant mass, where $tWH$ production and
$\ttbar H$--$tWH$ interference effects play an important role.

The paper is organised as follows.
In \refse{Sec:Structure} we discuss the structure of the off-shell
$\ttbar H+tWH$ process in the 4FNS and that of the on-shell $\ttbar H$ and
$tWH$ processes in the 5FNS, including the diagram-removal prescription
and the treatment of top-quark decays in the on-shell generators.
In \refse{Sec:POWHEG-res} we review the \POWHEGRES{} method, and we
present the resonance histories, the matrix-element--based projectors,
the inverse-width expansion and the generation of semileptonic events in
\bbfourlH{}.
The computational setup is specified in \refse{sec:setup}, and the
validation of the generators is presented in \refse{sec:validation}.
In \refse{sec:results} we present NLO and NLOPS predictions in the
dileptonic channel and a tuned comparison of off-shell and on-shell
generators in the dileptonic and semileptonic channels.
Our conclusions are presented in \refse{se:conclusions}.

\section{Structure of off-shell 4FNS and on-shell 5FNS production }
\label{Sec:Structure}

In this section we discuss the structure of the off-shell $\ttbar H+tWH$
process and of the on-shell $\ttbar H$ and $tWH$ processes that are
implemented in the generators presented in this paper.
All generators describe the equivalent final states, which in the dileptonic
channel consists of two $b$ quarks, two oppositely charged leptons, two
neutrinos and a Higgs boson, but they do so at a different level of
approximation. In $tWH$ production one of the $b$ quarks may arise from QCD radiation.
In the off-shell \bbfourlH{} generator, this final state is described in
terms of exact di-leptonic matrix elements, and $\ttbar H$ and $tWH$ production
emerge as different resonant regions of a single gauge-invariant process
(\refse{sec:offshell_structure}).
In contrast, the on-shell generators treat top quarks as stable particles
at the level of the hard process. This requires a prescription for the
separation of $tWH$ and $\ttbar H$ production beyond LO
(\refse{sec:onshell_structure}), while top-quark decays need to be
included a posteriori (\refse{Sec:Top decay}).
The comparison of the two approaches makes it possible to assess
off-shell and non-resonant effects, $\ttbar H$--$tWH$ interference
effects and QCD corrections to top-quark decays, as well as the
dependence on the flavour-number scheme.
 
\subsection[Off-shell $t\bar tH+tWH$ production in the 4FNS]{Off-shell \boldmath$\ttbar H+tWH$ production in the 4FNS}
\label{sec:offshell_structure}

\begin{figure}[tbp]
  \centering
  \begin{subfigure}{0.3\linewidth}
    \centering
    \includegraphics[width=\linewidth]{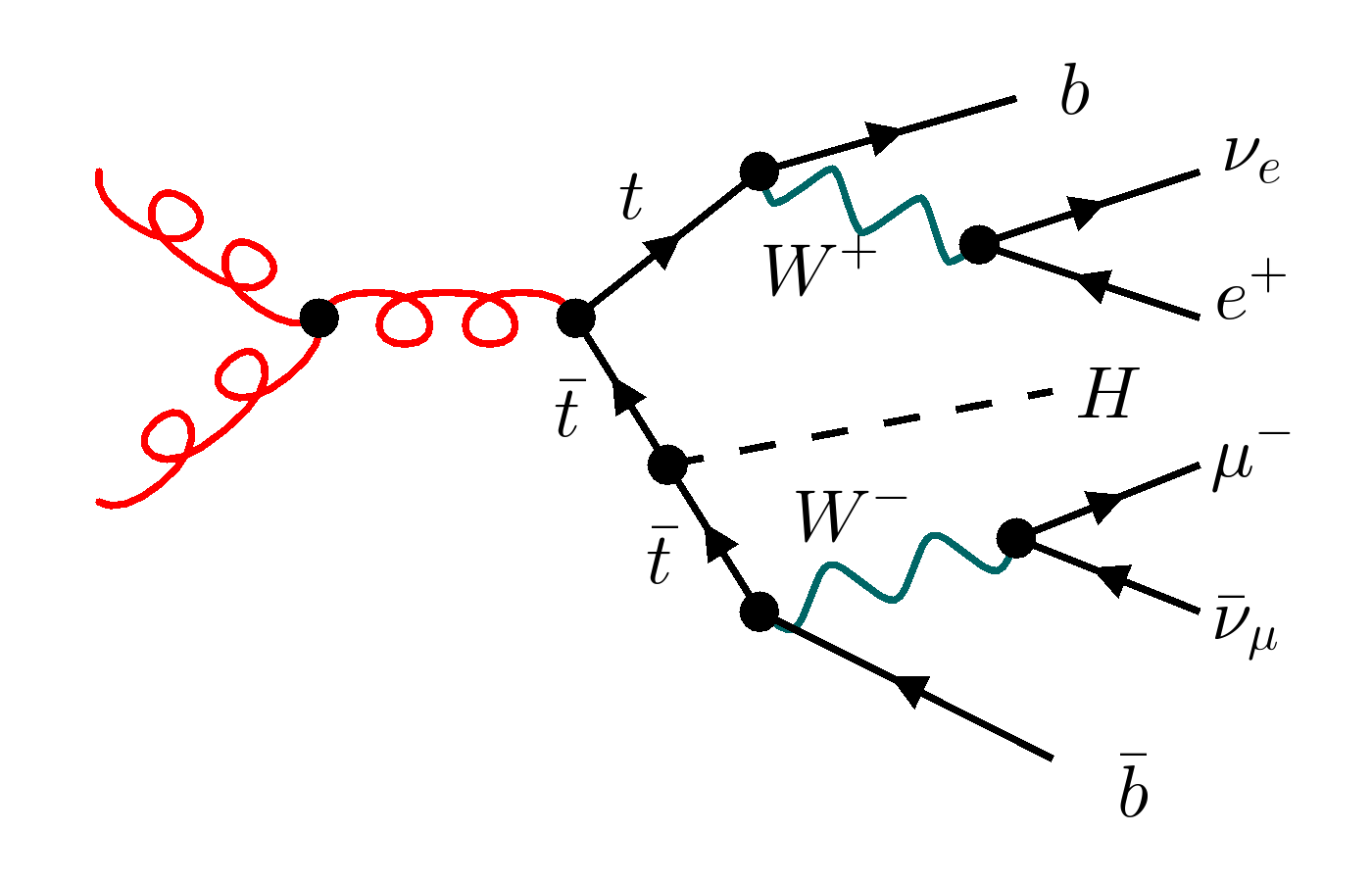}
    \caption{$\ttbar H$, $gg$ $s$-channel}
  \end{subfigure}
  \hfill
  \begin{subfigure}{0.3\linewidth}
    \centering
    \includegraphics[width=\linewidth]{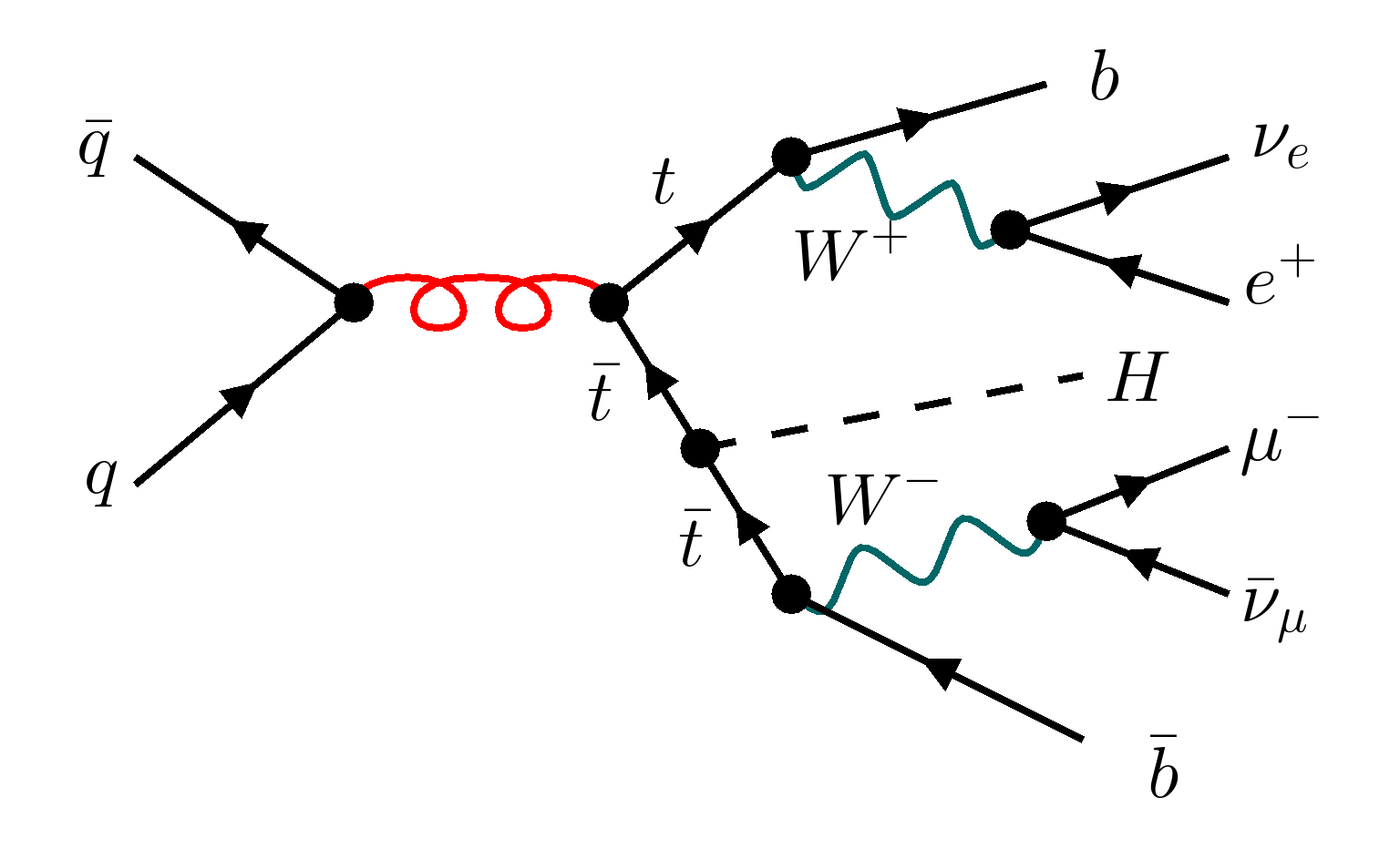}
    \caption{$\ttbar H$, $q\bar{q}$ channel}
  \end{subfigure}
  \hfill
  \begin{subfigure}{0.3\linewidth}
    \centering
    \includegraphics[width=\linewidth]{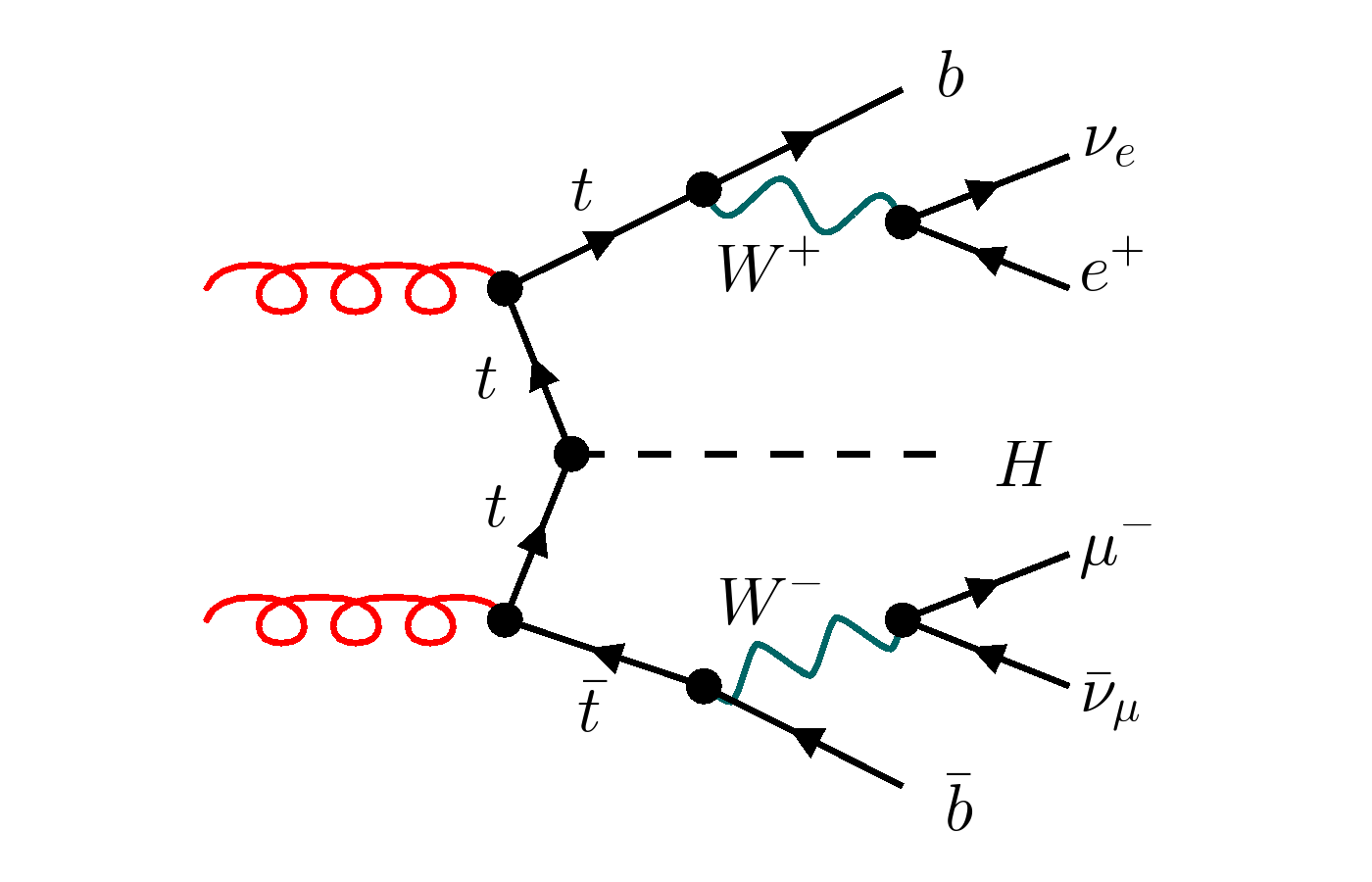}
    \caption{$\ttbar H$, $gg$ $t$-channel}
  \end{subfigure}
 
  \vspace{1em}
 
  \begin{subfigure}{0.3\linewidth}
    \centering
    \includegraphics[width=1.1\linewidth]{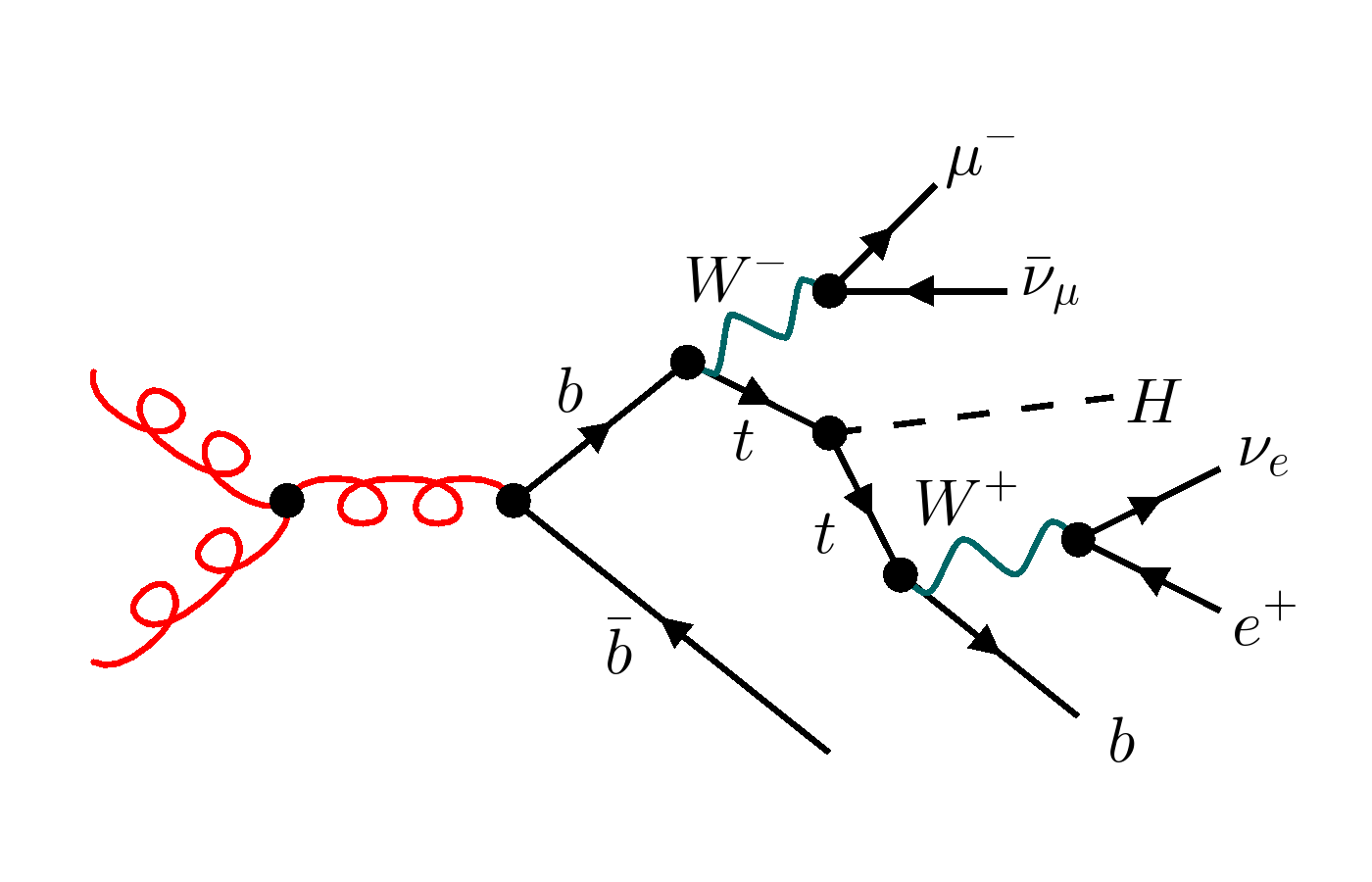}
    \caption{$tW^-H$, $gg$ $s$-channel}
  \end{subfigure}
  \hfill
  \begin{subfigure}{0.3\linewidth}
    \centering
    \includegraphics[width=1.1\linewidth]{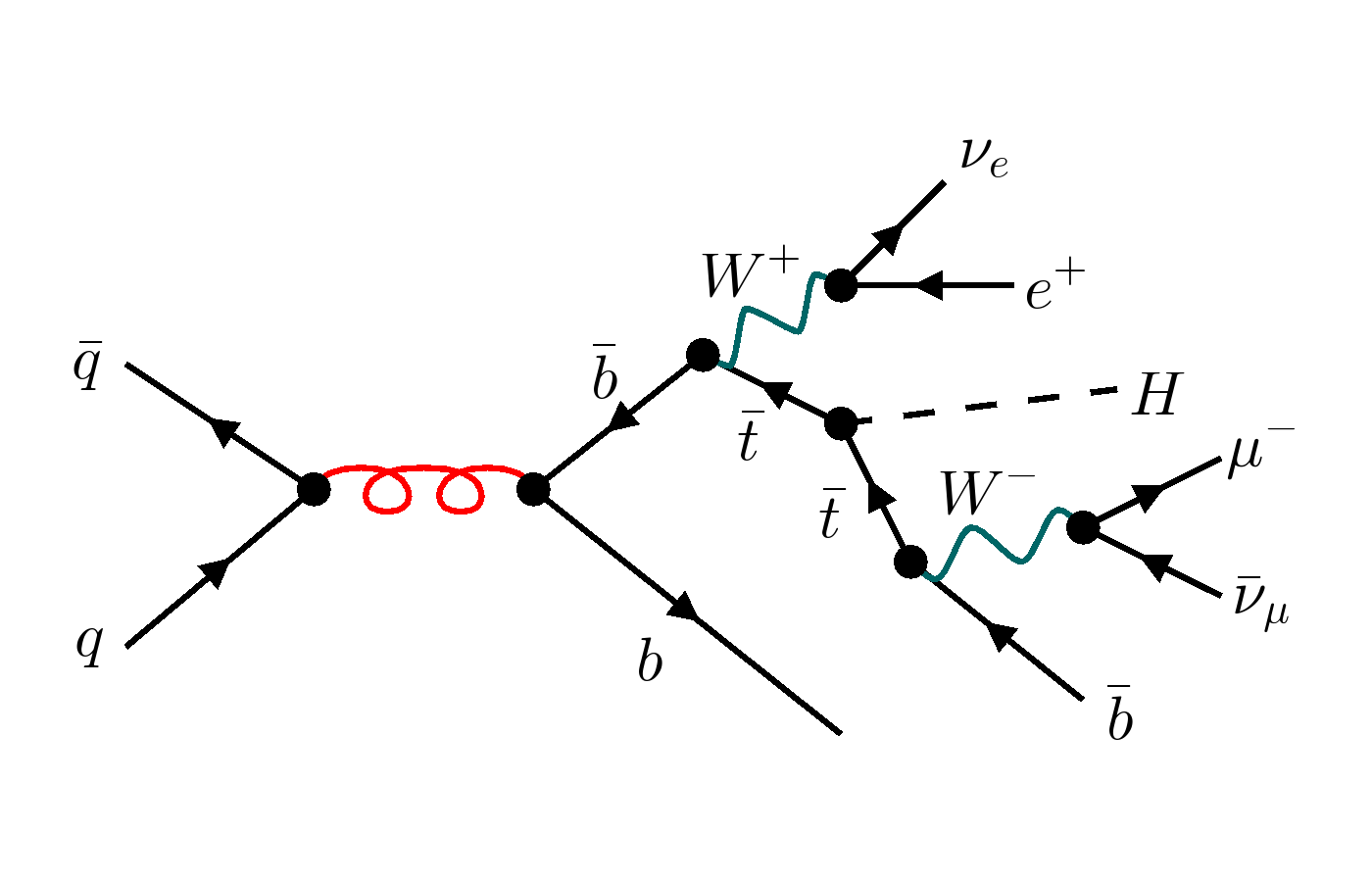}
    \caption{$\bar{t}W^+H$, $q\bar{q}$ channel}
  \end{subfigure}
  \hfill
  \begin{subfigure}{0.3\linewidth}
    \centering
    \includegraphics[width=0.9\linewidth]{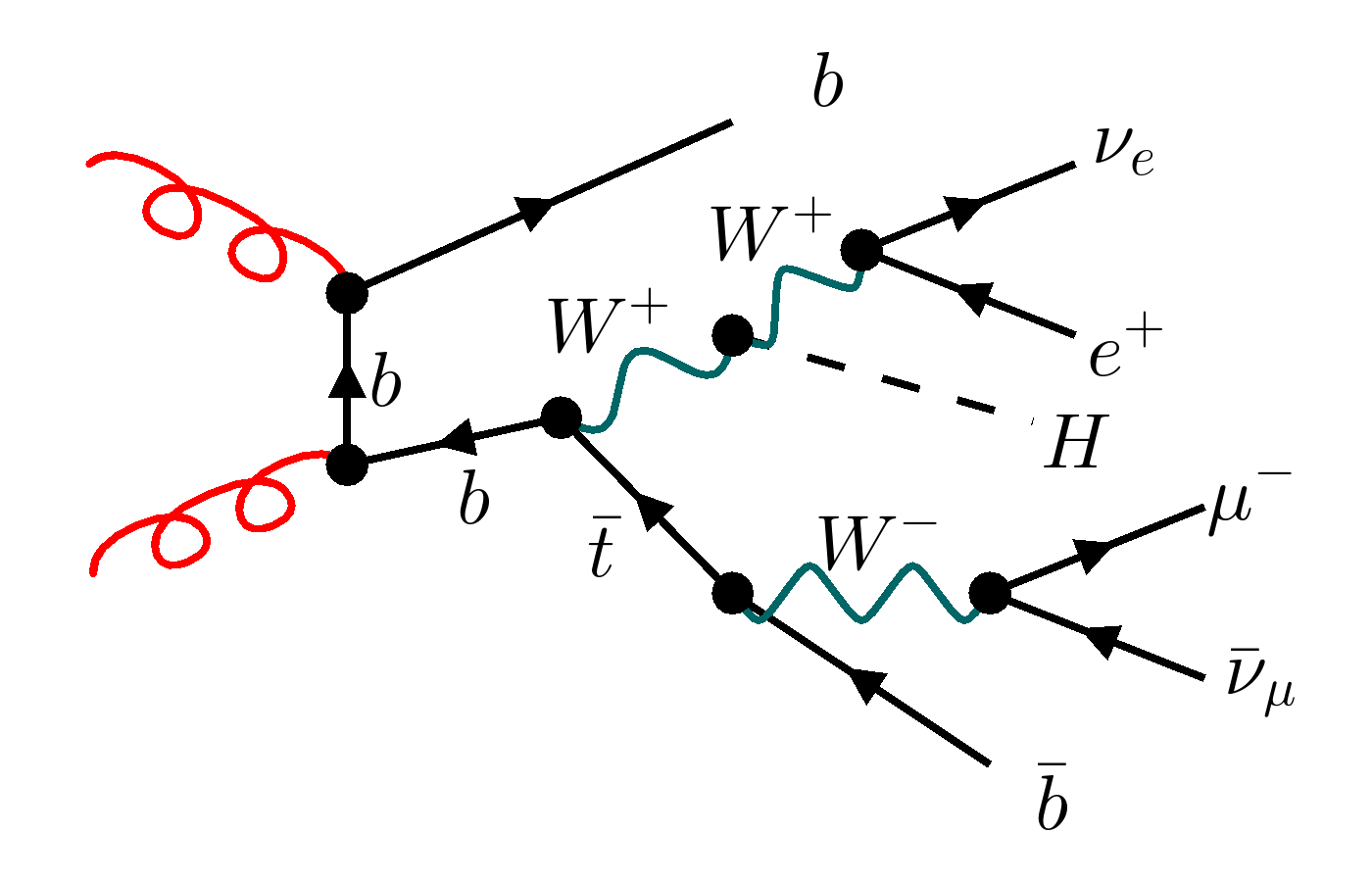}
    \caption{$\bar{t}W^+H$, $gg$ $t$-channel}
  \end{subfigure}
 
  \vspace{1em}
 
  \begin{subfigure}{0.3\linewidth}
    \centering
    \includegraphics[width=1.2\linewidth]{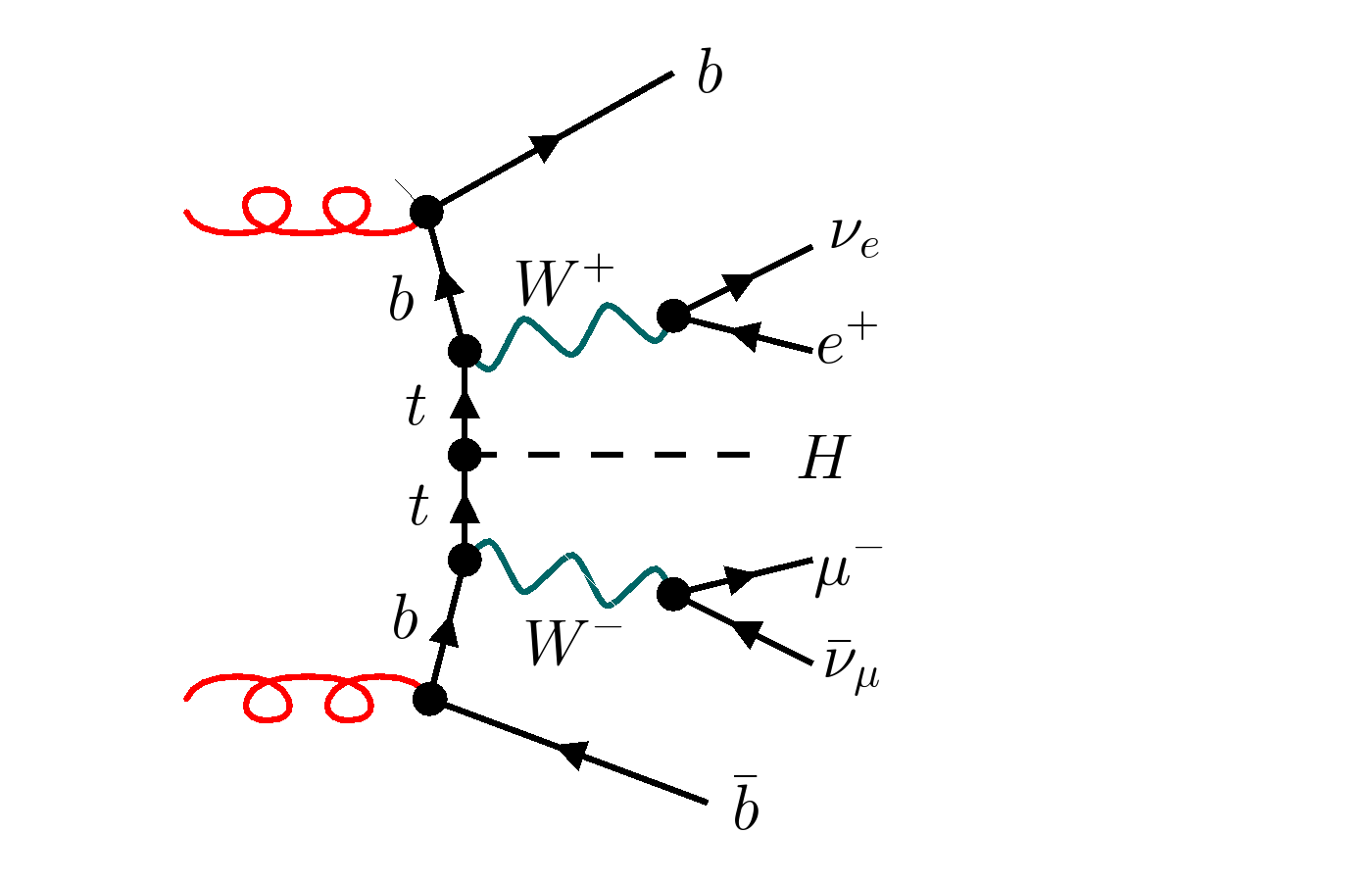}
    \caption{non-resonant, $gg$}
  \end{subfigure}
  \hfill
  \begin{subfigure}{0.3\linewidth}
    \centering
    \includegraphics[width=\linewidth]{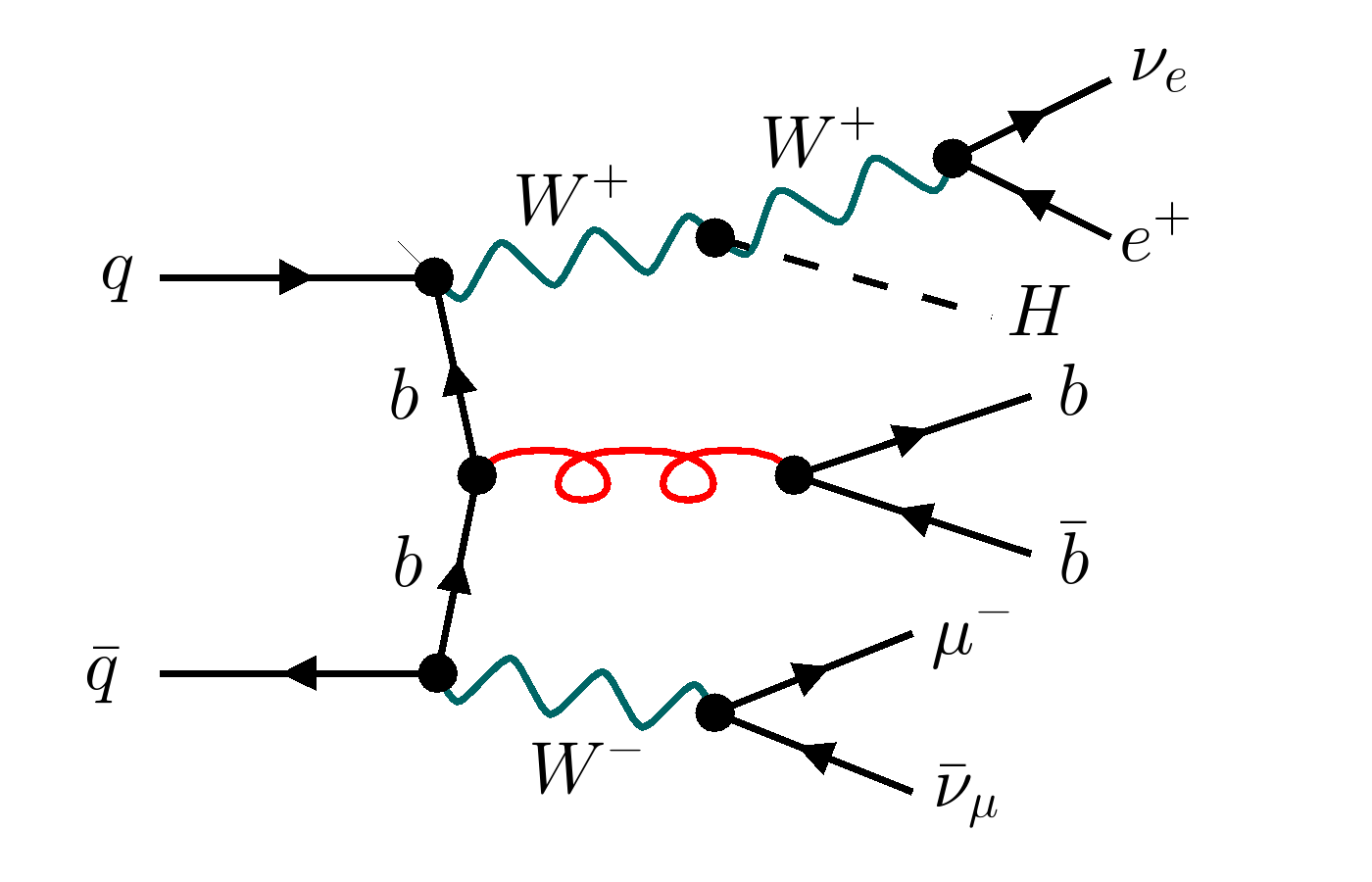}
    \caption{non-resonant, $q\bar q$}
  \end{subfigure}
  \hfill
  \begin{subfigure}{0.3\linewidth}
    \centering
    \includegraphics[width=0.9\linewidth]{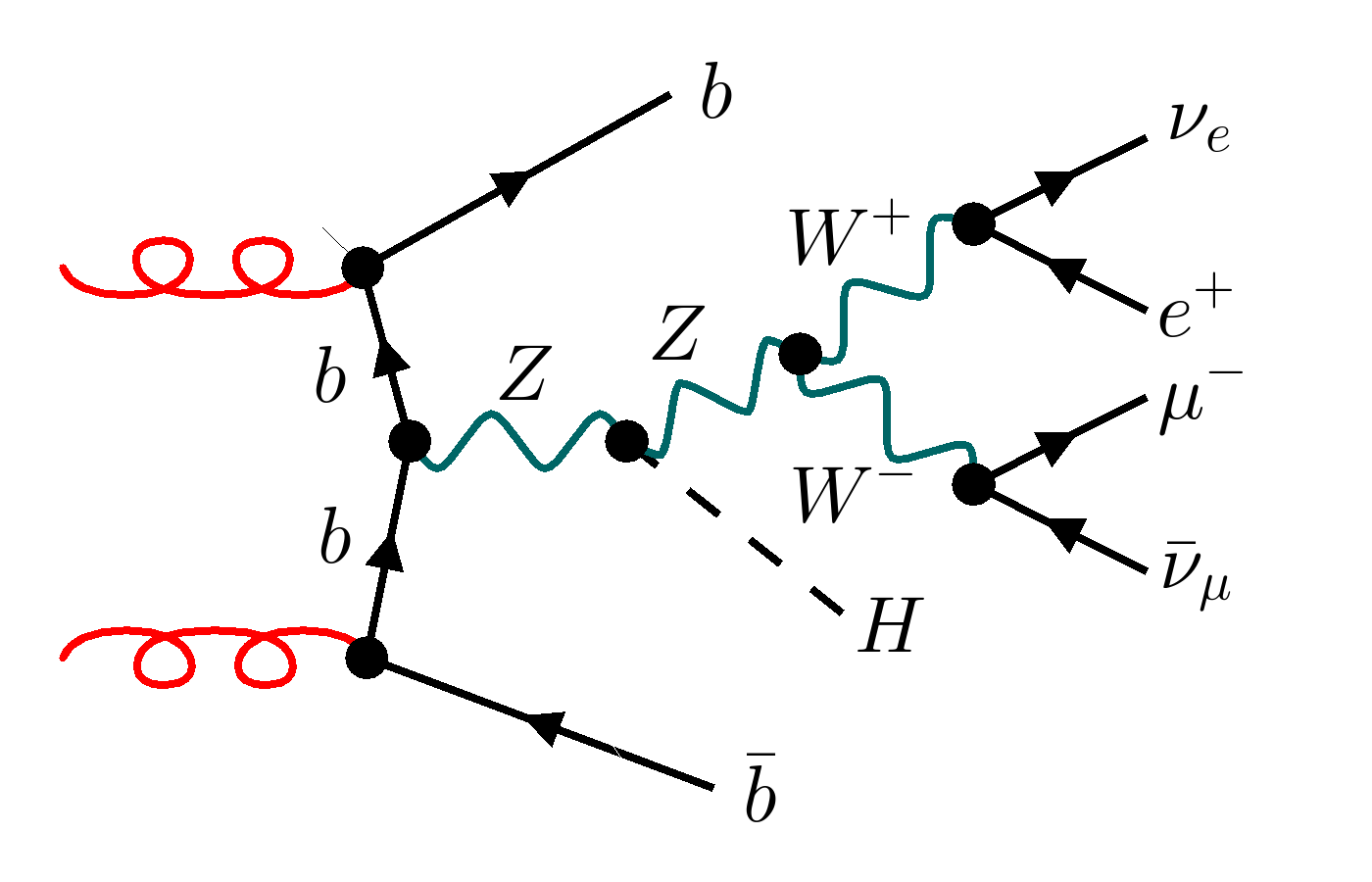}
    \caption{non-resonant, $gg$}
  \end{subfigure}
  \caption{Representative tree-level diagrams for
    $pp\to b\bar b\,e^+\nu_e\,\mu^-\bar\nu_\mu\,H$ with two (a--c), one
    (d--f) and no (g--i) resonant top-quark propagators, where
    $q\in\{u,d,c,s\}$. All Feynman diagrams in this paper are drawn with
    FeynGame~\cite{Bundgen:2025utt}.}
  \label{fig:Tree Level Diags}
\end{figure}

We consider the dileptonic channel with different-flavour leptons,
\begin{equation}
  \label{eq:Process}
  pp \to b\bar{b}\,e^+\nu_e\,\mu^-\bar{\nu}_\mu\,H\,,
\end{equation}
which at leading-order (LO) arises at $\mathcal{O}(\alpha_s^2\alpha^5)$.
The corresponding Feynman diagrams can be classified according to the
number of $s$-channel top-quark propagators, as illustrated in \reffi{fig:Tree Level Diags}.
Doubly-resonant diagrams (\reffi{fig:Tree Level Diags}a--c) involve a
$\ttbar$ pair that is produced in association with the Higgs boson and
decays via $t\to W^+b\to e^+\nu_e\,b$ and
$\bar t\to W^-\bar b\to \mu^-\bar\nu_\mu\,\bar b$. These diagrams
correspond to $\ttbar H$ production.
Singly-resonant diagrams (\reffi{fig:Tree Level Diags}d--f) involve only
one top-quark decay, while the second $W$ boson and $b$ quark are
produced directly in the hard scattering. These diagrams correspond to
$tW^-H$ and $\bar tW^+H$ production, where in the 4FNS the spectator $b$
quark is present in the final state already at LO.
Finally, non-resonant diagrams (\reffi{fig:Tree Level Diags}g--i) do not
involve any top-quark propagator that can go on shell.
 
Since $m_t<m_b+m_W+m_H$, on-shell top quarks cannot decay into an
on-shell $W$ boson and a Higgs boson, and contributions where the Higgs
boson is emitted from a far off-shell $W$ boson in a top-quark decay are
strongly suppressed. Thus, in doubly-resonant configurations the Higgs boson is
effectively always emitted at the production stage.
In fact, in doubly-resonant contributions the Higgs boson couples exclusively to
top quarks, such that $\ttbar H$ production provides a direct probe of
the (squared) top-Yukawa coupling.
In singly-resonant contributions the Higgs boson can be emitted from the
top quark or from the $W$ boson, and the corresponding amplitudes, which
are proportional to the top-Yukawa and to the $HWW$ coupling,
respectively, interfere destructively in the SM (see \refse{se:intro}).
The off-shell process thus comprises, within a single calculation, the
dominant $\ttbar H$ production mode as well as $tWH$ production, which is
sensitive to the relative sign of the two couplings.

For inclusive observables, the cross section is dominated by
doubly-resonant contributions, while singly-resonant and non-resonant
contributions are suppressed.
However, the latter can become relevant in specific regions of phase
space, in particular when one of the two $b$ quarks is soft or outside
the acceptance, as is typical for $tWH$ production, or when the
invariant masses of the top-decay products are far from the top-quark
resonance, e.g.\ in the tails of $b$-jet--lepton invariant-mass
distributions beyond the on-shell LO endpoint
$m_{b\ell}=\sqrt{m_t^2-m_W^2}$.
Since $\ttbar H$ and $tWH$ production lead to the same final state, they
interfere with each other, and their separation is meaningful only in an
approximate sense.
Moreover, any separation at the amplitude level, such as the omission of
doubly-resonant diagrams, violates gauge invariance.
In the present calculation no such separation is required: all diagrams are included, gauge invariance is
preserved, and the $\ttbar H$--$tWH$ interference is taken into account
exactly, without any need for diagram removal or subtraction.
A separation into $\ttbar H$-like and $tWH$-like contributions is
introduced in a probabilistic way only for the matching to the parton shower. 

The calculation is performed in the four-flavour number scheme (4FNS),
where $b$ quarks are massive and do not contribute to the initial state.
In fact, $b$-quark effects are included throughout, which includes also $b$-quark Yukawa couplings, and thus diagrams where the Higgs is radiated off the final state. However, such contributions are strongly suppressed.
Configurations that are described by $gb\to tW^-H$ in the five-flavour
number scheme (5FNS), where the incoming $b$ quark results from a
$g\to b\bar b$ splitting inside the proton, are described in terms of
$gg\to tW^-\bar bH$ and $q\bar q\to tW^-\bar bH$ in the 4FNS, including
the exact kinematics of the spectator $\bar b$ quark.
Since the collinear $g\to b\bar b$ singularity is regularised by the
$b$-quark mass, the entire $b$-quark phase space can be integrated over,
including the region where the spectator $b$ quark is soft or collinear
to the beam.
This allows for a unified description of $\ttbar H$ and $tWH$ production
for arbitrary requirements on the number of $b$-jets.
Resulting large logarithms of the ratio of $m_b$ and the hard
scale in the 4FNS, which result from initial-state $g\to b\bar b$ splittings, are
avoided through the use of 5FNS PDFs and appropriate matching factors~\cite{Cacciari:1998it,Jezo:2023rht}.

Top quarks and massive vector bosons are treated as off-shell particles
throughout, and their propagators are described in the complex-mass
scheme~\cite{Denner:2005fg}, which preserves gauge invariance in the
presence of finite widths.
In contrast, the Higgs boson, whose width $\Gamma_H\approx 4$\,MeV is
negligible as compared to its mass, is treated as a stable on-shell
particle in the hard matrix elements.
Its decays can subsequently be simulated by the parton shower, such that
different Higgs-boson decay channels can be studied with the same sample
of Les Houches events (LHEs).

\begin{figure}[tbp]
  \centering
  \begin{subfigure}{0.3\linewidth}
    \centering
    \includegraphics[width=1.1\linewidth]{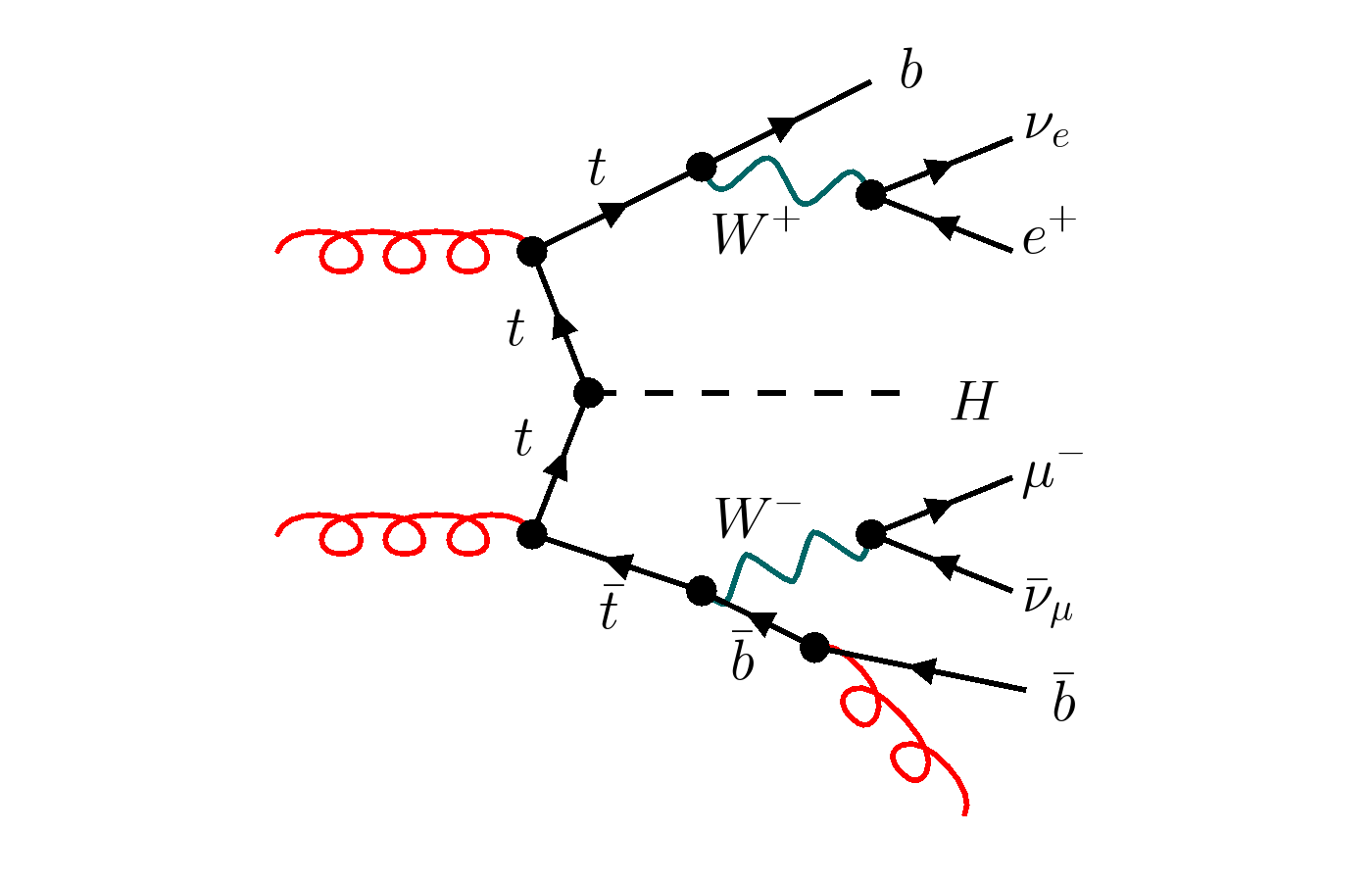}
    \caption{$gg$ channel}
  \end{subfigure}
  \hfill
  \begin{subfigure}{0.3\linewidth}
    \centering
    \includegraphics[width=1.1\linewidth]{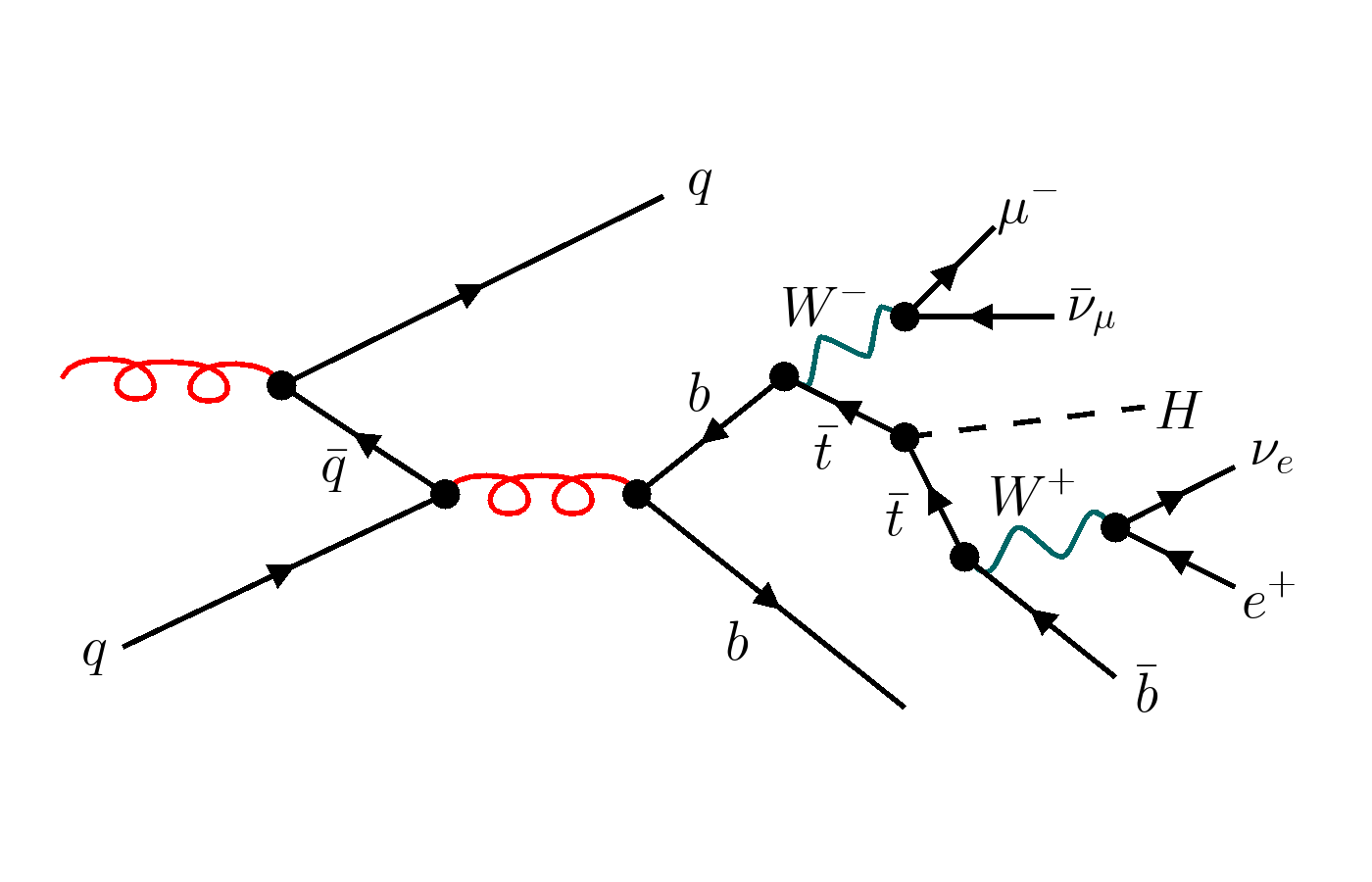}
    \caption{$gq$ channel}
  \end{subfigure}
  \hfill
  \begin{subfigure}{0.3\linewidth}
    \centering
    \includegraphics[width=\linewidth]{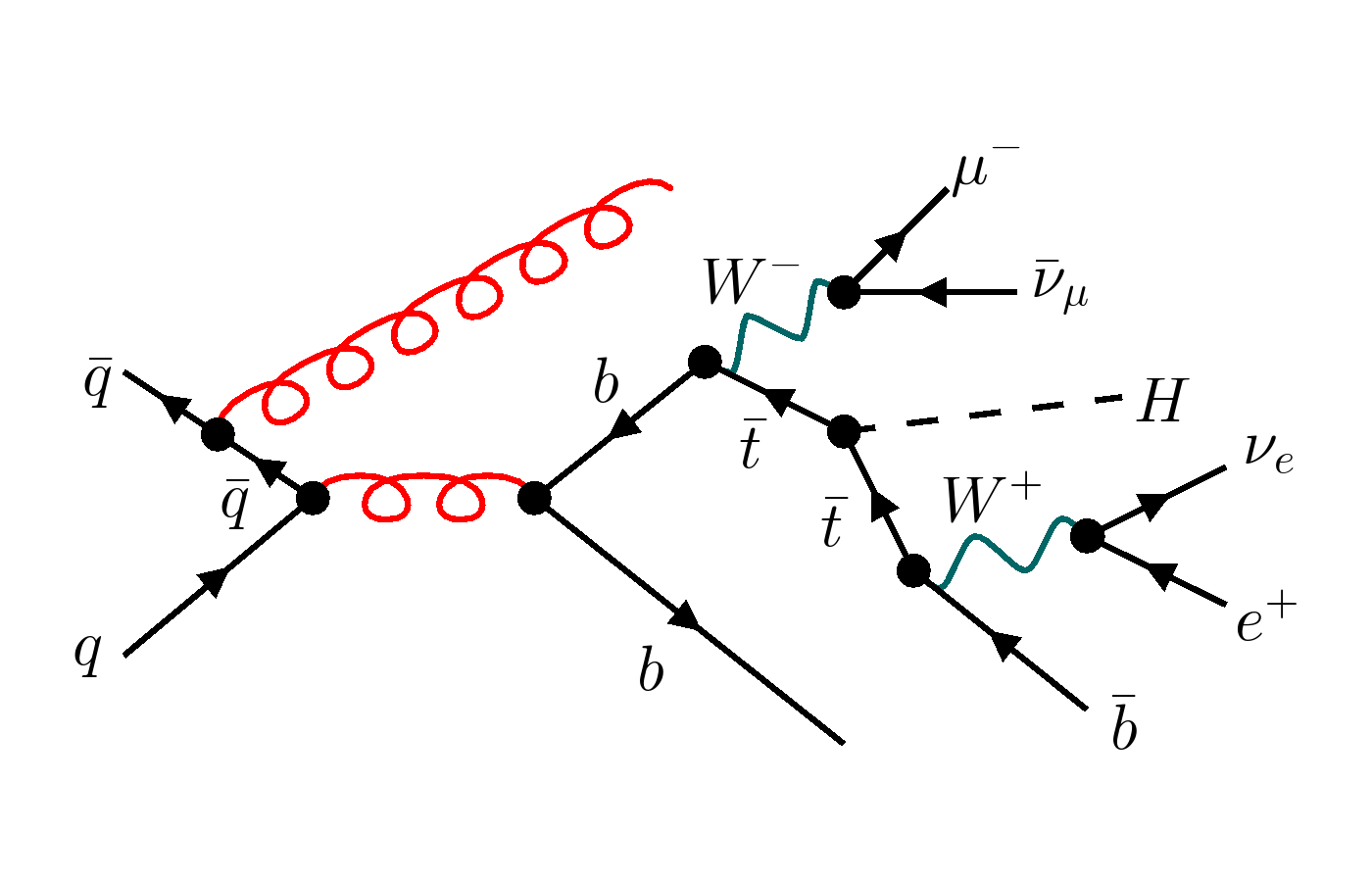}
    \caption{$q\bar{q}$ channel}
  \end{subfigure}
  \caption{Representative real-emission diagrams for
    $pp\to b\bar b\,e^+\nu_e\,\mu^-\bar\nu_\mu\,H\,j$ with doubly- and
    singly-resonant top-quark propagators, where $q\in\{u,d,c,s\}$.}
  \label{fig:Real Diags}
\end{figure}

We compute NLO QCD corrections to \refeq{eq:Process} at
$\mathcal{O}(\alpha_s^3\alpha^5)$, retaining the full off-shell structure
discussed above.
The virtual corrections comprise the one-loop amplitudes for all partonic
channels, while the real corrections,
$pp\to b\bar b\,e^+\nu_e\,\mu^-\bar\nu_\mu\,H\,j$, involve gluon emission
in the $gg$ and $q\bar q$ channels, as well as new quark--gluon channels
with $q\in\{u,d,c,s\}$ (see \reffi{fig:Real Diags}).
In total there are about 28k/14k virtual $2 \to 7$ diagrams contributing in the $gg/q\bar q$ channels with up to octagon topologies. 

Gluons are radiated at the production stage as well as from the $b$
quarks that result from top-quark decays. Thus NLO corrections to
top-quark decays and non-factorisable corrections that connect production
and decays are included on the same footing as the corrections to the
production process.
Soft singularities and collinear singularities associated with massless
initial-state partons are handled with the FKS subtraction
method~\cite{Frixione:1995ms,Frixione:1997np} as implemented in
\POWHEG{}, while collinear singularities associated with final-state $b$
quarks are regularised by the $b$-quark mass.

\subsection[On-shell $t\bar tH$ and $tWH$ production in the 5FNS]{On-shell \boldmath$\ttbar H$ and \boldmath$tWH$ production in the 5FNS}
\label{sec:onshell_structure}

Besides \bbfourlH{}, we setup two generators for on-shell $\ttbar H$ and
$tWH$ production, referred to as \ttH and \tWH generators.
In these generators, top quarks are treated as stable particles at the
level of the hard process, NLO QCD corrections are applied only to the
production process, and top-quark decays are included at LO in a second
step (\refse{Sec:Top decay}).
Apart from the smearing of top-quark virtualities, this corresponds to
the NWA with NLO production and LO decays.
In contrast to \bbfourlH{}, non-resonant contributions,
$\ttbar H$--$tWH$ interference effects and QCD corrections to top-quark
decays are not included at matrix-element level, and the latter are
accounted for only through the parton shower.
Both generators are formulated in the 5FNS, where $b$ quarks are massless, and they are based on the
same framework and on the same matrix-element provider as \bbfourlH{}.
In this way, differences with respect to off-shell predictions can be
traced back to the above physics approximations and to the flavour
scheme, rather than to technical aspects. The \ttH generator is formally equivalent to the one of \citere{Hartanto:2015uka}.
 
\begin{figure}[tbp]
  \centering
  \begin{subfigure}{0.3\linewidth}
    \centering
    \includegraphics[width=0.9\linewidth]{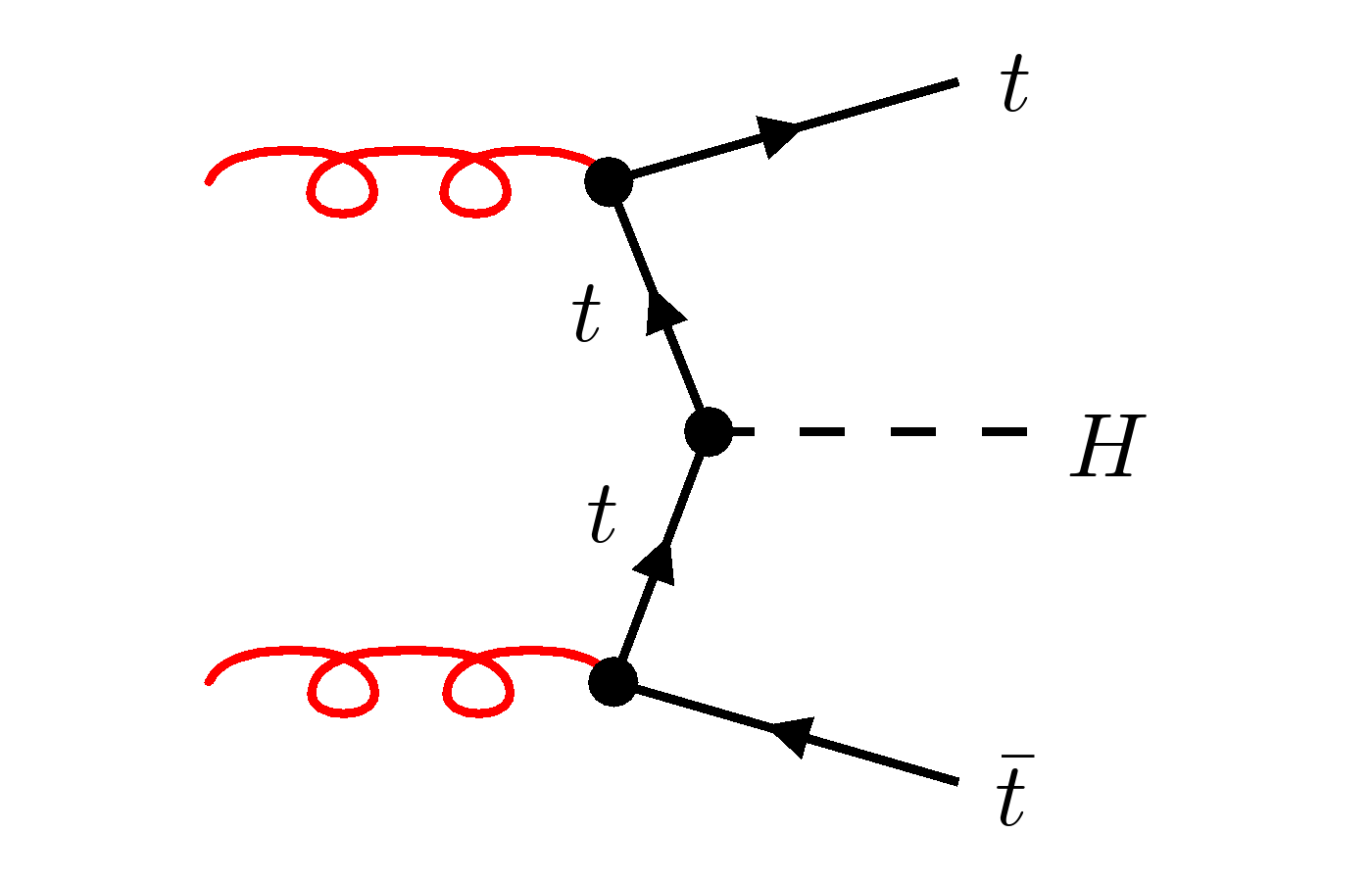}
    \caption{Born}
  \end{subfigure}
  \hfill
  \begin{subfigure}{0.3\linewidth}
    \centering
    \includegraphics[width=1.1\linewidth]{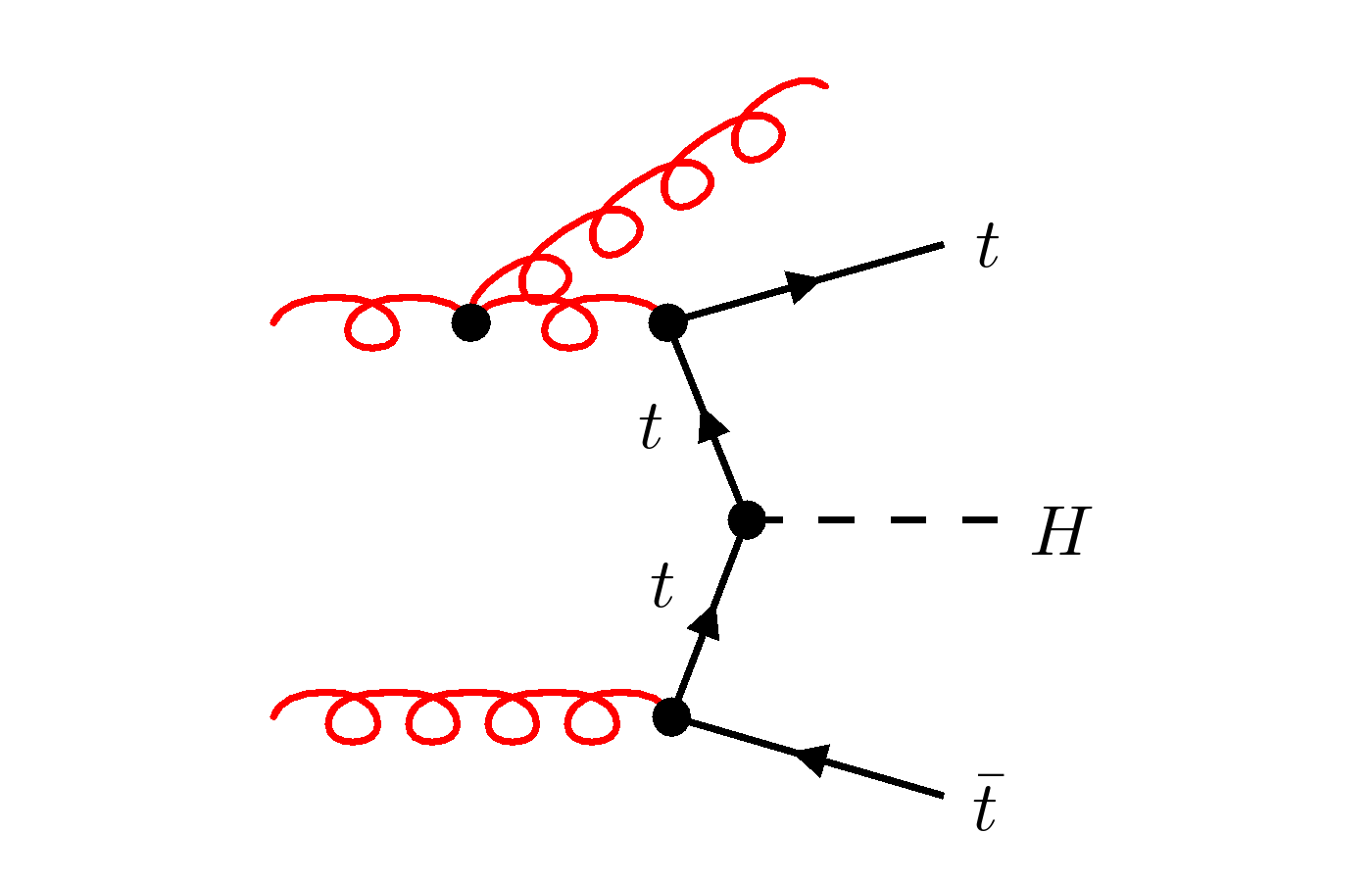}
    \caption{real emission}
  \end{subfigure}
  \hfill
  \begin{subfigure}{0.3\linewidth}
    \centering
    \includegraphics[width=\linewidth]{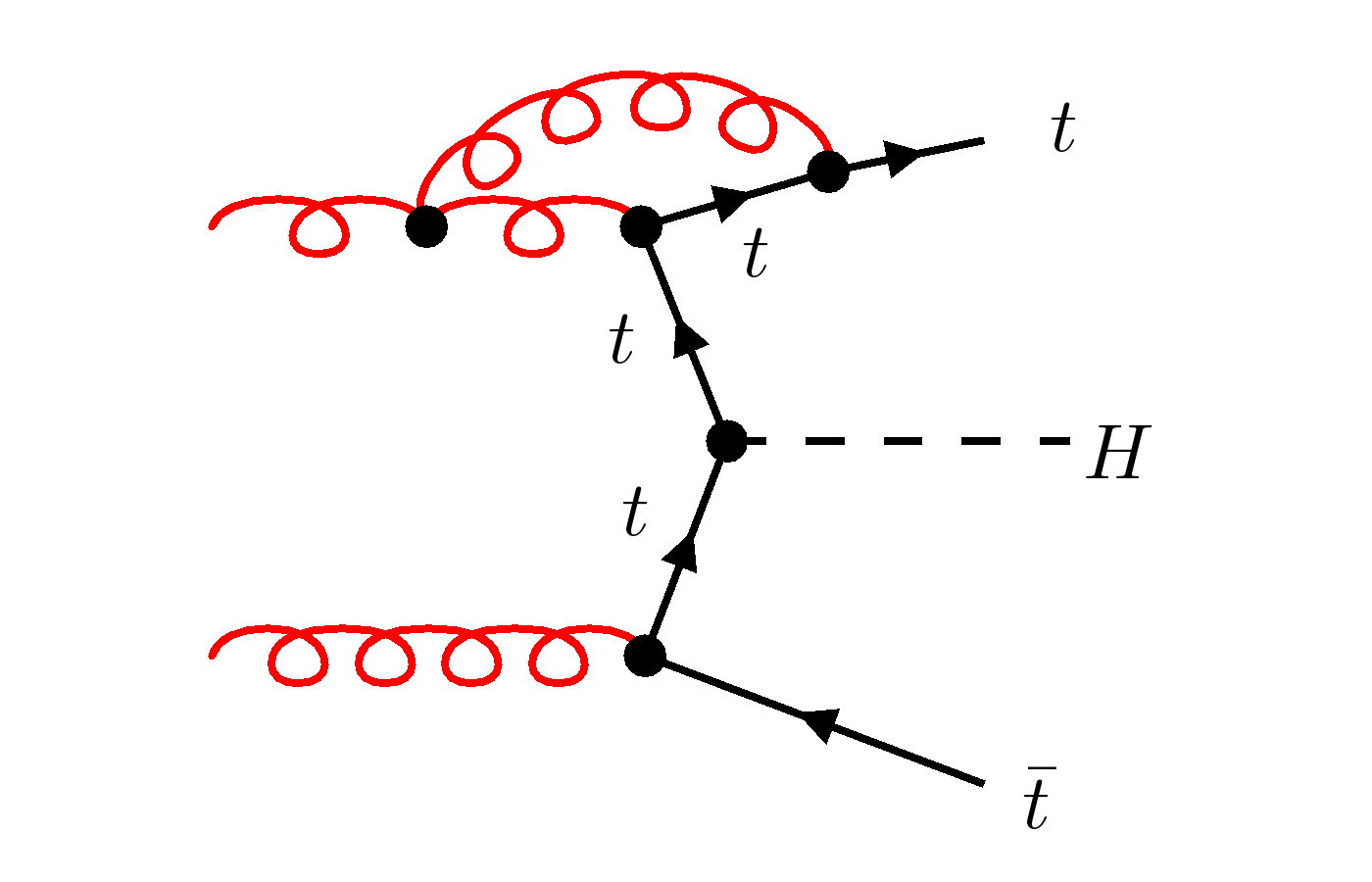}
    \caption{one loop}
  \end{subfigure}
  \caption{Representative Born (a), real-emission (b) and one-loop (c)
    diagrams for on-shell $pp\to\ttbar H$ production.}
  \label{fig:ttH Diags}
\end{figure}

In the case of $\ttbar H$ production, the LO proceeds at $\mathcal{O}(\alpha_s^2\alpha)$ through $gg$ fusion and
$q\bar q$ annihilation, with the Higgs boson emitted from the top-quark
line (see \reffi{fig:ttH Diags}).
At NLO, the real corrections involve also quark--gluon channels.
Since top quarks are external particles, this process does not involve
intermediate resonances, and no resonance histories are required. In the 5FNS $b$-initiated channels contribute, however, they are strongly suppressed at inclusive level.
 
\begin{figure}[tbp]
  \centering
  \begin{subfigure}{0.3\linewidth}
    \centering
    \includegraphics[width=\linewidth]{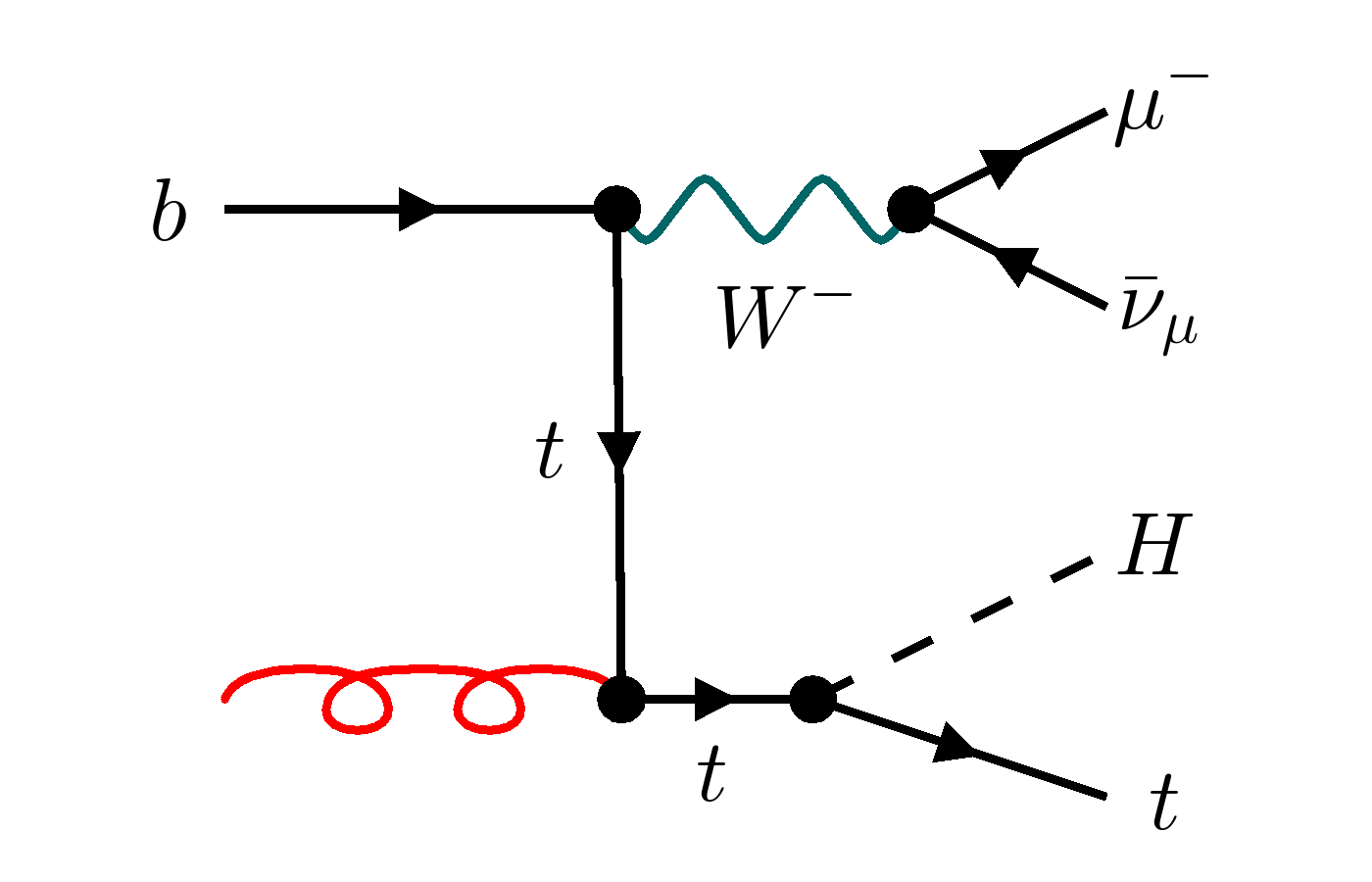}
    \caption{$tW^-H$}
  \end{subfigure}
  \hspace{0.05\linewidth}
  \begin{subfigure}{0.3\linewidth}
    \centering
    \includegraphics[width=\linewidth]{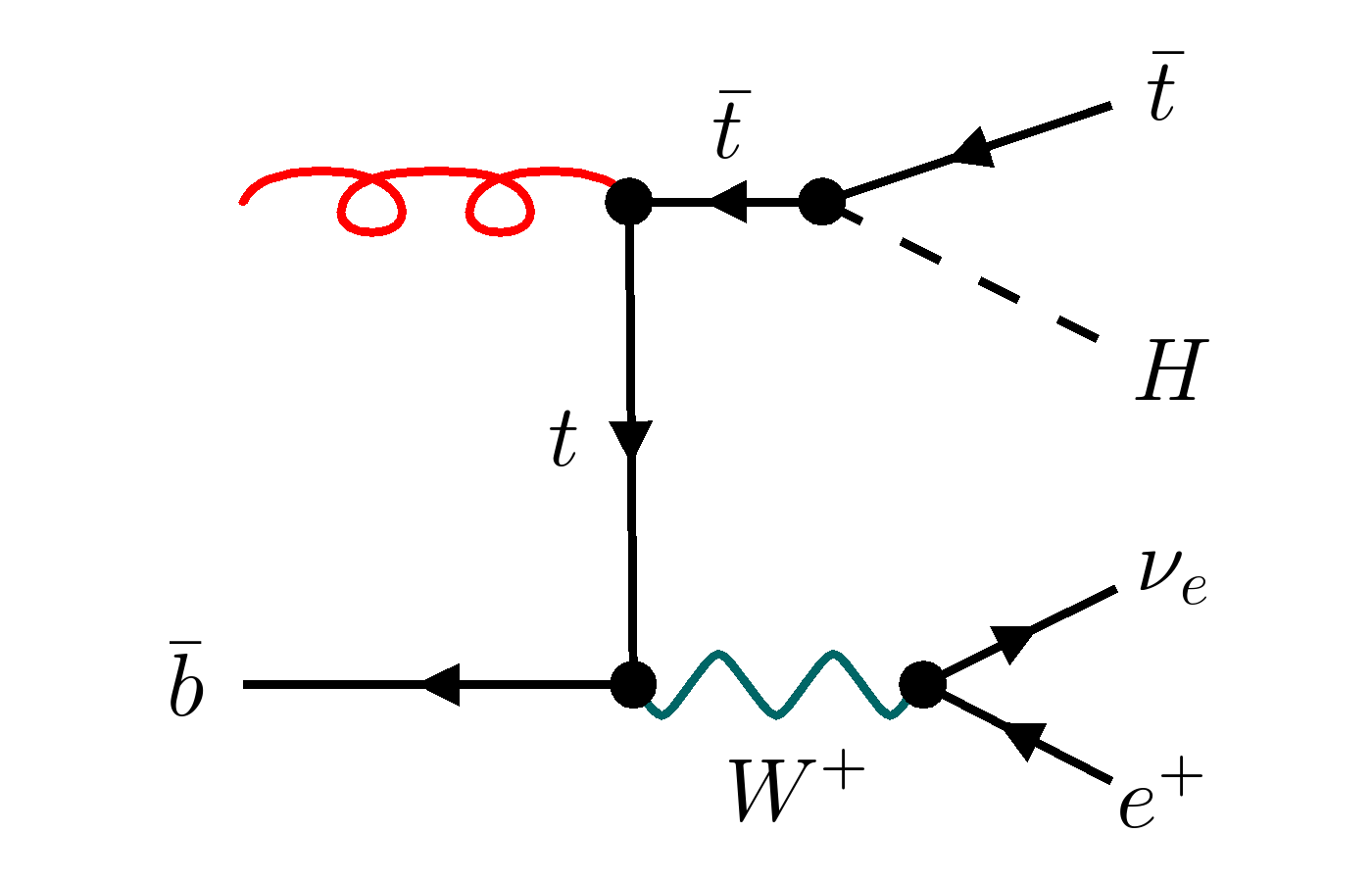}
    \caption{$\bar tW^+H$}
  \end{subfigure}
  \caption{Representative LO diagrams for on-shell $tWH$ production in
    the 5FNS with leptonic $W$-boson decays,
    $gb\to t\,\ell^-\bar\nu_\ell\,H$ (a) and
    $g\bar b\to\bar t\,\ell^+\nu_\ell\,H$ (b).}
  \label{fig:tWH LO Diags}
\end{figure}

The situation is more subtle in the case of $tWH$ production, which
proceeds through $gb\to tW^-H$ and $g\bar b\to\bar tW^+H$ at LO in the
5FNS (see \reffi{fig:tWH LO Diags}).
In our \tWH generator the $W$ boson that is produced in association with the top quark is
decayed leptonically at matrix-element level, i.e.\ we consider the processes\
$gb\to t\,\mu^-\bar\nu_\mu H$ at $\mathcal{O}(\alpha_s\alpha^3)$. The top quark
is kept on shell.
Similarly as in the off-shell process, the Higgs boson is emitted from
the top quark or from the $W$ boson, and the two contributions interfere
destructively in the SM.
 
\begin{figure}[tbp]
  \centering
  \begin{subfigure}{0.3\linewidth}
    \centering
    \includegraphics[width=1.1\linewidth]{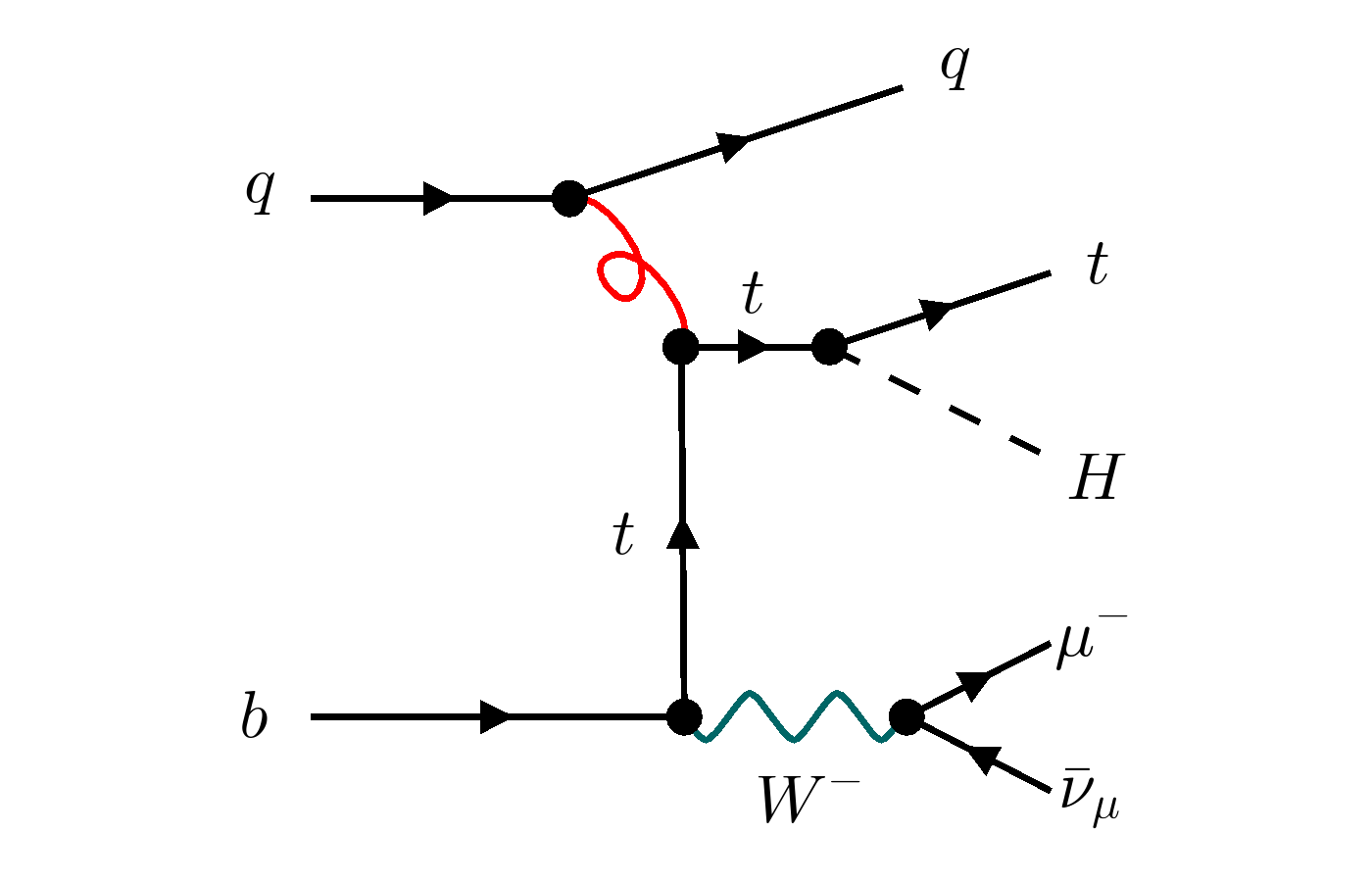}
    \caption{$\mathcal{A}_{1t}$}
  \end{subfigure}
  \hspace{0.05\linewidth}
  \begin{subfigure}{0.3\linewidth}
    \centering
    \includegraphics[width=\linewidth]{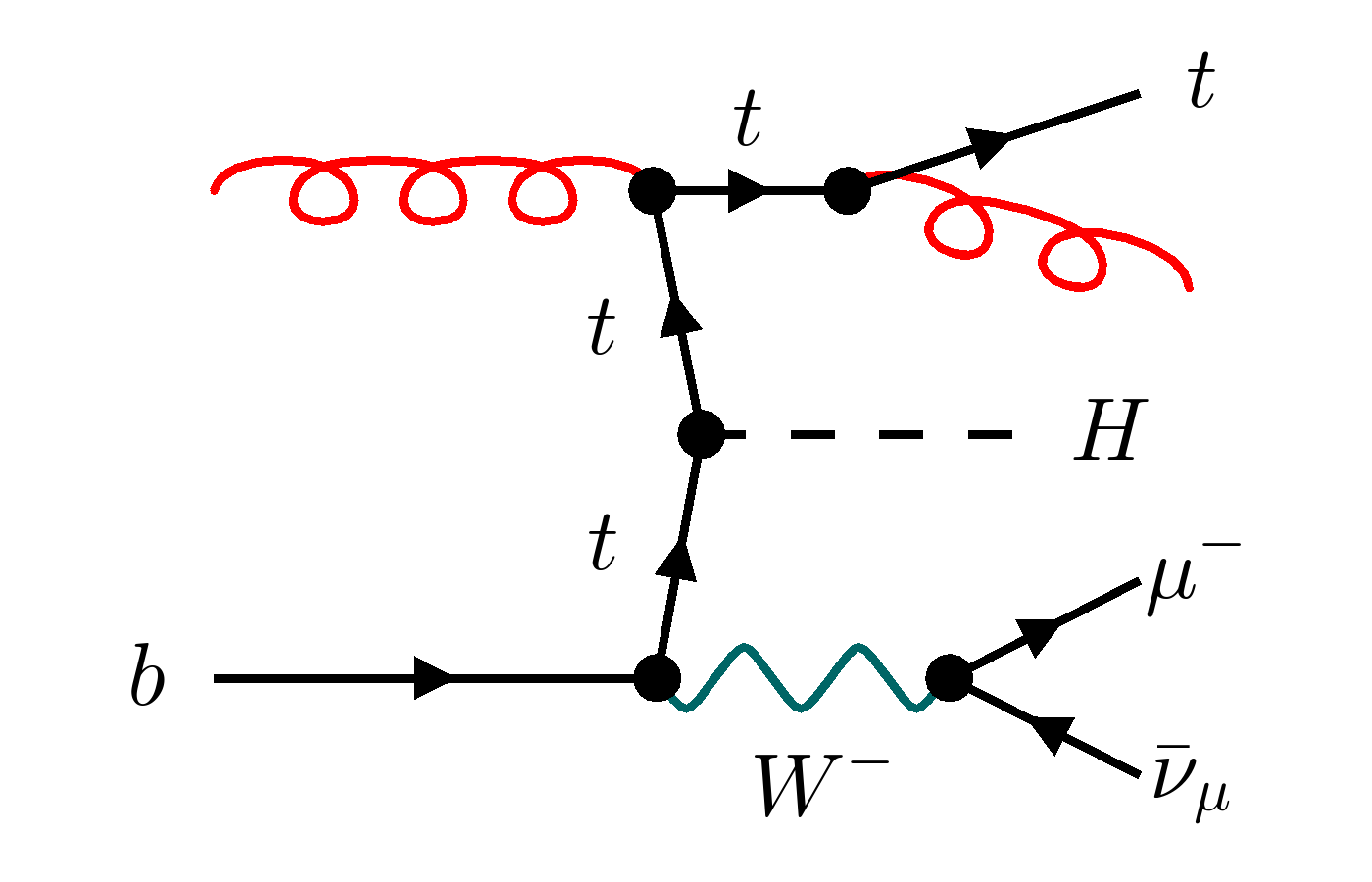}
    \caption{$\mathcal{A}_{1t}$}
  \end{subfigure}
 
  \vspace{1em}
 
  \begin{subfigure}{0.3\linewidth}
    \centering
    \includegraphics[width=\linewidth]{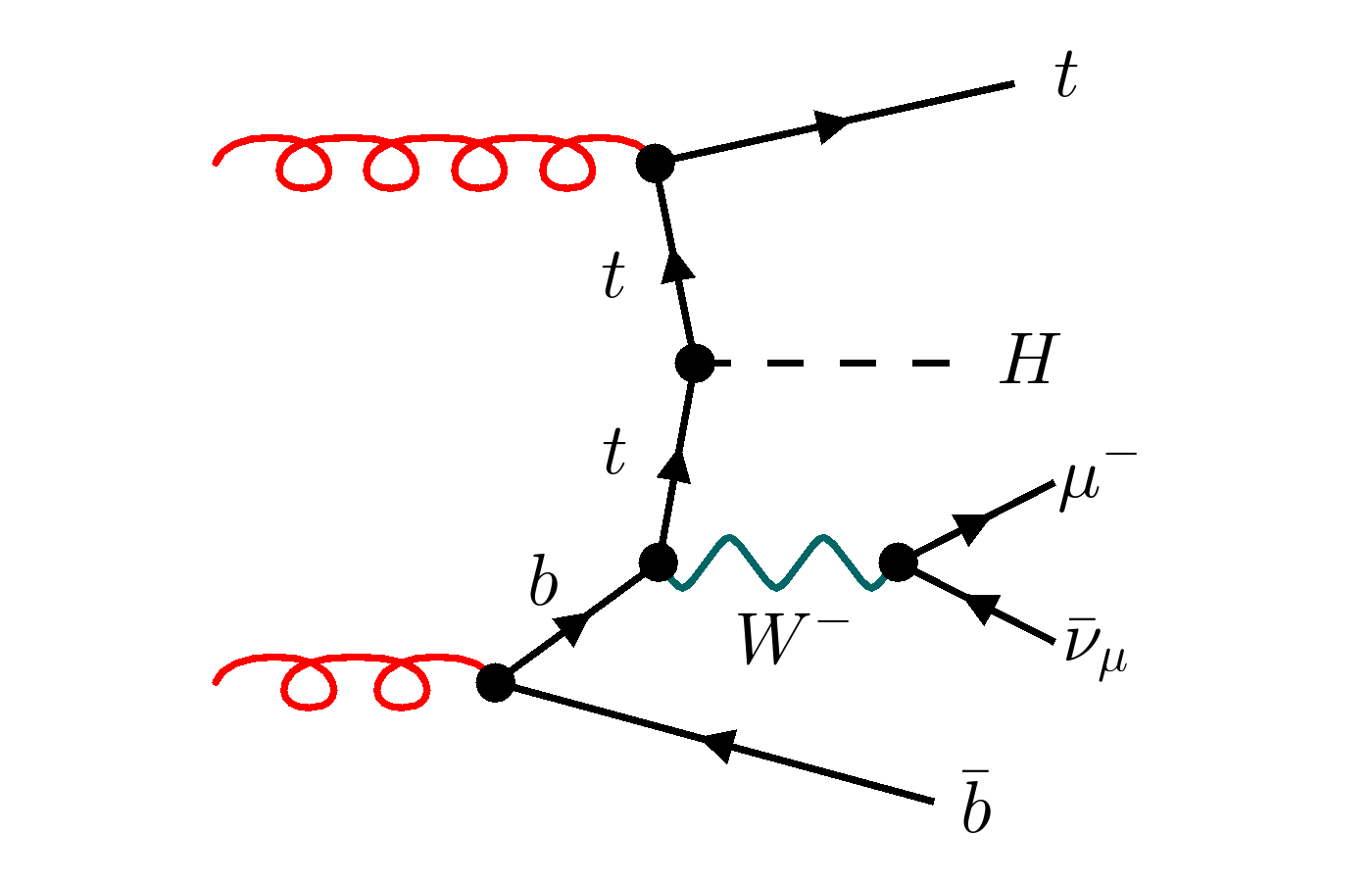}
    \caption{$\mathcal{A}_{1t}$}
  \end{subfigure}
  \hspace{0.05\linewidth}
  \begin{subfigure}{0.3\linewidth}
    \centering
    \includegraphics[width=\linewidth]{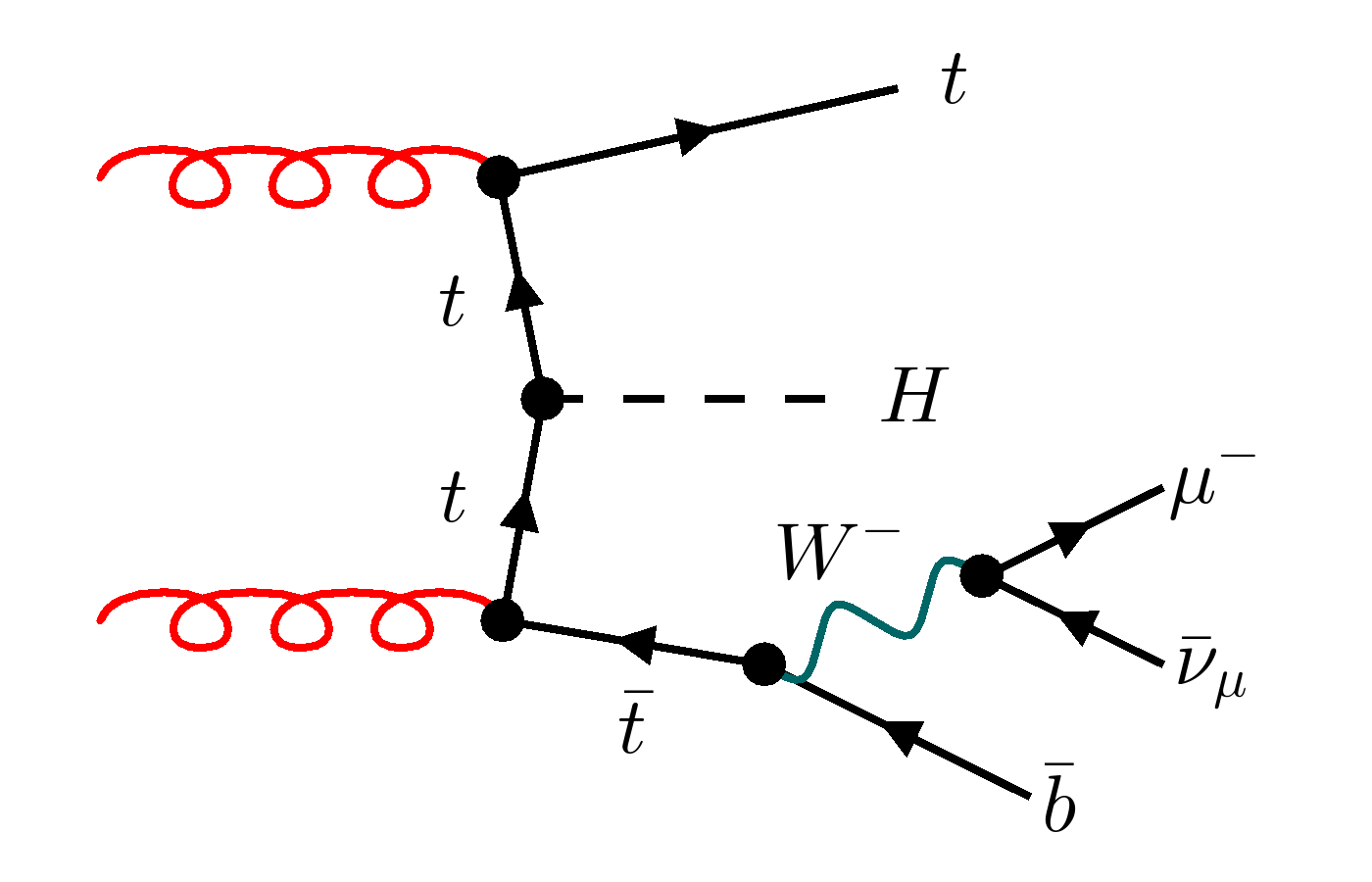}
    \caption{$\mathcal{A}_{2t}$}
  \end{subfigure}
  \caption{Representative real-emission diagrams for on-shell $tW^-H$
    production in the 5FNS, with $q\in\{u,d,c,s,b\}$. Diagrams (a)--(c)
    are genuine real corrections to $tWH$ production and contribute to
    $\mathcal{A}_{1t}$, while diagram (d) contains an intermediate
    anti-top quark that can become resonant and contributes to
    $\mathcal{A}_{2t}$.}
  \label{fig:Real tWH with res}
\end{figure}

Beyond LO, $tWH$ production cannot be defined in isolation.
The real corrections $gg\to tW^-\bar bH$ and $q\bar q\to tW^-\bar bH$
involve, besides genuine real-emission corrections to $tWH$ production
(\reffi{fig:Real tWH with res}a--c), also diagrams where the $W^-\bar b$
pair results from the decay of an intermediate anti-top quark
(\reffi{fig:Real tWH with res}d).
Such diagrams correspond to $\ttbar H$ production with subsequent
$\bar t\to W^-\bar b$ decay, and they are resonantly enhanced when the
$W^-\bar b$ invariant mass approaches $m_t$.
Upon integration over the resonance, they yield
$\sigma_{\ttbar H}\times\mathrm{BR}(\bar t\to W^-\bar b)$, which exceeds
the LO $tWH$ cross section by more than an order of magnitude.
Without an appropriate treatment, NLO corrections to $tWH$ production
would thus be dominated by $\ttbar H$ contributions. Since the two generators generally need to be combined to provide a realistic description of relevant observables, these contributions would be double counted.

Denoting by $\mathcal{A}_{1t}$ the contribution of diagrams that involve
only the external top quark, and by $\mathcal{A}_{2t}$ the contribution
of diagrams with an additional anti-top propagator that can become
resonant, the squared real-emission amplitude can be written
as
\begin{equation}
  \label{eq:tWHb amp}
  \big|\mathcal{A}_{tWHb}\big|^2
  = \big|\mathcal{A}_{1t}+\mathcal{A}_{2t}\big|^2
  = \big|\mathcal{A}_{1t}\big|^2
  + 2\,\mathrm{Re}\big(\mathcal{A}_{1t}^*\mathcal{A}_{2t}\big)
  + \big|\mathcal{A}_{2t}\big|^2\,.
\end{equation}
Various prescriptions for the removal of the $\ttbar H$ contribution from
\refeq{eq:tWHb amp} are available, and we refer to
\citere{Demartin:2016axk} for a detailed discussion.
%
%
%
%
Fully local such removal can be achieved, by means of diagram
removal (DR) or diagram subtraction (DS)~\cite{Frixione:2008yi}, as in
the \POWHEG{} generator for $tW$ production~\cite{Re:2010bp}.
 
For all results presented in this paper we adopt the DR1 scheme, where
the doubly-resonant diagrams are removed from the real-emission
amplitudes by setting $\mathcal{A}_{2t}=0$, i.e.\
\begin{equation}
  \label{eq:DR1 amp}
  \big|\mathcal{A}_{tWHb}\big|^2_{\rm DR1} = \big|\mathcal{A}_{1t}\big|^2\,.
\end{equation}
This removes the $\ttbar H$ contribution $|\mathcal{A}_{2t}|^2$, which is
described by the $\ttbar H$ generator, as well as the interference term,
which is in this scheme included in neither of the two on-shell generators.
The alternative DR2 scheme, which retains the interference term,
\begin{equation}
  \label{eq:DR2 amp}
  \big|\mathcal{A}_{tWHb}\big|^2_{\rm DR2} = \big|\mathcal{A}_{1t}\big|^2
  + 2\,\mathrm{Re}\big(\mathcal{A}_{1t}^*\mathcal{A}_{2t}\big)\,,
\end{equation}
is also available in the \tWH generator and can be used to estimate
uncertainties related to the treatment of interference effects in $tWH$
production.
As for any removal of diagrams, the DR1 and DR2 prescriptions formally violate
gauge invariance. For $tW$ and $tWH$ production, the resulting gauge
dependence was found to be numerically small in covariant
gauges~\cite{Frixione:2008yi,Demartin:2016axk}, such as the Feynman
gauge, which we also employ for this contribution in the \tWH generator.
%

\subsection{Top-quark decays in the on-shell generators}
\label{Sec:Top decay}

In order to compare on-shell generators and \bbfourlH{} at the level of
the physical final state, and with identical fiducial cuts, in the on-shell generators top quarks
are decayed at the level of the LHEs, i.e.\ before the parton shower.
In the dileptonic channel the decay chains of the \ttH and \tWH generators read respectively,
\begin{align}
  \label{eq:decay-chain-ttH}
  \ttbar H &\;\to\; \big(b\,e^+\nu_e\big)\big(\bar b\,\mu^-\bar\nu_\mu\big)\,H\,,\\
  \label{eq:decay-chain-tWH}
  t\,\mu^-\bar\nu_\mu\,H &\;\to\; \big(b\,e^+\nu_e\big)\,\mu^-\bar\nu_\mu\,H\,,
\end{align}
and similarly for $\bar t\,e^+\nu_e\,H$, where top quarks decay via an
intermediate $W$ boson.
Thus, at LO the decayed $tWH$ final state involves a single $b$ quark,
while the second $b$ quark of \refeq{eq:Process} arises only through
real emission or through initial-state parton-shower evolution.
 
Top-quark decays are generated with the method of
\citere{Frixione:2007zp}, as implemented in \POWHEG{} for
$t\bar t$ production~\cite{Frixione:2007nw},
single-top production~\cite{Alioli:2009je,Re:2010bp},
$\ttbar j$ production~\cite{Alioli:2011as} and 
$\ttbar b\bar b$ production~\cite{Jezo:2018yaf}.
This approach retains spin correlations between production and decay at
tree level, since the decay kinematics are distributed according to the
tree-level matrix element of the decayed process, without the need for
explicit spin-density matrices.
Moreover, finite-width effects are included in an approximate way, through
Breit--Wigner distributions of the top-quark and $W$-boson virtualities.
Finally, top-quark decays are described at LO, such that QCD radiation
off the decay products is generated only by the parton shower.
Non-resonant contributions, interferences between different resonant
structures and NLO corrections to top-quark decays are thus absent by
construction. These are precisely the effects that are captured by
\bbfourlH{}.
 
Starting from the undecayed kinematics $\Phi_{\rm undec}$ of a \POWHEG{}
event, the top-quark virtualities $M^2$ are generated through the
replacement
\begin{equation}
  \label{eq:top-BR}
  \delta(M^2-m_t^2)\;\to\;\frac{1}{\pi}\,
  \frac{m_t\Gamma_t}{(M^2-m_t^2)^2+m_t^2\Gamma_t^2}
\end{equation}
in the on-shell phase space, and the momenta of the undecayed event are
reshuffled accordingly.
The $W$-boson virtualities are generated in the same way.
In the following we describe the procedure for $\ttbar H$ production,
while for $tWH$ production it is applied to the single top quark.
The decayed phase space factorises as
\begin{equation}
  \label{eq:tdec psp}
  d\Phi_{\rm dec} = d\Phi_{\rm undec}\,d\Phi_{t\to b\ell\nu}\,d\Phi_{\bar t\to\bar b\ell'\nu'}\,,
\end{equation}
where $\Phi_{t\to b\ell\nu}$ and $\Phi_{\bar t\to\bar b\ell'\nu'}$
denote the phase spaces of the top-quark decay products.
 
The decay kinematics are distributed according to the ratio of the
squared tree-level matrix elements of the decayed and undecayed
processes, $\mathcal{M}_{\rm dec}$ and $\mathcal{M}_{\rm undec}$, which
depend on the type of event.
For LHEs without \POWHEG{} emission, these are the Born matrix elements
of $pp\to\ttbar H$ and of the corresponding decayed process.
For LHEs with a \POWHEG{} emission, $\mathcal{M}_{\rm undec}$ is the
real-emission matrix element of $pp\to\ttbar H+j$, where $j$ denotes the
emitted parton, and $\mathcal{M}_{\rm dec}$ is the real-emission matrix
element of the decayed process,
$pp\to b\bar b\,e^+\nu_e\,\mu^-\bar\nu_\mu\,H+j$, evaluated at the same
production kinematics.
The latter is computed in the pole approximation retaining only Feynman diagrams with
the relevant number of resonant top-quark propagators, i.e.\ in a double pole approximation
(DPA) for $ttH$ production and a single pole approximation
(SPA) for $tWH$ production.
In this way, $\mathcal{M}_{\rm dec}$ describes the decays of the top
quarks of the undecayed event, including spin correlations in the
presence of the extra emission.
In the NWA, the decayed and undecayed matrix elements are related by
\begin{equation}
  \label{eq:undec->dec amp}
  \mathcal{M}_{\rm undec}\,\mathrm{BR}(t\to b\ell\nu)\,\mathrm{BR}(\bar t\to\bar b\ell'\nu')
  = \int \mathcal{M}_{\rm dec}\,d\Phi_{t\to b\ell\nu}\,d\Phi_{\bar t\to\bar b\ell'\nu'}\,,
\end{equation}
where $\mathrm{BR}(t\to b\ell\nu)$ and $\mathrm{BR}(\bar t\to\bar b\ell'\nu')$
are the leptonic top-quark branching ratios.
The decay kinematics are thus generated with the conditional probability
\begin{equation}
  \label{eq:decay probability}
  \begin{split}
  dP\big(\Phi_{\rm dec}\,\big|\,\Phi_{\rm undec}\big)
  = {}& \frac{1}{\mathrm{BR}(t\to b\ell\nu)\,\mathrm{BR}(\bar t\to\bar b\ell'\nu')}\,
  \frac{\mathcal{M}_{\rm dec}(\Phi_{\rm dec})}{\mathcal{M}_{\rm undec}(\Phi_{\rm undec})}\\
  &\times d\Phi_{t\to b\ell\nu}\,d\Phi_{\bar t\to\bar b\ell'\nu'}\,,
  \end{split}
\end{equation}
which is normalised to unity as a result of \refeq{eq:undec->dec amp}.
In this way the weight of the undecayed event is left unchanged, while
the dependence on the decay kinematics, including spin correlations, is
entirely encoded in the ratio $\mathcal{M}_{\rm dec}/\mathcal{M}_{\rm undec}$.
 
The conditional probability in \refeq{eq:decay probability} is sampled
with a hit-or-miss algorithm based on the upper-bounding
function~\cite{Alioli:2011as}
\begin{equation}
  \label{eq:Udec}
  U_{\rm dec}(\Phi_{\rm dec}) = \mathcal{N}\,
  \frac{\mathcal{M}_{t\to bW}\big(M_t^2,M_{\ell\nu}^2\big)}{\big(M_t^2-m_t^2\big)^2+m_t^2\Gamma_t^2}\,
  \frac{\mathcal{M}_{W\to\ell\nu}\big(M_{\ell\nu}^2\big)}{\big(M_{\ell\nu}^2-m_W^2\big)^2+m_W^2\Gamma_W^2}
  \;\times\;(t\leftrightarrow\bar t)\,,
\end{equation}
where $M_t^2$ and $M_{\ell\nu}^2$ are the top-quark and $W$-boson
virtualities, $\mathcal{M}_{t\to bW}$ and $\mathcal{M}_{W\to\ell\nu}$ are
the squared matrix elements of the corresponding $1\to2$ decays, and
$\mathcal{N}$ is a normalisation factor.
Trial decay kinematics are generated from $\Phi_{\rm undec}$ with a
density proportional to $U_{\rm dec}$, and a random number $r$ is drawn
uniformly in $[0,U_{\rm dec}(\Phi_{\rm dec})]$.
The trial configuration is accepted if
$r<\mathcal{M}_{\rm dec}/\mathcal{M}_{\rm undec}$ and regenerated
otherwise. This requires
\begin{equation}
  \label{eq:upper bound inequality}
  \mathcal{M}_{\rm dec}(\Phi_{\rm dec})\le
  \mathcal{M}_{\rm undec}(\Phi_{\rm undec})\,U_{\rm dec}(\Phi_{\rm dec})\,,
\end{equation}
and the normalisation $\mathcal{N}$ is determined adaptively, i.e.\ it is
increased whenever the condition in \refeq{eq:upper bound inequality} is
violated.
Since a larger value of $\mathcal{N}$ reduces the acceptance rate, the
efficiency of the decay step, and thus the event-generation time,
depends on how tightly $U_{\rm dec}$ bounds the ratio of decayed and
undecayed matrix elements. The actual NLO event weight remains unchanged.
 
Finally, for LHEs with a \POWHEG{} emission the DPA matrix element
$\mathcal{M}_{\rm dec}$ involves also diagrams where the emitted gluon is
attached to a $b$ quark from a top-quark decay.
For massless $b$ quarks, $\mathcal{M}_{\rm dec}$ becomes singular in the
limit where the two partons are collinear, and no finite upper bound
exists.
For this reason, $\mathcal{M}_{\rm dec}$ is evaluated with a finite
$b$-quark mass, as specified in \refse{sec:setup}, which regularises this
singularity.
As a consequence, in the generated LHE files, the final state $b$ quarks originating from top decay are massive.
Consistently with this, and with massive splitting kernels in \PythiaEight{}, also the momenta of $b$ quarks originating in the production process are reshuffled such that they are on-shell with the same finite $b$-quark mass.

\section{Resonance-aware \POWHEG matching} 
\label{Sec:POWHEG-res}

The matching of \bbfourlH{} to parton showers is based on the
resonance-aware \POWHEG{} method~\cite{Jezo:2015aia}, as implemented in
the \bbfourl{} generator for off-shell $\ttbar+tW$
production~\cite{Jezo:2016ujg}, and includes the methodological
improvements introduced in \citere{Jezo:2023rht}.
In \refse{sec:powheg-res-summary} we briefly summarise the method, restricting
ourselves to the ingredients and the notation needed in the following.
For derivations and a detailed discussion we refer to
Refs.~\cite{Jezo:2015aia,Jezo:2023rht}.
%

\subsection{Summary of the \POWHEGRES{} method}
\label{sec:powheg-res-summary}
 
In the \POWHEG{} method~\cite{Nason:2004rx,Frixione:2007nw,Alioli:2010xd},
events are generated starting from Born configurations $\Phi_{\rm B}$ with
NLO weights
\begin{equation}
  \label{eq:POWHEG-Born weight}
  \bar{B}(\Phi_{\rm B}) = B(\Phi_{\rm B}) + V(\Phi_{\rm B})
  + \sum_{c\in\mathcal{C}}\int R_c(\Phi_{{\rm R},c})\,{\rm d}\Phi_{{\rm rad},c}\,,
\end{equation}
where $B$ and $V$ denote the Born and virtual contributions. The
real-emission weight is split into collinear sectors, $R=\sum_c R_c$ with
$R_c=\omega_c R$, where the projectors $\omega_c$ isolate the singularity
associated with a specific collinear splitting $c\in\mathcal{C}$. In each
sector, the real-emission phase space is constructed from $\Phi_{\rm B}$
and the radiation variables $\Phi_{{\rm rad},c}$ through an FKS mapping
$\Phi_{{\rm R},c}(\Phi_{\rm B},\Phi_{{\rm rad},c})$. The hardest emission
is then generated according to
\begin{equation}
  \label{eq:POWHEG-xsec}
  {\rm d}\sigma = \bar{B}(\Phi_{\rm B})\,{\rm d}\Phi_{\rm B}
  \left[\Delta(q_{\rm cut})
  + \sum_{c\in\mathcal{C}}\Delta(k_{{\rm T},c})\,
  \frac{R_c(\Phi_{{\rm R},c})}{B(\Phi_{\rm B})}\,{\rm d}\Phi_{{\rm rad},c}\right],
\end{equation}
with the Sudakov form factor
\begin{equation}
  \label{eq:Sudakov Form Factor}
  \Delta(q_{\rm T}) = \exp\left[-\sum_{c\in\mathcal{C}}\int
  \theta\big(k_{{\rm T},c}-q_{\rm T}\big)\,
  \frac{R_c(\Phi_{{\rm R},c})}{B(\Phi_{\rm B})}\,{\rm d}\Phi_{{\rm rad},c}\right],
\end{equation}
where $k_{{\rm T},c}$ is the transverse momentum of the splitting and
$q_{\rm cut}$ an infrared cutoff.
 
The emission probability $R_c/B$ is well behaved only if the mappings are
consistent with collinear factorisation, i.e.\ if
$R_c(\Phi_{{\rm R},c})\to B(\Phi_{\rm B})\,P_c(z)/k_{{\rm T},c}^2$ for
small $k_{{\rm T},c}$. In the presence of an intermediate unstable particle
this requirement can be violated.
In fact, for top quarks 
this condition is violated by all but the softest emissions, which results in
distorted resonance line shapes and an unphysical radiation pattern.
 
The \POWHEGRES{} method~\cite{Jezo:2015aia} avoids this problem by means of
mappings that preserve all resonance virtualities. 
Since the mappings depend on the resonance structure, events are assigned
to \emph{resonance histories} $h\in\mathcal{H}$, i.e.\ to specific
combinations of production and decay subprocesses. At Born level this is
done via
\begin{equation}
  \label{eq:Born res history}
  B(\Phi_{\rm B}) = \sum_{h\in\mathcal{H}}B_h(\Phi_{\rm B})\,,
  \qquad
  B_h(\Phi_{\rm B}) = \omega^{(\text{hist})}_h(\Phi_{\rm B})\,B(\Phi_{\rm B})\,,
\end{equation}
and analogously for $\bar{B}(\Phi_{\rm B})$, with projectors
\begin{equation}
  \label{eq:Born res proj}
  \omega^{(\text{hist})}_h(\Phi_{\rm B}) =
  \rho^{(\text{hist})}_h(\Phi_{\rm B})
  \left[\sum_{h'\in\mathcal{H}}\rho^{(\text{hist})}_{h'}(\Phi_{\rm B})\right]^{-1}.
\end{equation}
The resonance weights $\rho^{(\text{hist})}_h$ should reflect the relative
probabilities of the different histories and, in the zero-width limit,
reduce to
\begin{equation}
  \label{eq:Born res limit}
  \lim_{\Gamma_r\to0}\rho^{(\text{hist})}_h(\Phi_{\rm B})
  \propto\prod_{r\in\mathcal{R}(h)}\delta\big(q_{{\rm B},r}^2-M_r^2\big)\,,
\end{equation}
where $\mathcal{R}(h)$ is the set of resonances in the history $h$, such
that the NWA is recovered.
The real-emission weight is split into histories and collinear sectors,
$R=\sum_{h}\sum_{c\in\mathcal{C}(h)}R_{h,c}$ with
$R_{h,c}=\omega_{h,c}R$ and
\begin{equation}
  \label{eq:real res proj}
  \begin{split}
  \omega_{h,c}(\Phi_{\rm R}) &= \rho_{h,c}(\Phi_{\rm R})
  \left[\sum_{h'\in\mathcal{H}}\sum_{c'\in\mathcal{C}(h')}
  \rho_{h',c'}(\Phi_{\rm R})\right]^{-1},\\
  \rho_{h,c}(\Phi_{\rm R}) &= \rho^{(\text{hist})}_{h,c}(\Phi_{\rm R})\,
  \rho^{(\text{coll})}_c(\widetilde\Phi_{{\rm rad},c})\,.
  \end{split}
\end{equation}
Here the sectors $c\in\mathcal{C}(h)$ correspond to the collinear
splittings in the production and decay subprocesses of $h$, and
$\rho^{(\text{coll})}_c$ are the standard collinear
weights~\cite{Frixione:2007nw}, evaluated in the rest frame of the
decaying resonance for splittings within a decay.
%
 
In the narrow-width limit, QCD radiation factorises into independent
contributions from the production and decay subprocesses. In the
\texttt{allrad} mode~\cite{Campbell:2014kua,Jezo:2015aia}, which we use
throughout, this is exploited by generating one \POWHEG{} emission per
subprocess $s\in\mathcal{S}(h)$ of the history $h$.
\begin{equation}
  \label{eq:POWHEG-RES allrad xsec}
  {\rm d}\sigma = \sum_{h\in\mathcal{H}}\bar{B}_h(\Phi_{\rm B})\,{\rm d}\Phi_{\rm B}
  \prod_{s\in\mathcal{S}(h)}\left[\Delta_s(q_{\rm cut})
  + \sum_{c\in\mathcal{C}(s)}\Delta_s(k_{{\rm T},c})\,
  \frac{R_{h,c}(\Phi_{{\rm R},c})}{B_h(\Phi_{\rm B})}\,{\rm d}\Phi_{{\rm rad},c}\right]
\end{equation}
$\Delta_s$ is defined as in \eqref{eq:Sudakov Form Factor}, but
restricted to the sectors $\mathcal{C}(s)\subset\mathcal{C}(h)$ of the
subprocess $s$ and with emission probabilities $R_{h,c}/B_h$.
As usual, the real-emission weight is split into a singular and a hard
part by means of the \texttt{hdamp} parameter. Eq.~\eqref{eq:POWHEG-RES allrad xsec}
applies to the singular part, while the hard remnant is generated as a
single fixed-order emission~\cite{Jezo:2023rht}.
In the subsequent parton shower, the final-state partons of each event are
assigned to the production and decay subprocesses of its history, the
shower preserves the resonance virtualities, and shower emissions within
each subprocess are vetoed against the hardness of the corresponding
\POWHEG{} emission~\cite{Jezo:2015aia,Jezo:2023rht}.
 
\subsection{Resonance histories}
\label{Sec:res histories}

\begin{figure}[tbp]
  \centering
  \begin{subfigure}{0.3\linewidth}
    \centering
    \includegraphics[width=\linewidth]{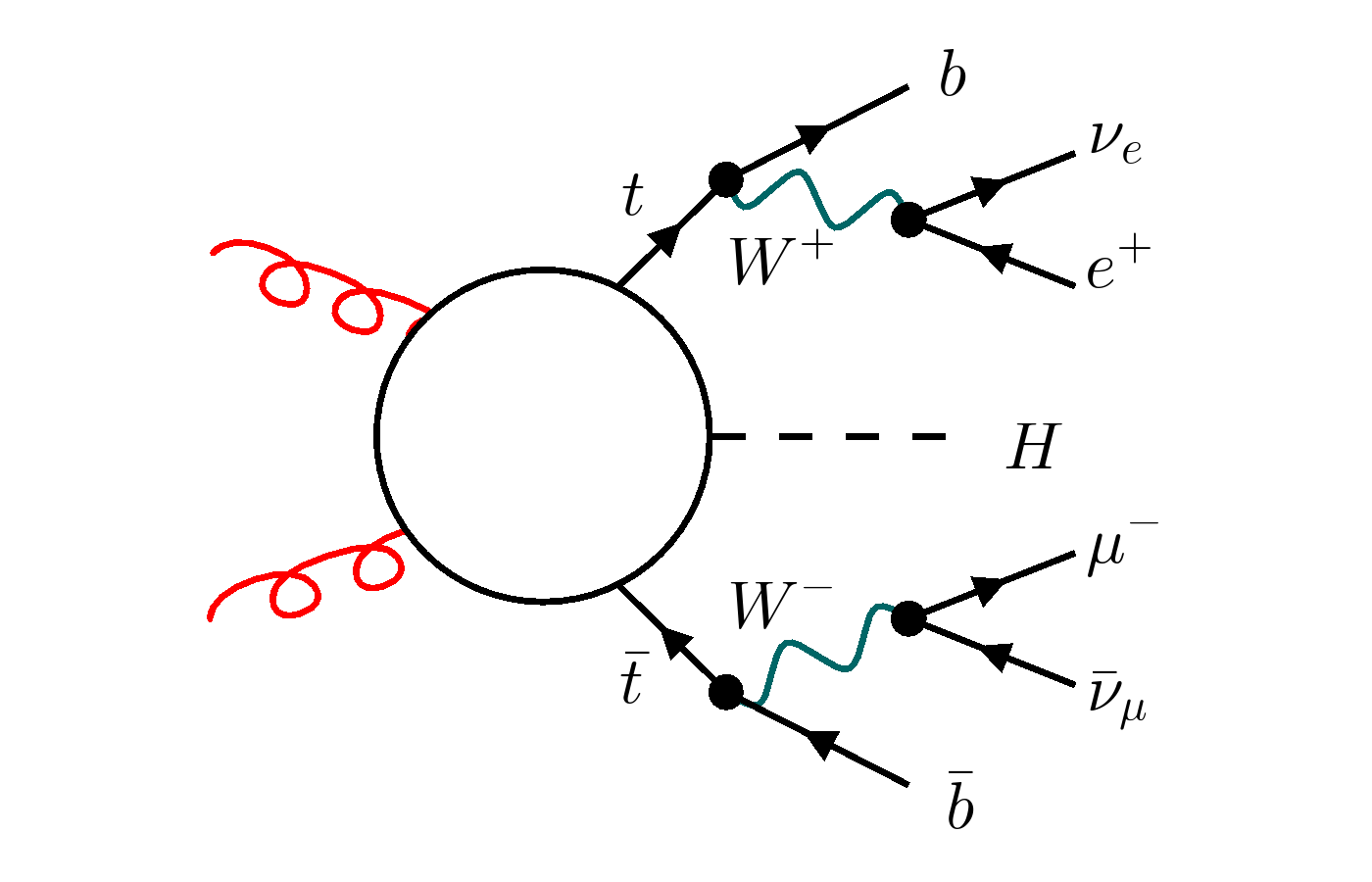}
    \caption{$\ttbar H$ Born}
  \end{subfigure}
  \hfill
  \begin{subfigure}{0.3\linewidth}
    \centering
    \includegraphics[width=1.1\linewidth]{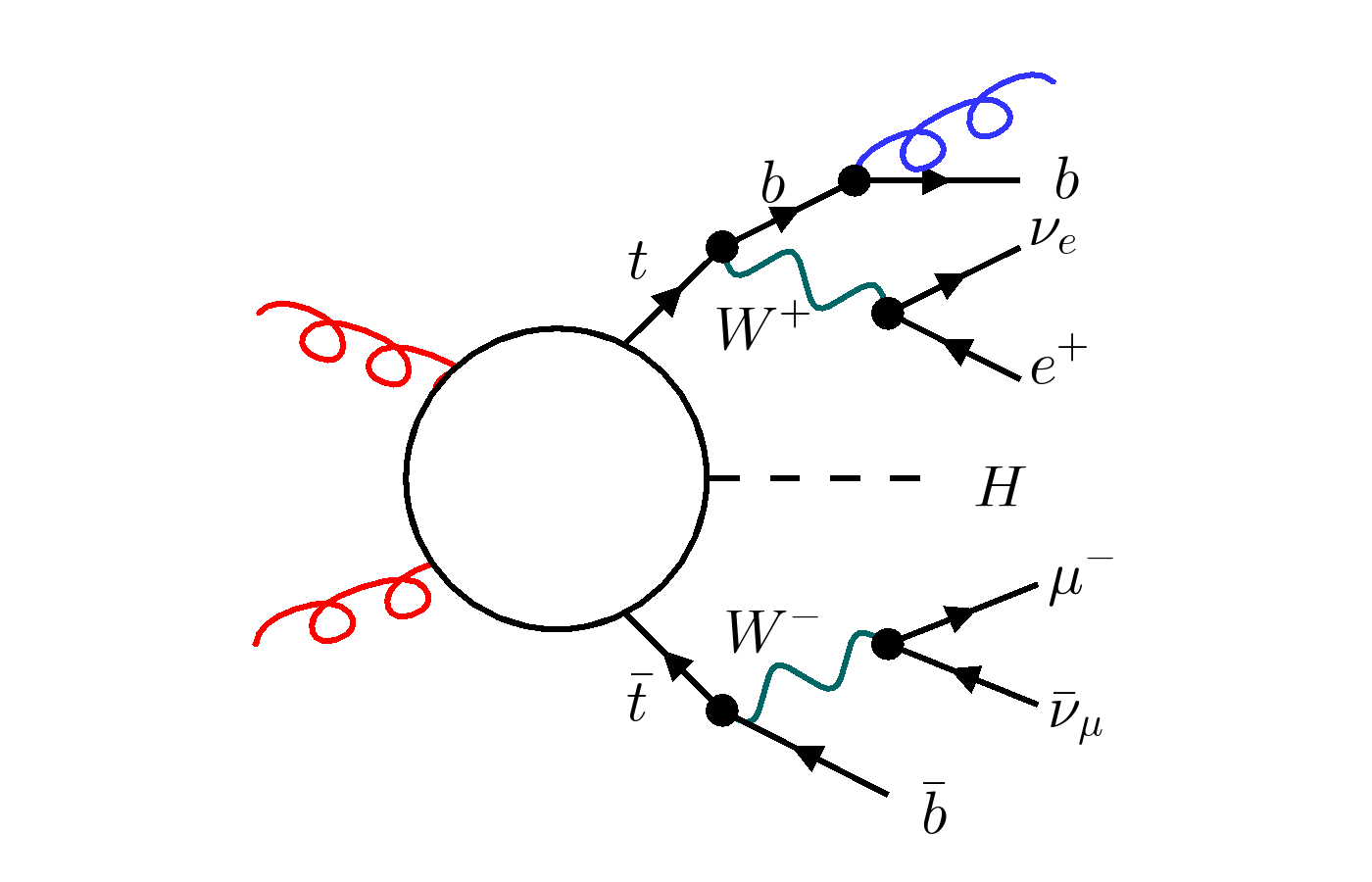}
    \caption{$\ttbar H$ real}
  \end{subfigure}
  \hfill
  \begin{subfigure}{0.3\linewidth}
    \centering
    \includegraphics[width=1.2\linewidth]{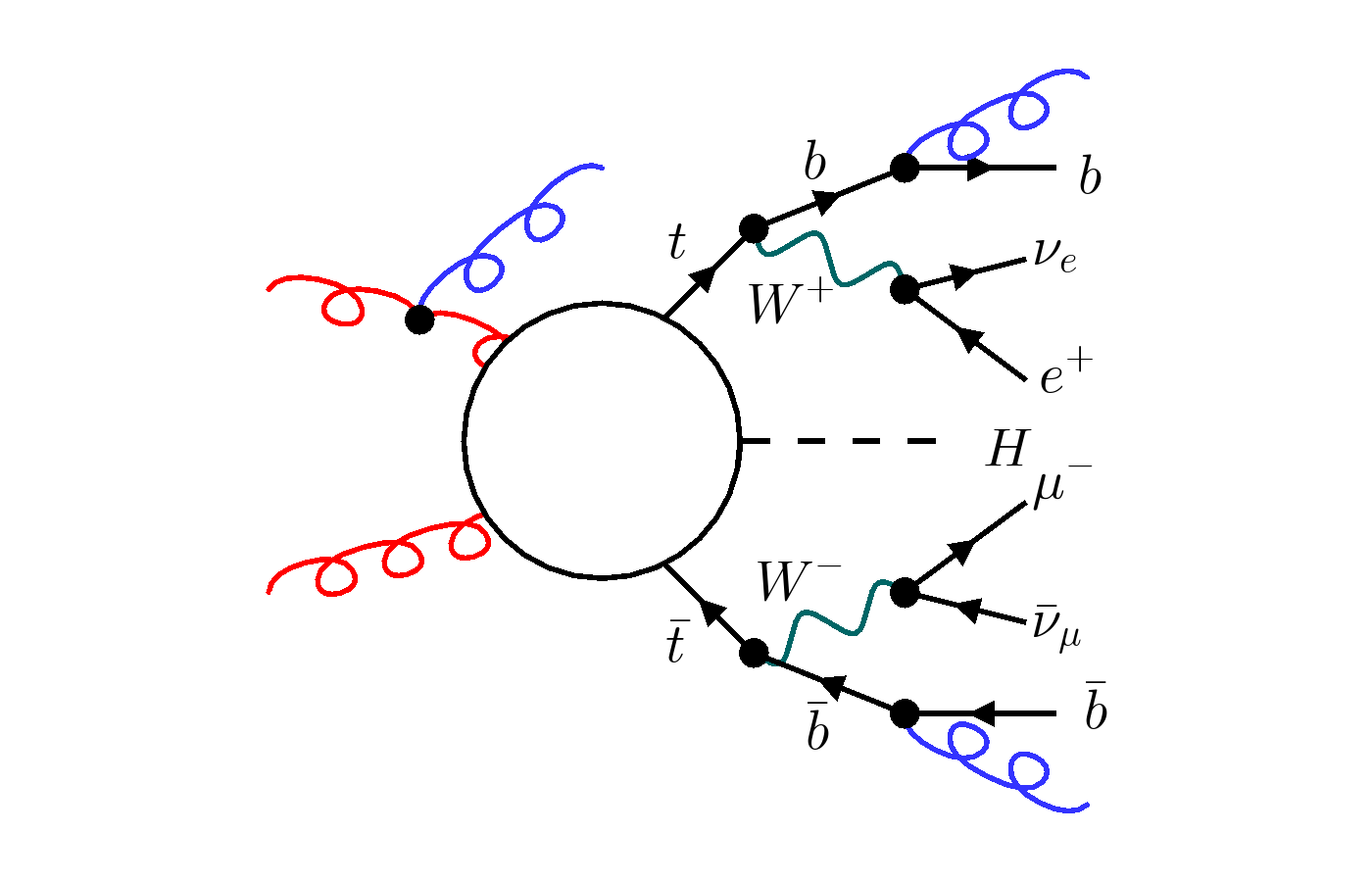}
    \caption{$\ttbar H$ real, \texttt{allrad}}
  \end{subfigure}
 
  \vspace{1em}
 
  \begin{subfigure}{0.3\linewidth}
    \centering
    \includegraphics[width=\linewidth]{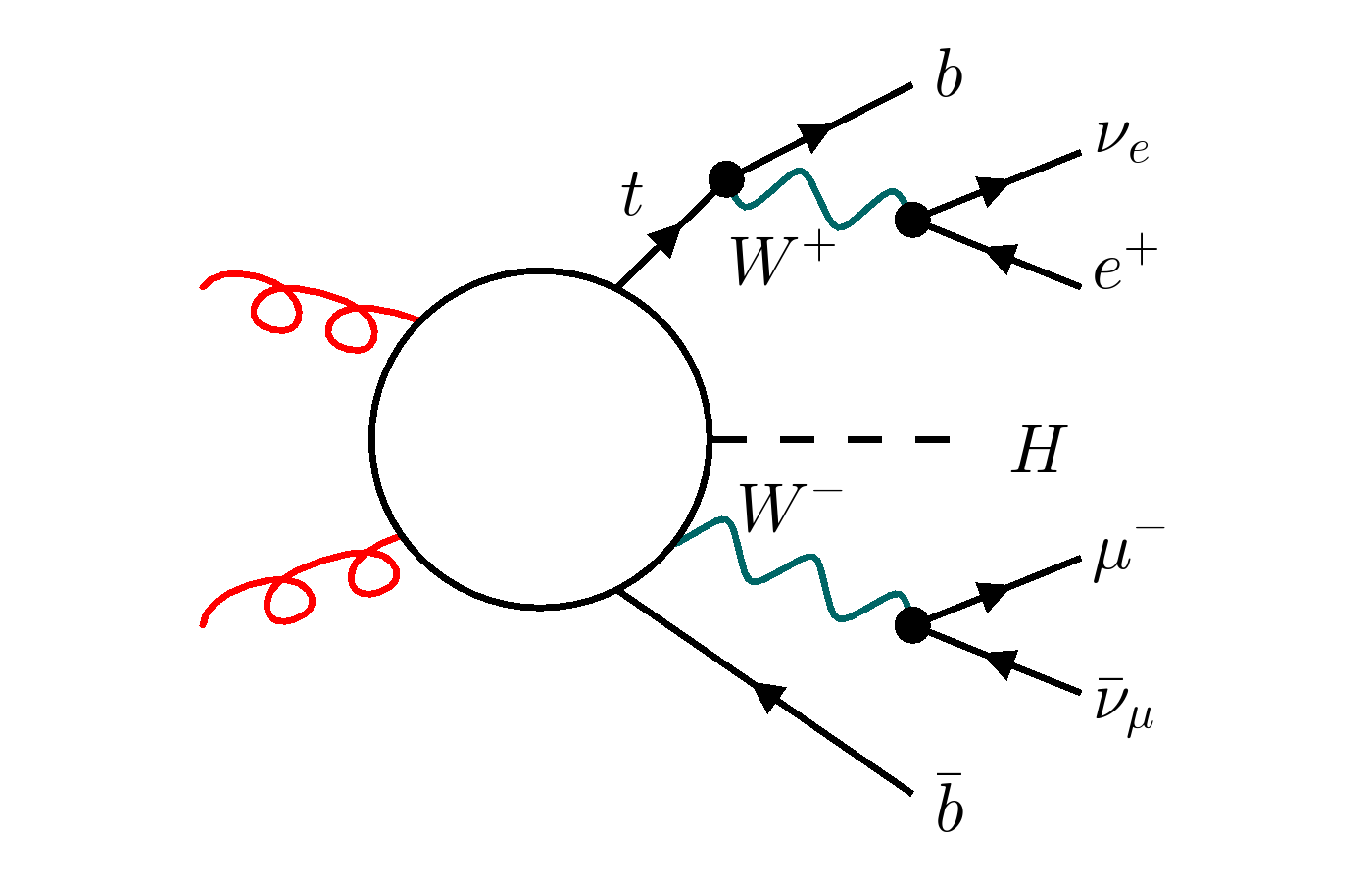}
    \caption{$tW^-H$ Born}
  \end{subfigure}
  \hfill
  \begin{subfigure}{0.3\linewidth}
    \centering
    \includegraphics[width=1.15\linewidth]{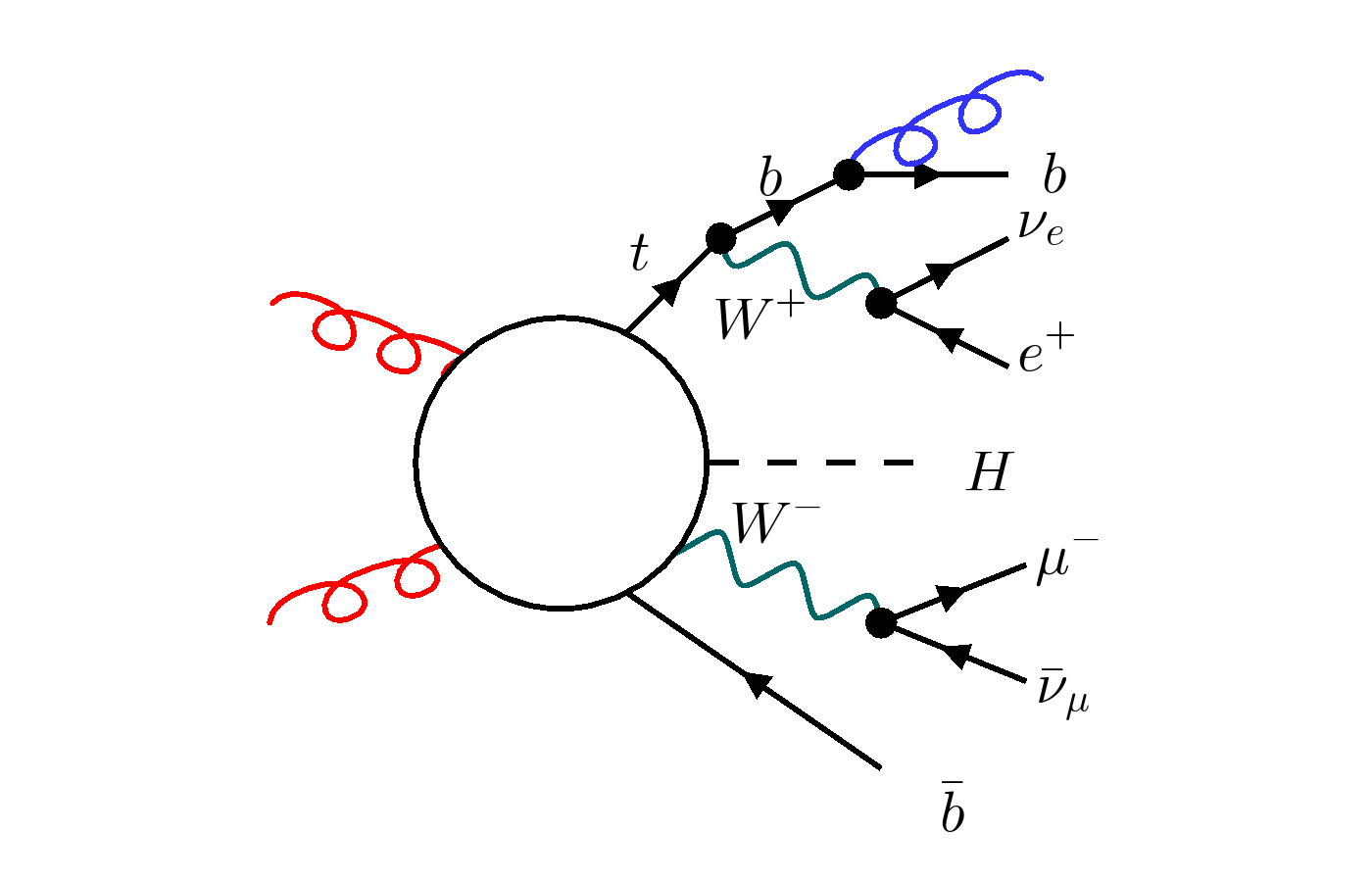}
    \caption{$tW^-H$ real}
  \end{subfigure}
  \hfill
  \begin{subfigure}{0.3\linewidth}
    \centering
    \includegraphics[width=1.1\linewidth]{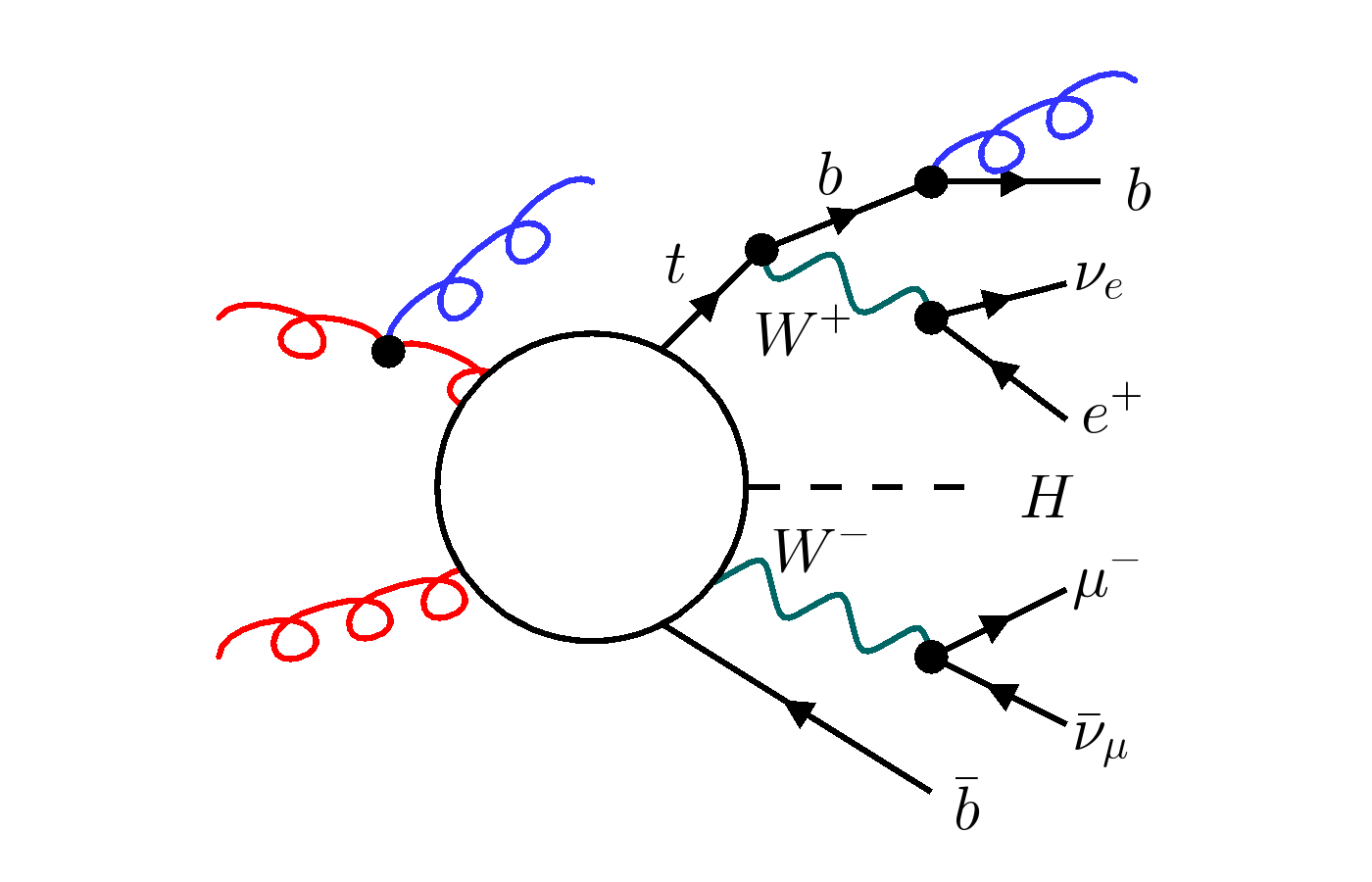}
    \caption{$tW^-H$ real, \texttt{allrad}}
  \end{subfigure}
 
  \vspace{1em}
 
  \begin{subfigure}{0.3\linewidth}
    \centering
    \includegraphics[width=\linewidth]{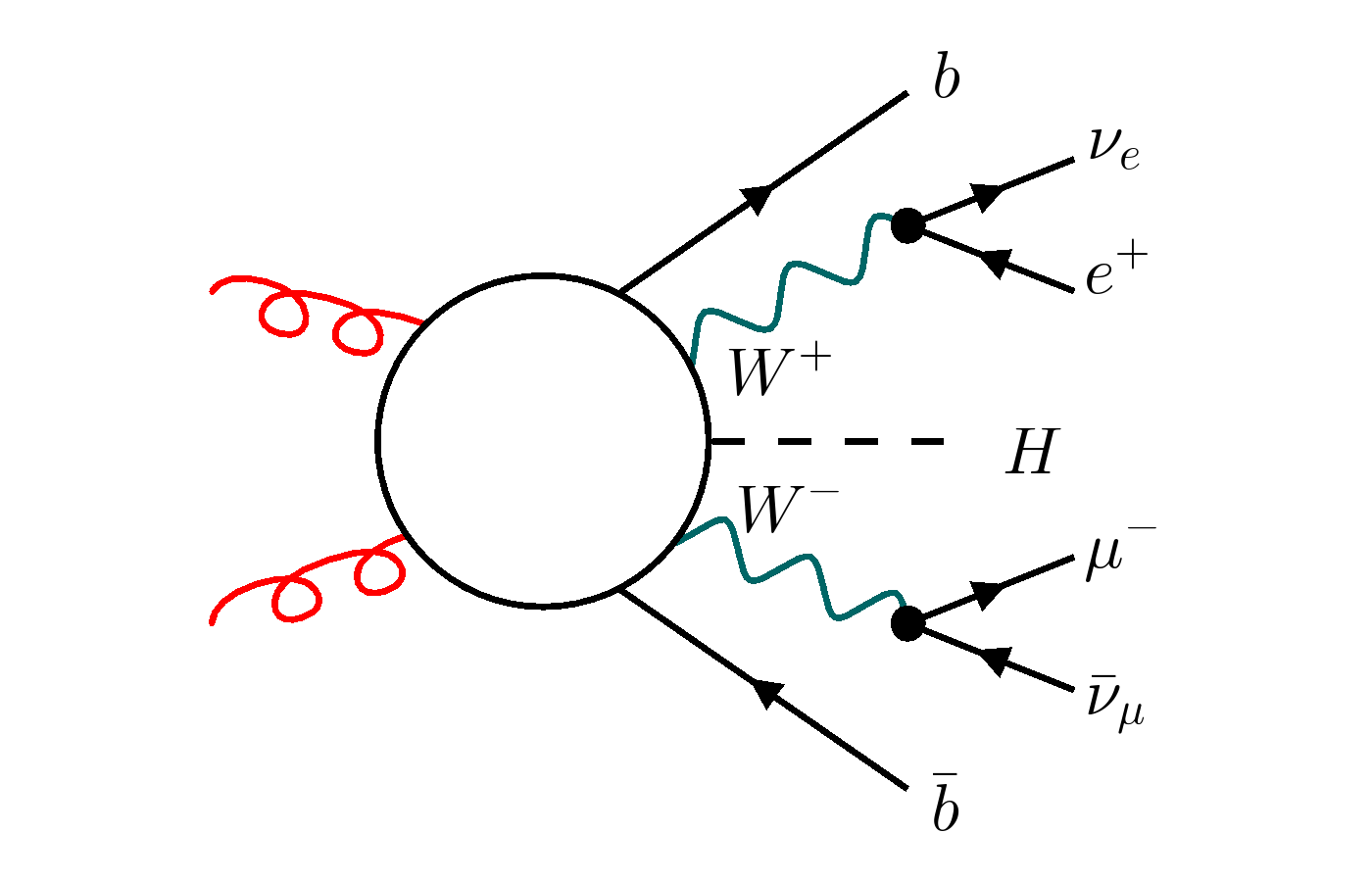}
    \caption{rem Born}
  \end{subfigure}
  \hfill
  \begin{subfigure}{0.3\linewidth}
    \centering
    \includegraphics[width=\linewidth]{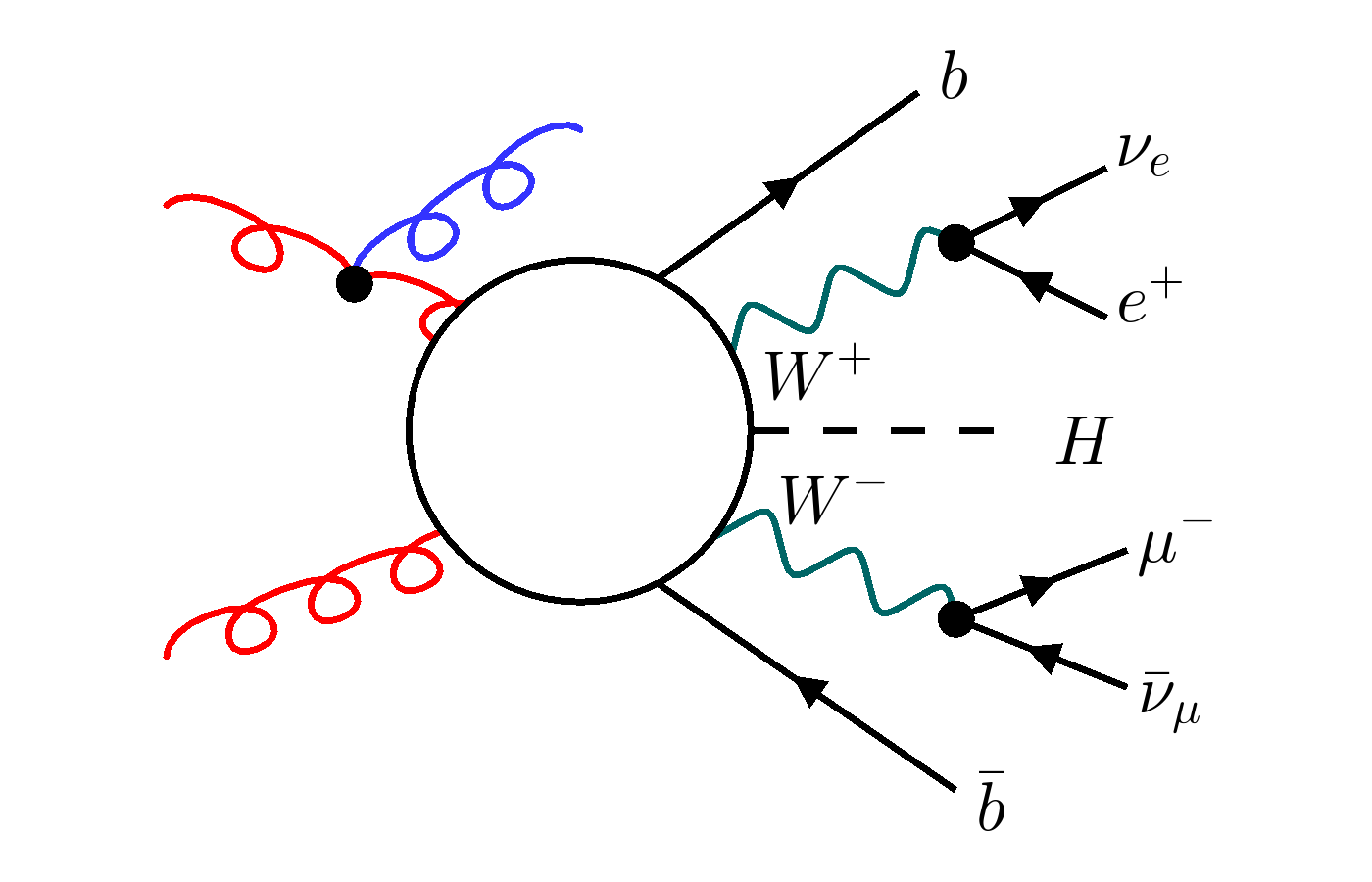}
    \caption{rem real}
  \end{subfigure}
  \hfill
  \begin{subfigure}{0.3\linewidth}
    \centering
    \includegraphics[width=\linewidth]{ProcessesDiags/res_hist/diag_bb4lH_no_res_real.png}
    \caption{rem real, \texttt{allrad}}
  \end{subfigure}
  \caption{Examples of resonance histories for
    $pp\to b\bar b\,e^+\nu_e\,\mu^-\bar\nu_\mu\,H$: $\ttbar H$ (a--c),
    $tW^-H$ (d--f), where analogous $\bar tW^+H$ histories are not shown,
    and the optional remainder history (g--i). The columns show Born
    configurations, a single real emission, and the multiple emissions
    generated in the \texttt{allrad} mode. Blobs indicate the
    decomposition into production and decay subprocesses, and blue lines
    denote \POWHEG{} emissions.}
  \label{fig:res histories}
\end{figure}

The resonance histories of \bbfourlH{} follow directly from the resonance
structure discussed in \refse{sec:offshell_structure} and are listed in
Tab.~\ref{tab:res histories} and illustrated in Fig.~\ref{fig:res histories}.
The $\ttbar H$ history consists of the production subprocess
$pp\to\ttbar H$ and the decays $t\to W^+b$ and $\bar t\to W^-\bar b$. The
two $tWH$ histories consist of the production subprocesses
$pp\to tW^-\bar bH$ and $pp\to\bar tW^+bH$, respectively, followed by a
single top or anti-top decay. Leptonic $W$-boson decays are part of all
histories but do not radiate at $\mathcal{O}(\alpha_s)$.
The Higgs boson is always part of the production subprocess (see
\refse{sec:offshell_structure}). Being stable and colour neutral, it does
not radiate and only takes part in the recoil of production-stage
emissions. The history structure is therefore identical to the one of
\bbfourl{}~\cite{Jezo:2023rht}, with the Higgs boson attached to the
production subprocess. Optionally, a remainder history can be added that
collects interference and non-resonant contributions (see
\refse{Sec:ME res histories}).
 
\begin{table}[tbp]
  \centering
  \begin{tabular}{llll}
    \hline
    history & production & decays & emissions\\
    \hline
    $\ttbar H$     & $pp\to\ttbar H$       & $t\to W^+b$,\; $\bar t\to W^-\bar b$ & $\le 3$\\
    $tW^-H$        & $pp\to tW^-\bar b H$  & $t\to W^+b$                          & $\le 2$\\
    $\bar tW^+H$   & $pp\to\bar tW^+b H$   & $\bar t\to W^-\bar b$                & $\le 2$\\
    rem (optional) & $pp\to W^+W^-b\bar bH$ & ---                                 & $\le 1$\\
    \hline
  \end{tabular}
  \caption{Born resonance histories of \bbfourlH{} with their decomposition
    into production and decay subprocesses, and the maximum number of
    \POWHEG{} emissions per event in the \texttt{allrad} mode (last column). The leptonic
    decays $W^+\to e^+\nu_e$ and $W^-\to\mu^-\bar\nu_\mu$ are part of all
    histories and are kept implicit. The remainder history (rem) is
    present only with the ME$''$ projectors of
    \refse{Sec:ME res histories}.}
  \label{tab:res histories}
\end{table}
 
In the $\ttbar H$ history, the mappings for production-stage radiation
treat the $W^+b$ and $W^-\bar b$ systems as on-shell top quarks with
modified masses, while for radiation within a top decay the four-momentum
of the $Wbg$ system is kept fixed. In the $tWH$ histories, the
non-resonant $W^-\bar b$ ($W^+b$) pair is handled as part of the
production subprocess, and only the virtuality of the resonant top quark
is preserved~\cite{Jezo:2023rht}.
As a result, in the \texttt{allrad} mode events of $\ttbar H$ type receive
up to three \POWHEG{} emissions, one from the production and one from
each top decay, while events of $tWH$ type receive at most two. In
particular, the spectator $b$ quark of $tWH$ events radiates as part of
the production subprocess, such that its radiation is consistently
ordered with all other production-stage emissions. Treating such events
as $\ttbar H$ histories would instead generate a third emission from an
off-shell $Wb$ pair, which is not supported by the factorisation
properties of QCD radiation.

\begin{figure}[tbp]
  \centering
  \begin{subfigure}{0.3\linewidth}
    \centering
    \includegraphics[width=\linewidth]{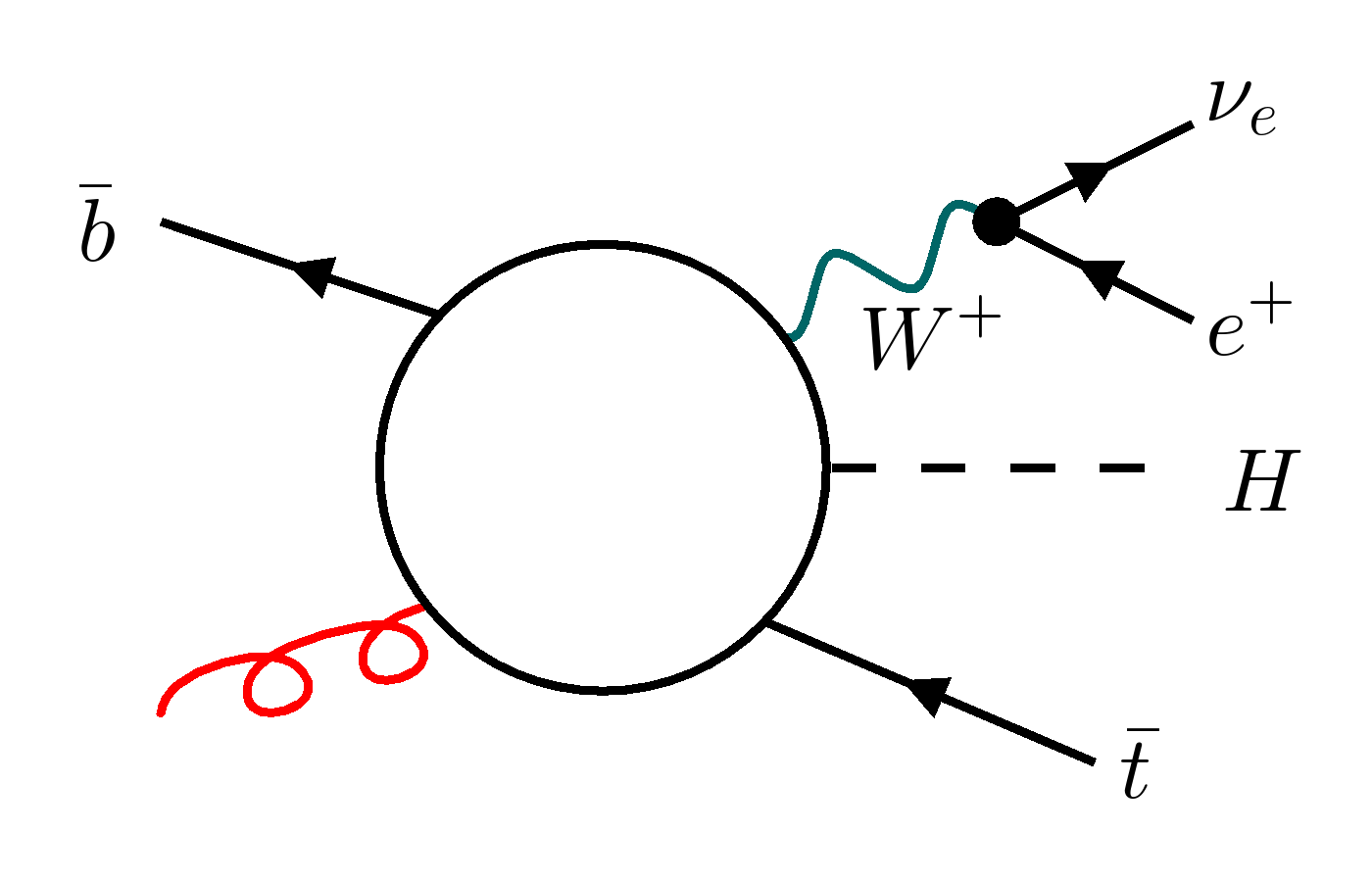}
    \caption{Born}
  \end{subfigure}
  \hspace{0.05\linewidth}
  \begin{subfigure}{0.3\linewidth}
    \centering
    \includegraphics[width=1.1\linewidth]{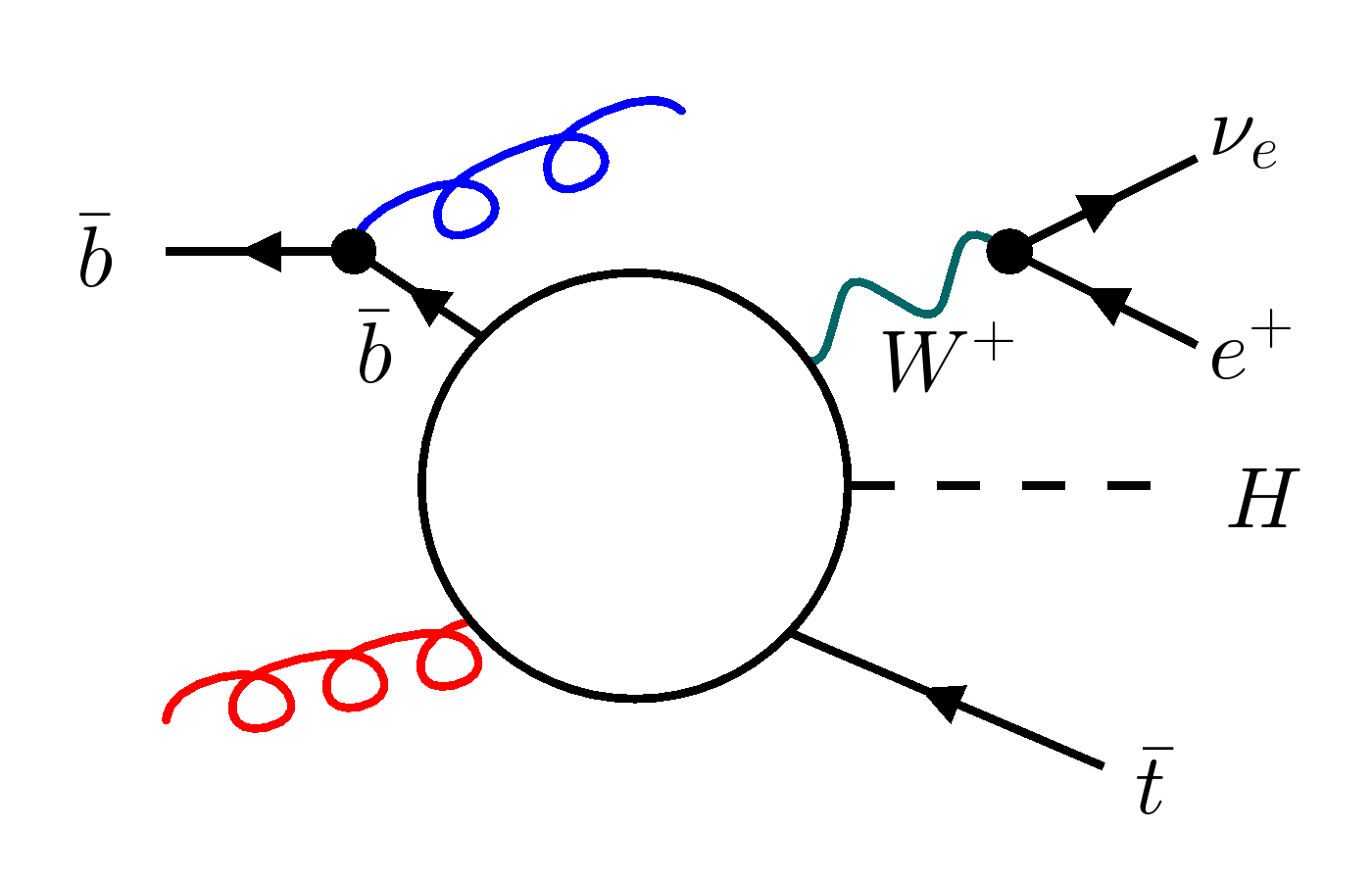}
    \caption{real}
  \end{subfigure}
  \caption{Examples of resonance histories for on-shell $tWH$ production in the 5FS,$pp\to t\,\ell\nu\,H$, at Born level (a) and with real emission (b). The blob separates the production subprocess from the $W\to\ell\nu$ decay, and blue lines denote \POWHEG{} emissions.}
  \label{fig:tWH res histories}
\end{figure}
 
\subsection{Matrix-element--based resonance projectors}
\label{Sec:ME res histories}
 
The resonance weights determine how the cross section is shared among the
histories and, in particular, how events in the transition region between
$\ttbar H$- and $tWH$-dominated phase space are treated. 
Following \citere{Jezo:2023rht}, we use history weights based on matrix
elements. To this end, the Born amplitudes are split according to their
resonance structure,
\begin{equation}
  \label{eq:amp res breakdown}
  \mathcal{A}_{\rm full} = \mathcal{A}_{\ttbar H} + \mathcal{A}_{tW^-H}
  + \mathcal{A}_{\bar tW^+H} + \mathcal{A}_{\rm rem}\,,
\end{equation}
where the terms on the rhs are the subsets of Feynman diagrams with two
top resonances, a single top resonance ($tW^-H$), a single anti-top
resonance ($\bar tW^+H$), and no top resonance.
This splitting is not gauge invariant in general, but it provides a prescription for a 
gauge-invariant separation in the phase-space regions that are dominated
by on-shell $\ttbar H$ or $tWH$ production.
The simplest choice of weights is
\begin{equation}
  \label{eq:ME based weigths 1}
  \begin{split}
  \rho^{(\text{hist})}_{\ttbar H}\Big|_{\rm ME} &= \big|\mathcal{A}_{\ttbar H}\big|^2,\\
  \rho^{(\text{hist})}_{tW^-H}\Big|_{\rm ME} &= \big|\mathcal{A}_{tW^-H}\big|^2,\\
  \rho^{(\text{hist})}_{\bar tW^+H}\Big|_{\rm ME} &= \big|\mathcal{A}_{\bar tW^+H}\big|^2.
  \end{split}
\end{equation}
Here interference and non-resonant contributions, which are suppressed
close to the resonances but can be relevant in the off-shell region
between $\ttbar H$- and $tWH$-dominated phase space, are shared among the
histories in an uncontrolled way. As our default we therefore use the
variant
\begin{equation}
  \label{eq:ME based weigths 2}
  \begin{split}
  \rho^{(\text{hist})}_{\ttbar H}\Big|_{\rm ME'} &=
  \big|\mathcal{A}_{\rm full}\big|^2-\big|\mathcal{A}_{tW^-H}\big|^2-\big|\mathcal{A}_{\bar tW^+H}\big|^2,\\
  \rho^{(\text{hist})}_{tW^-H}\Big|_{\rm ME'} &= \big|\mathcal{A}_{tW^-H}\big|^2,\\
  \rho^{(\text{hist})}_{\bar tW^+H}\Big|_{\rm ME'} &= \big|\mathcal{A}_{\bar tW^+H}\big|^2,
  \end{split}
\end{equation}
which assigns all interference and non-resonant contributions to the
$\ttbar H$ history. Alternatively, the weights~\eqref{eq:ME based weigths 1}
can be supplemented by a separate remainder history with
\begin{equation}
  \label{eq:ME based weights 3}
  \rho^{(\text{hist})}_{\rm rem}\Big|_{\rm ME''} =
  \big|\mathcal{A}_{\rm full}\big|^2-\big|\mathcal{A}_{\ttbar H}\big|^2
  -\big|\mathcal{A}_{tW^-H}\big|^2-\big|\mathcal{A}_{\bar tW^+H}\big|^2.
\end{equation}
The spread between these variants quantifies the ambiguity in the
assignment of interference effects. For $\ttbar+tW$ production this
ambiguity was found to be small, and the ME-based separation agrees well
with a gauge-invariant $\Gamma_t\to0$ extrapolation~\cite{Jezo:2023rht}.
We note that the single-resonant amplitudes $\mathcal{A}_{tW^-H}$ and
$\mathcal{A}_{\bar tW^+H}$ contain the diagrams with the Higgs boson
radiated both off the top quark and off the $W$ boson. The destructive
interference between the two, which suppresses $tWH$ production in the
SM, is therefore built into the $tWH$ resonance weights.
 
For the separation of real-emission contributions, the Born weights are
extended to the real-emission phase space as
\begin{equation}
  \label{eq:real ME based weights}
  \rho^{(\text{hist})}_{h,c}(\Phi_{\rm R})\Big|_{\rm ME'} =
  \rho^{(\text{hist})}_{h}\big(\widetilde{\Phi}_{{\rm B},c}\big)\Big|_{\rm ME'}\,,
  \qquad
  \widetilde{\Phi}_{{\rm B},c}\equiv\widetilde{\Phi}_{{\rm B},c}(\Phi_{\rm R})\,,
\end{equation}
where $\widetilde{\Phi}_{{\rm B},c}$ is obtained through dedicated
real-to-Born mappings. Inverting the standard collinear mappings is not
adequate here, since it can strongly distort the virtualities of the
enhanced $t$-channel propagators associated with initial-state
$g\to b\bar b$ splittings in the $tWH$ histories, and thus the relative
$\ttbar H$ and $tWH$ probabilities. We use the mappings of
\citere{Jezo:2023rht}, which preserve the virtualities of the resonant top
quarks and $W$ bosons and, as far as possible, the transverse energies of
the $b$-quark emitters, with the Higgs boson treated as part of the
production system. These mappings enter only the resonance weights and
not the generation of radiation.

\subsection{Spurious width effects and inverse-width expansion}
\label{sec:inverse-width}
 
In the zero-width limit, integrating the cross section of an off-shell process over the
decay phase space and summing over decay channels must reproduce the
cross section of the production subprocess, since all decay probabilities
add up to one. However, off-shell NLO calculations with NLO widths in the
propagators violate this property.\footnote{The corresponding violation is formally beyond NLO, but can be numerically sizeable} In the language of the NWA,
the unexpanded factors $1/\Gamma_{r,{\rm NLO}}$ associated with each
resonance $r \in \mathcal{R}$, where $\mathcal{R}$ is the set of intermediate unstable particles, generates a term $-{\rm d}\sigma_1\,\Gamma_{r,1}/\Gamma_{r,{\rm NLO}}$,
where ${\rm d}\sigma_1$ is the NLO correction to the production subprocess
and $\Gamma_{r,1}=\Gamma_{r,{\rm NLO}}-\Gamma_{r,0}$ is the correction to the decay width~\cite{Melnikov:2009dn,Campbell:2012uf,Hollik:2012rc}. This
$\mathcal{O}(\alpha_s^2)$ term is cancelled by the product of NLO
corrections to production and decay at NNLO. It is therefore not a genuine
higher-order uncertainty but a spurious artefact of the truncation, which
distorts the NLO corrections to the decays~\cite{Jezo:2023rht}.
Its relative size is approximately
$\delta\kappa_{\rm spurious}\simeq-({\rm d}\sigma_1/{\rm d}\sigma_0)\sum_r\Gamma_{r,1}/\Gamma_{r,0}$,
which for two top resonances with $\Gamma_{t,1}/\Gamma_{t,0}\approx-0.09$
amounts to about $+17\%$ of the relative NLO correction to the production
process~\cite{Jezo:2023rht}. For $\ttbar H$ production, with its NLO corrections of about $25\%$, this corresponds to a shift of several percent, which is larger than the entire inclusive $tWH$ contribution.
 
Following \citere{Jezo:2023rht}, we avoid these spurious terms through an
inverse-width expansion that is applied history by history. With all
ingredients $B$, $V$ and $R$ computed as usual with NLO widths, the
$\bar B_h$ weights in~\eqref{eq:POWHEG-RES allrad xsec} are replaced by
\begin{equation}
  \label{eq:Bbar exp}
  \bar{B}_h(\Phi_{\rm B})\Big|_{\rm exp} =
  \left(\prod_{r\in\mathcal{R}(h)}\frac{\Gamma_{r,{\rm NLO}}}{\Gamma_{r,0}}\right)
  \left[\bar{B}_h(\Phi_{\rm B})
  -\left(\sum_{r\in\mathcal{R}(h)}\frac{\Gamma_{r,1}}{\Gamma_{r,0}}\right)B_h(\Phi_{\rm B})\right],
\end{equation}
and the hard remnant is rescaled accordingly,
\begin{equation}
  \label{eq:hard remnant exp}
  R^{(\rm hard)}_{h,c}(\Phi_{\rm R})\Big|_{\rm exp} =
  \left(\prod_{r\in\mathcal{R}(h)}\frac{\Gamma_{r,{\rm NLO}}}{\Gamma_{r,0}}\right)
  R^{(\rm hard)}_{h,c}(\Phi_{\rm R})\,,
\end{equation}
while the emission probabilities in~\eqref{eq:POWHEG-RES allrad xsec},
which do not affect the normalisation, are left unchanged. In the
zero-width limit this reproduces the NLO production cross section exactly.

In \bbfourlH{} the expansion is restricted to top-quark resonances, i.e.\
\eqref{eq:Bbar exp} involves factors
$(\Gamma_{t,{\rm NLO}}/\Gamma_{t,0})^{n_t(h)}$ and
$n_t(h)\,\Gamma_{t,1}/\Gamma_{t,0}$, with $n_t=2$ for the $\ttbar H$
history, $n_t=1$ for the $tWH$ histories, and $n_t=0$ for the optional
remainder history. Leptonic $W$ decays receive no $\mathcal{O}(\alpha_s)$
corrections, so no spurious terms arise from the $W$ width, and the NLO
$W$ width is used without expansion.
%

\subsection{Semileptonic channels}
\label{sec:semileptonic}
 
Semileptonic events can be obtained from corresponding dileptonic \bbfourlH{} events
following the approach of \citere{Jezo:2023rht}. The semileptonic
$\ttbar H+tWH$ process is defined through the Feynman diagrams of the
dileptonic process~\eqref{eq:Process} with one lepton--neutrino pair
replaced by a $q\bar q'$ pair with the same weak-isospin quantum numbers.
This definition is gauge invariant and separates $\ttbar H+tWH$
production from other contributions to the same final state, which can be
described by independent generators.
Since the $q\bar q'$ line exchanges no colour with the rest of the
process, the additional QCD corrections associated with it do not
interfere with those inherited from the dileptonic calculation. They are
included as an extra decay subprocess in the \texttt{allrad}
formula~\eqref{eq:POWHEG-RES allrad xsec},
\begin{equation}
  \label{eq:semilep}
  {\rm d}\sigma_{\rm sl} = {\rm d}\sigma_{\rm dl}\,K_{W_{\rm had}}
  \left[\Delta_{W_{\rm had}}(q_{\rm cut})
  + \sum_{c\in\mathcal{C}(W_{\rm had})}\Delta_{W_{\rm had}}(k_{{\rm T},c})\,
  \frac{R_{\rm DPA}(\Phi_{{\rm R},c})}{B_{\rm DPA}(\Phi_{\rm B})}\,{\rm d}\Phi_{{\rm rad},c}\right],
\end{equation}
where ${\rm d}\sigma_{\rm dl}$ denotes the dileptonic events reinterpreted
as semileptonic ones. The emission probability for $W\to q\bar q'(g)$ is
evaluated in a $W$-DPA that retains only
topologies with two resonant $W$ bosons, with exact spin correlations.
The finite part of the virtual corrections to $W\to q\bar q'$ is
accounted for by the normalisation factor
\begin{equation}
  \label{eq:K Whad}
  K_{W_{\rm had}} = \frac{{\rm BR}(W\to jj)}{{\rm BR}(W\to\ell\nu)}\,,
\end{equation}
with branching ratios consistent with the NLO $W$ width,
\begin{equation}
  \label{eq:BR W}
  {\rm BR}(W\to\ell\nu) = \left[9+\frac{6}{\pi}\,\alpha_s(m_W)\right]^{-1},
  \qquad
  {\rm BR}(W\to jj) = 1-3\,{\rm BR}(W\to\ell\nu)\,.
\end{equation}
Events of $\ttbar H$ type thus receive up to four \POWHEG{} emissions.
The inverse-width expansion of \refse{sec:inverse-width} is inherited from
the dileptonic events. For the matching to the parton shower, the hadronic
$W$ decay is handled as an additional decay subprocess, on the same
footing as the top-quark decays.

This approach has a limited range of validity.
In the full $\ell\nu_\ell\,jj\,b\bar b\,H$ final state, light jets can
also arise from topologies where the light quarks do not stem from a
$W$-boson decay, such as single-top or vector-boson-fusion topologies, as
well as from QCD-induced contributions of different order in $\alpha_s$.
These contributions, and their interference with $\ttbar H+tWH$
production, are not included in the semileptonic \bbfourlH{} generator.
Moreover, the QCD corrections to the hadronic $W$ decay are described
through the $W$-DPA and the constant normalisation
factor~\eqref{eq:K Whad}, which are designed for resonant $W$ bosons.
Thus, semileptonic \bbfourlH{} predictions are reliable only in
phase-space regions where the invariant mass of the light-jet pair
associated with the hadronic $W$ decay is close to the $W$-boson mass,
$m_{jj}\sim m_W$.
For $\ttbar+tW$ production, it was shown in \citere{Jezo:2023rht} that,
at LO, the corresponding semileptonic process reproduces the full
$\ell\nu_\ell\,jj\,b\bar b$ final state with sub-percent precision for
$|m_{jj}-m_W|<30\,\GeV$, where $jj$ denotes the two hardest light jets,
while other contributions become significant only for
$m_{jj}\gtrsim 130\,\GeV$.
For the real-emission process with an additional jet, agreement at the
percent level was found for technical cuts on the transverse momentum of
the extra jet down to $1\,\GeV$.
We expect a similar behaviour for $\ttbar H+tWH$ production.
Such regions are typically selected in experimental analyses through the
kinematic reconstruction of the hadronically decaying $W$ boson.

\section{Computational setup} 
\label{sec:setup}

In this section we specify the input parameters, scale choices and
technical settings that are used for the NLO and NLOPS predictions of
\refses{sec:validation}{sec:results}.
Unless stated otherwise, the same setup is employed for the off-shell
\bbfourlH{} generator and for the on-shell $\texttt{ttH}$ and $\texttt{tWH}$
generators, such that differences between their predictions can be
attributed to the physics approximations.
All predictions refer to proton--proton collisions at
$\sqrt{s}=13\,\TeV$.
 
\subsection{Input parameters and PDFs}
\label{sec:input}
 
For the masses of the heavy quarks and of the massive bosons we use
\begin{equation}
  \label{eq:setup_masses}
  \begin{aligned}
  m_t &= 172.5\,\GeV, & m_b &= 4.75\,\GeV, & M_H &= 125\,\GeV,\\
  M_W &= 80.385\,\GeV, & M_Z &= 91.1876\,\GeV, &&
  \end{aligned}
\end{equation}
while all other quarks and leptons are treated as massless, and the
CKM matrix is assumed to be diagonal.
Unstable particles are treated in the complex-mass
scheme~\cite{Denner:2005fg}, where the masses of top quarks and of $W$,
and $Z$ are replaced throughout by the complex-valued parameters
\begin{equation}
\mu_i^2=M_i^2-{\rm i}\,\Gamma_iM_i\,.
\end{equation}
The electroweak couplings are derived from the complex $W$- and $Z$-boson
masses and from the Fermi constant, $G_\mu=1.16637\times10^{-5}\,\GeV^{-2}$, in the
$G_\mu$ scheme, i.e.\
\begin{equation}
  \label{eq:setup_alpha}
  \alpha = \frac{\sqrt2\,G_\mu}{\pi}\left|\mu_W^2\left(1-\frac{\mu_W^2}{\mu_Z^2}\right)\right|,
  \qquad
  \cos\theta_{\rm w} = \frac{\mu_W}{\mu_Z}\,,
\end{equation}
which yields $\alpha^{-1}=132.1712893$.
For the widths of the $W$ and $Z$ bosons we use the NLO QCD values
\begin{equation}
  \label{eq:setup_widths}
  \Gamma_W^{\rm NLO} = 2.0928\,\GeV,\qquad
  \Gamma_Z^{\rm NLO} = 2.4952\,\GeV.
\end{equation}
The top-quark width is computed at LO and NLO QCD, consistently with the
above input parameters, from the off-shell three-body decays
$t\to f\bar f' b$ with a massive $b$ quark based on the MCFM routine of~\citere{Campbell:2012uf}, which results in
\begin{equation}
  \label{eq:setup_topwidth}
  \Gamma_t^{\rm LO} = 1.4576\,\GeV,\qquad
  \Gamma_t^{\rm NLO} = 1.3387\,\GeV\,.
\end{equation}
In \bbfourlH{} all matrix elements are evaluated with
$\Gamma_t^{\rm NLO}$, while $\Gamma_t^{\rm LO}$ enters only through the
inverse-width expansion of \refse{sec:inverse-width}.
In the on-shell generators, the Breit--Wigner smearing of top-quark and
$W$-boson virtualities in the decay procedure of \refse{Sec:Top decay} is
based on $\Gamma_t=\Gamma_t^{\rm NLO}$ and
$\Gamma_W=\Gamma_W^{\rm NLO}$.
In all generators the Higgs boson is treated as a stable particle at the
level of the hard matrix elements and for all numerical predictions considered in this paper.
 
In \bbfourlH{} we work in the 4FNS, i.e.\ $b$ quarks are massive and do
not enter the initial state.
Closed $b$- and $t$-quark loops are included in the virtual corrections,
and $b$ quarks are treated as active flavours in the
renormalisation of $\alpha_s$. The bottom-quark Yukawa coupling is included throughout.
In the on-shell generators we work in the 5FNS, with massless $b$ quarks
in the production matrix elements, while a finite $b$-quark mass is used
only in the decay matrix elements (see \refse{Sec:Top decay}).
 
For the parton distributions we use the $\text{NNPDF40\_nlo\_as\_01180}$~\cite{NNPDF:2021njg} obtained in variable flavour number scheme with 5 active flavours with $\alpha_s(M_Z)=0.11800$, as provided by
LHAPDF6~\cite{Buckley:2014ana} (LHAPDF ID 331700). 
The running of
$\alpha_s$ is assumed at two-loops as provided by \POWHEGBOX.
The different treatment of $b$ quarks in the PDF and the \bbfourlH matrix elements, due to a different choice of number of active flavours, is compensated by the following $\mathcal{O}(\alpha_s)$ matching factors~\cite{Cacciari:1998it,Jezo:2023rht}, which are applied to the $\bar B$ weights in
the $q\bar q$ and $gg$ channels,
\begin{align}
  \label{eq:setup_4fmatch_qq}
  \bar B_{q\bar q}(\Phi_{\rm B}) &\;\to\; \bar B_{q\bar q}(\Phi_{\rm B})
  - \frac{4}{3}\,T_F\,\frac{\alpha_s}{2\pi}\,
  \ln\!\left(\frac{Q_{\rm R}^2}{m_b^2}\right) B_{q\bar q}(\Phi_{\rm B})\,,\\
  \label{eq:setup_4fmatch_gg}
  \bar B_{gg}(\Phi_{\rm B}) &\;\to\; \bar B_{gg}(\Phi_{\rm B})
  - \frac{4}{3}\,T_F\,\frac{\alpha_s}{2\pi}
  \left[\ln\!\left(\frac{Q_{\rm R}^2}{m_b^2}\right)
  + \ln\!\left(\frac{m_b^2}{\mu_{\rm F}^2}\right)\right] B_{gg}(\Phi_{\rm B})\,.
\end{align}
The logarithm of $m_b^2/\mu_{\rm F}^2$ removes the $b$-quark contribution to the evolution
of the gluon density, and the logarithm of $Q_{\rm R}^2/m_b^2$ removes the
$b$-quark contribution to the running of $\alpha_s$ between $m_b$ and
$Q_{\rm R}$.
Since $b$-quark loops are included in the matrix elements and
treated as active in the running of $\alpha_s$, we set
$Q_{\rm R}=m_b$ as in Ref.~\cite{Jezo:2023rht}.

The renormalisation and factorisation scales are set to
$\mu_{\rm R}=\xi_{\rm R}\,\mu_0$ and $\mu_{\rm F}=\xi_{\rm F}\,\mu_0$,
with a central fixed scale
\begin{equation}
  \label{eq:setup_scale}
  \mu_0 = m_t + m_H/2\,.
\end{equation}
Alternative dynamical scales evaluated on the underlying Born kinematics
can be defined independently for different resonance histories allowing different scale choices for ttH- and tWH-type events.
In the on-shell generators, $\mu_0$ is set to the same value.
Scale uncertainties are estimated through the usual 7-point variations,
\begin{equation}
  \label{eq:setup_scalevar}
  (\xi_{\rm R},\xi_{\rm F}) = (2,2),\,(2,1),\,(1,2),\,(1,1),\,(1,0.5),\,(0.5,1),\,(0.5,0.5)\,.
\end{equation}

\subsection{\POWHEG{} and parton-shower settings}
\label{sec:setup_powheg}
 
All generators are based on the \POWHEGRES{} framework, and all tree-level and one-loop matrix elements are
evaluated with \OpenLoops2~\cite{Cascioli:2011va,Buccioni:2017yxi,Buccioni:2019sur}.
In \OpenLoops, matrix elements are computed in terms of
Feynman diagrams, whose structures are here for the first time generated based on a novel interface with
 \texttt{QGraf}~\cite{Nogueira:1991ex}.

In \POWHEG the separation of real radiation into singular and hard parts is
controlled by the damping parameter, which we set to
$h_{\rm damp}=m_t$.
In \bbfourlH{} \POWHEG{} emissions
are generated in the \texttt{allrad} mode described in
\refse{sec:powheg-res-summary} with up to three emissions per event.
Semileptonic events are obtained from dileptonic \bbfourlH{} events as
described in \refse{sec:semileptonic}.
%
%
%
%
%
In the dileptonic channel we consider the decays
$t\to b\,e^+\nu_e$ and $\bar t\to\bar b\,\mu^-\bar\nu_\mu$.
The decayed matrix elements are evaluated with $m_b=4.75\,\GeV$.

For the parton shower we use \Pythia ~\cite{Bierlich:2022pfr}.
The top-quark resonances, and the intermediate $W$ bosons, are
written to the LHEs, such that their virtualities are preserved by the
shower.
The matching of shower and \POWHEG{} emissions in \bbfourlH{} is based on modified
\texttt{PowhegHooksBB4L.h} user
hooks~\cite{Jezo:2023rht}, where shower emissions are attributed to the top-quark decays or, in the semileptonic
channel, to the hadronic $W$ decay, and are vetoed against the hardness
of the \POWHEG{} emission from the same subprocess. 
In the on-shell generators and in \bbfourlH{}, shower emissions from the production are
vetoed by means of \texttt{PowhegHooks.h}, while radiation in top-quark
decays is generated by the shower.
QED radiation is switched off, as are hadronisation and
multi-parton interactions. Matrix-element corrections (MEC) in the decay are switched on throughout.

\subsection{Physics objects and fiducial cuts}
\label{sec:setup_cuts}
 
The event analysis is carried out with \RivetThree~\cite{Bierlich:2019rhm},
and physics objects are defined at Monte Carlo truth level.
Charged leptons are identified by their flavour and charge, and the missing transverse momentum,
$p_{\rm T}^{\rm miss}$, is defined as the vector sum of the neutrino
transverse momenta.
Jets are reconstructed with the anti-$k_{\rm T}$
algorithm~\cite{Cacciari:2008gp} as implemented in
\textsc{FastJet}~\cite{Cacciari:2011ma}, with jet radius
$R=0.4$.
Jets are $b$-tagged if they contain a $b$ quark at parton level, and are
otherwise classified as light jets.
For the reconstruction of top-quark observables (reconstructed top quark invariant mass, invariant mass of the lepton and the associated $b$-jet) we distinguish $b$-jets from $\bar b$-jets according to flavour content of the jets and discard any jets containing  $b\bar b$ pairs.
 
In the dileptonic channel we require exactly two charged leptons,
$e^+\mu^-$, with
\begin{equation}
  \label{eq:setup_cuts_dl_leptons}
  p_{{\rm T},\ell} > 20\,\GeV,\qquad |\eta_\ell| < 2.5\,,
\end{equation}
at least 2 $b$-jets with
\begin{equation}
  \label{eq:setup_cuts_dl_jets}
  p_{{\rm T},{j_b}} > 25\,\GeV,\qquad |\eta_{j_b}| < 2.5\,,
\end{equation}
and $p_{\rm T}^{\rm miss}>20\,\GeV$.

In the semileptonic channel we require exactly one charged lepton (a positron $e^+$), at
least 2 $b$-jets that fulfil the same acceptance cuts as above, together with the same missing transverse momentum cut as above, as well as at least 2 light jets with
\begin{equation}
  \label{eq:setup_cuts_sl_jets}
  p_{{\rm T},j} > 25\,\GeV,\qquad |\eta_j| < 2.5\,.
\end{equation}

\section{Validation} 
\label{sec:validation}
We now move on to the discussion of validation of our implementations. All generators are compared to existing results, where available. For the off-shell generator we also perform a consistency check based on zero-width extrapolation.

\subsection{Literature results at fixed-order}

In this section we validate the LO and NLO cross sections of the three
generators against independent fixed-order calculations, where available.
These results are summarised in \refta{tab:FO validation}.

\begin{table}[tbp]
  \centering
  \begin{tabular}{llllll}
    \hline
    process & order & \POWHEGBOX{} [fb] & reference [fb] && rel. difference\\
    \hline
    $pp \to \ttbar H$ & NLO & 498.5(2) & 498.7(5) &\cite{LHCHiggsCrossSectionWorkingGroup:2016ypw} & $4 \cdot 10^{-4}$\\
    \hline
    \multirow{2}{*}{$pp\to tWH$ (DR1)} & LO  & 1.7020(16) & 1.70327 &  \multirow{2}{*}{\cite{Demartin:2016axk}} & $8 \cdot 10^{-4}$\\
                                 & NLO & 2.351(9)   & 2.34598& & $2\cdot 10^{-3}$\\
    \hline
    \multirow{2}{*}{$pp \to b\bar{b}\,e^+\nu_e\,\mu^-\bar{\nu}_\mu\,H$} & LO  & 2.2400(10)   & 2.2401&   \multirow{2}{*}{\cite{Denner:2015yca,Stremmer:2021bnk}} & $1 \cdot 10^{-4}$\\
                                 & NLO & 2.609(8)   & 2.6330& & $9\cdot 10^{-3}$\\
    \hline
  \end{tabular}
  \caption{LO and NLO cross sections of the on-shell \ttH and \tWH
    generators and of \bbfourlH{} compared to reference results in the
    literature. All comparisons are performed in the 5FNS, following the
    setup of the respective reference. For $tWH$ production, the DR1
    results of \citere{Demartin:2016axk}, where the $W$ boson is stable,
    are multiplied by ${\rm BR}(W\to\ell\nu)=0.108$. The last column
    shows the relative difference with respect to the reference cross section. Numbers in
    parentheses indicate Monte Carlo uncertainties on the last digit(s).}
  \label{tab:FO validation}
\end{table}

An on-shell $\ttbar H$ generator is already available in the
\POWHEGBOX{}-V2 framework~\cite{Hartanto:2015uka}.
In order to guarantee a fully consistent comparison with the \tWH and
\bbfourlH{} generators, we have implemented a new $\ttbar H$ generator in
\POWHEGRES{}, based on \OpenLoops2{}
amplitudes~\cite{Cascioli:2011va,Buccioni:2017yxi,Buccioni:2019sur}.
This new \ttH generator has been validated against
\Sherpa~2.2.16~\cite{Gleisberg:2008ta,Sherpa:2019gpd} at LO and NLO yielding  sub-per-mille level agreement.
In the 5FNS at NLO we also compare against the reference cross section of
the LHC Higgs Cross Section Working
Group~\cite{LHCHiggsCrossSectionWorkingGroup:2016ypw}, using the
\texttt{PDF4LHC15\_nlo\_30} set, and we find agreement at the level of
$0.4$ per-mille.
 
For the validation of the \tWH generator we compare against the 5FNS
results of \citere{Demartin:2016axk} in the DR1 scheme, using the setup of
that reference with the \texttt{NNPDF30\_nlo\_as\_0118} PDF set.
Since the $W$ boson is treated as a stable particle in
\citere{Demartin:2016axk}, while it is decayed leptonically at
matrix-element level in our generator, the reference cross sections are
multiplied by ${\rm BR}(W\to\ell\nu)=0.108$.
At LO and NLO the two calculations agree within the statistical
uncertainties, and the resulting $K$-factor, $K=1.38$, coincides with
the one of \citere{Demartin:2016axk}.
Since the DR1 prescription formally violates gauge invariance, we note that this
comparison is performed in the Feynman gauge. 
 
Off-shell NLO calculations of $\ttbar H$ production with dileptonic
top-quark decays are available only in the 5FNS with massless $b$
quarks~\cite{Denner:2015yca,Stremmer:2021bnk}, while no corresponding
results exist in the 4FNS.
For a validation of \bbfourlH{} at fixed-order we thus employ the 5FNS setup  with massless $b$ quarks of Refs.~\cite{Denner:2015yca,Stremmer:2021bnk}, using the
\texttt{CT10nlo} PDF set with one-loop (two-loop) running of $\alpha_S$ at LO (NLO). As in Refs.~\cite{Denner:2015yca,Stremmer:2021bnk} $b$-initiated channels are discarded, and we require the following fiducial cuts for the two anti-$k_T$ $(R>0.4)$ $b$-jets 
\begin{equation}
  p_{{\rm T},j_b} > 25\,\GeV,\qquad |\eta_{j_b}| < 2.5\,,
\end{equation}
for the two leptons 
\begin{equation}
  p_{{\rm T},{\ell}} > 20\,\GeV,\qquad |\eta_{\ell}| < 2.5\,,
\end{equation}
and $p_{T,{\rm miss}}>20\,\GeV$.
For this comparison we do not include the inverse-width expansion correction factor of Section~\ref{sec:inverse-width}.
At LO the two calculations agree at the level of $0.1$ per-mille, while at
NLO we find agreement at the level of $0.9\%$, i.e.\ within about 2.69 sigma of the Monte Carlo uncertainty.

\subsection{Numerical narrow-width approximation} 
  
In the limit of vanishing top-quark width, the off-shell cross section
must reproduce the corresponding cross section in the NWA.
This provides a stringent test of the implementation of \bbfourlH{}, since
it probes the matrix elements, the phase-space integration and the
subtraction of infrared singularities in the presence of extremely narrow
resonances.
At NLO it also tests the consistent treatment of the top-quark width by
means of the inverse-width expansion of \refse{sec:inverse-width}.
To this end we compute the inclusive \bbfourlH{} cross section at LO and
NLO for increasingly smaller (and larger) values of the top-quark width, and compare the results to
the NWA for $pp\to\ttbar H$ at $\sqrt{s}=13\,\TeV$.

Doubly-resonant, singly-resonant and non-resonant contributions to the
off-shell cross section scale as $\Gamma_t^{-2}$, $\Gamma_t^{-1}$ and
$\Gamma_t^{0}$, respectively.
In order to obtain a finite limit, we keep the top-quark branching ratios
fixed as $\Gamma_t$ is varied, i.e.\ the off-shell cross section computed
with width $\Gamma_t$ is multiplied by $(\Gamma_t/\Gamma_t^{\rm phys})^2$,
and in the following $\sigma_{\bbfourlH{}}(\Gamma_t)$ denotes this rescaled
cross section.
Its deviation from the NWA can be parametrised as~\cite{Cascioli:2013wga,Buonocore:2025fqs}
\begin{equation}
  \label{eq:nwa_fit}
  \frac{\sigma_{\bbfourlH{}}(\Gamma_t)}{\sigma_{\rm NWA}} - 1
  = a\,\Gamma_t^2 + b\,\Gamma_t + c   + \mathcal{O}(\Gamma_t^3)\,,
\end{equation}
where the linear term receives contributions from singly-resonant $tWH$
topologies as well as from $\mathcal{O}(\Gamma_t/m_t)$ finite-width
corrections to $\ttbar H$ production, while the quadratic term receives
contributions from non-resonant topologies and from higher-order
finite-width effects.
Consistency with the NWA requires $c=0$.
For this test we employ a strict 4FNS setup in which the off-shell and NWA
predictions are both computed with four-flavour $\text{NNPDF40\_nlo\_nf\_4\_pdfas}$ PDFs and corresponding strong coupling,
without the matching factors in \refeq{eq:setup_4fmatch_qq} and
\refeq{eq:setup_4fmatch_gg}. The NWA predictions are obtained with \Sherpa2~\cite{Sherpa:2019gpd}. In this way,
off-shell and NWA predictions are based on identical ingredients.

\begin{figure}[tbp]
  \centering
  \begin{subfigure}{0.49\linewidth}
    \centering
            \includegraphics[width=\linewidth]{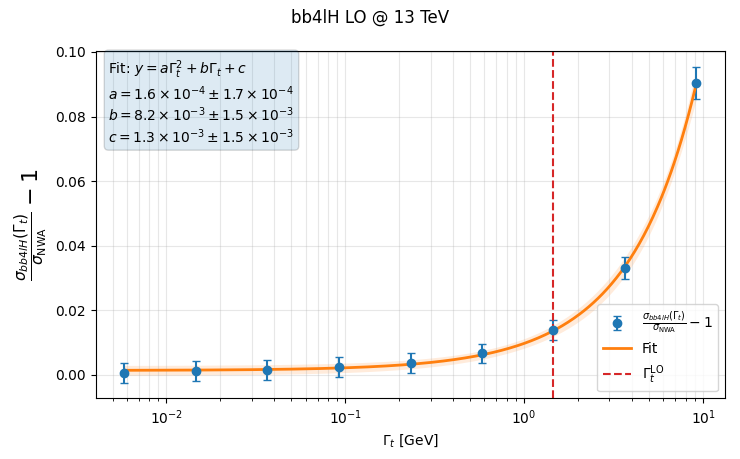}    \caption{LO}
    \label{fig:nwa_LO}
  \end{subfigure}
  \hfill
  \begin{subfigure}{0.49\linewidth}
    \centering
            \includegraphics[width=\linewidth]{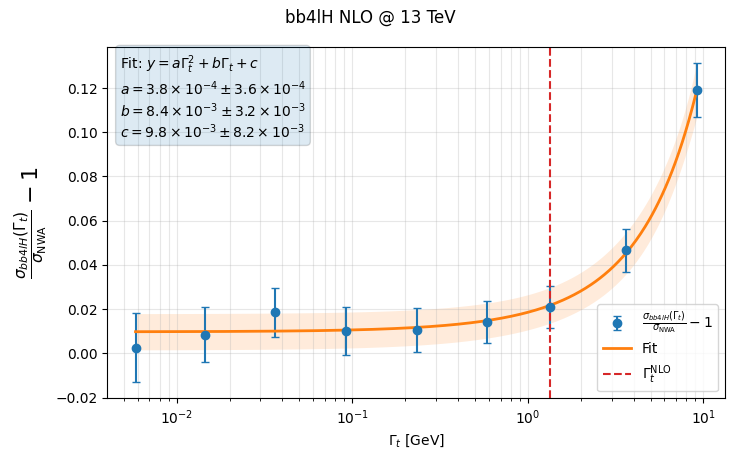}    \caption{NLO}
    \label{fig:nwa_NLO}
  \end{subfigure}
  \caption{Relative deviation of the inclusive \bbfourlH{} cross section
    from the NWA as a function of the top-quark width at LO (a) and NLO
    (b) for $pp$ collisions at $\sqrt{s}=13\,\TeV$. The off-shell cross
    sections are rescaled by $(\Gamma_t/\Gamma_t^{\rm phys})^2$, such
    that the top-quark branching ratios are kept fixed. Error bars
    indicate Monte Carlo uncertainties, while solid lines and bands
    correspond to fits of the form \refeq{eq:nwa_fit} and their
    $1\sigma$ uncertainties. The vertical dashed red line indicates the physical top-quark width.}
  \label{fig:nwa}
\end{figure}
 
The LO results are shown in \reffi{fig:nwa_LO}.
For $\Gamma_t\lesssim 0.1\,\GeV$ the off-shell cross section agrees with
the NWA at the per-mille level, and a quadratic fit yields
 $ c = (1.3\pm1.5)\times10^{-3}\,$,
i.e.\ the extrapolated offset $c$ is compatible with zero and the NWA is
reproduced with a precision of about $0.2\%$.
As $\Gamma_t$ increases, the deviation grows linearly, reaching about
$1.4\%$ at the physical LO width, while the quadratic term is compatible
with zero and becomes visible only for $\Gamma_t\gtrsim$ few GeV.
Thus non-resonant contributions play a negligible role for the inclusive
cross section at the physical width, where finite-width effects are
dominated by singly-resonant contributions and by $\mathcal{O}(\Gamma_t/m_t)$
corrections to $\ttbar H$ production.
 
NLO results with varying top quark width are shown in \reffi{fig:nwa_NLO}.
%
%
Within Monte Carlo  uncertainties the results are independent of $\Gamma_t$ for
$\Gamma_t\lesssim 0.6\,\GeV$, and the quadratic fit yields
\begin{equation}
  \label{eq:nwa_fit_NLO}
  a = (3.8\pm3.6)\times10^{-4}\,\GeV^{-2},\qquad
  b = (8.4\pm3.2)\times10^{-3}\,\GeV^{-1},\qquad
  c = (9.8\pm8.2)\times10^{-3}\,.
\end{equation}
The offset $c$ is compatible with zero at the level of $1.2$ standard
deviations, and the NWA is thus reproduced at NLO with percent-level
precision.
This test is particularly non-trivial, since without the inverse-width
expansion of \refse{sec:inverse-width} the zero-width limit would be
shifted by spurious $\mathcal{O}(\alpha_s^2)$ terms of relative size
$\delta\kappa_{\rm spurious}\simeq 0.17\,\sigma_1/\sigma_0\approx {4.0}\%$, see Eq.~(3.2) in \refse{sec:inverse-width}.
The linear coefficient is in good agreement with the one at LO, and at
the physical NLO width finite-width effects amount to
$b\,\Gamma_t+a\,\Gamma_t^2\approx 1.2\%$, as at LO.
This indicates that NLO corrections affect off-shell and singly-resonant
contributions in a similar way as the NWA cross section, and confirms
that finite-width effects in the inclusive cross section are of the
expected order $\Gamma_t/m_t$.
 
In summary, both at LO and NLO, the \bbfourlH{} cross section approaches
the NWA smoothly as $\Gamma_t\to0$, and the extrapolated results agree
with the NWA within the statistical uncertainties.
This constitutes a non-trivial test of the numerical robustness of the
calculation. 
 As the top-quark width decreases, the numerical integration becomes increasingly challenging due to the sharper resonance structures associated with real radiation emitted before and after the top-quark decay, whose treatment relies on the resonance-aware mappings described in \refse{sec:powheg-res-summary}. This is reflected in the growing statistical uncertainties at smaller widths, despite the fact that the smallest-width point, $\Gamma_t\approx 6\times10^{-3},\GeV$ ($\Gamma_t/m_t\approx 3\times10^{-5}$), was evaluated with 2.5 times the statistics of the physical-width point.
Moreover, non-factorisable corrections, which are enhanced by logarithms
of $\Gamma_t/m_t$ in the individual virtual and real contributions, need
to cancel numerically in the inclusive cross section.

\section{Numerical results} 
\label{sec:results}
 
Finally we present and discuss  predictions from our generators. First \bbfourlH{} in the dileptonic channel alone comparing fixed-order against LHE and NLOPS, later at NLOPS compared to the on-shell generators in both the dileptonic and the semileptonic channels.

\subsection{NLO and NLOPS predictions in the dileptonic channel}
\label{sec:results_dl_nlops}

In the following we present predictions of the \bbfourlH{} generator for
$pp\to e^+\nu_e\,\mu^-\bar\nu_\mu\,b\bar b\,H$ at $\sqrt{s}=13\,\TeV$ in
the dileptonic fiducial phase space defined in \refse{sec:setup_cuts}.
We investigate the impact of the \POWHEG{} matching comparing predictions at three different levels:
fixed-order NLO;
LHE level, i.e.\ \POWHEG{} events before showering; 
and NLOPS level, i.e.\ after showering with \Pythia{} as described in
\refse{sec:setup_powheg}.
Differences between NLO and LHE predictions are formally beyond NLO and
reflect the Sudakov resummation, the evaluation of $\alpha_s$ at the
transverse momentum of the emissions, and the generation of multiple
emissions in \POWHEG{} via the \allrad{} procedure, while differences between NLOPS and LHE predictions
quantify the effect of the subsequent emissions generated by the parton
shower.
In order to isolate matching and shower effects, NLO, LHE and NLOPS
predictions in this section are all computed without the inverse-width
expansion of \refse{sec:inverse-width}, and the impact of this expansion
is shown separately at the NLOPS level.
In the comparisons with on-shell generators in the following two sections,
\bbfourlH{} predictions include the inverse-width expansion.

\begin{figure}[tbp]
  \centering
  \begin{subfigure}{0.48\linewidth}
    \centering
    \includegraphics[width=\linewidth]{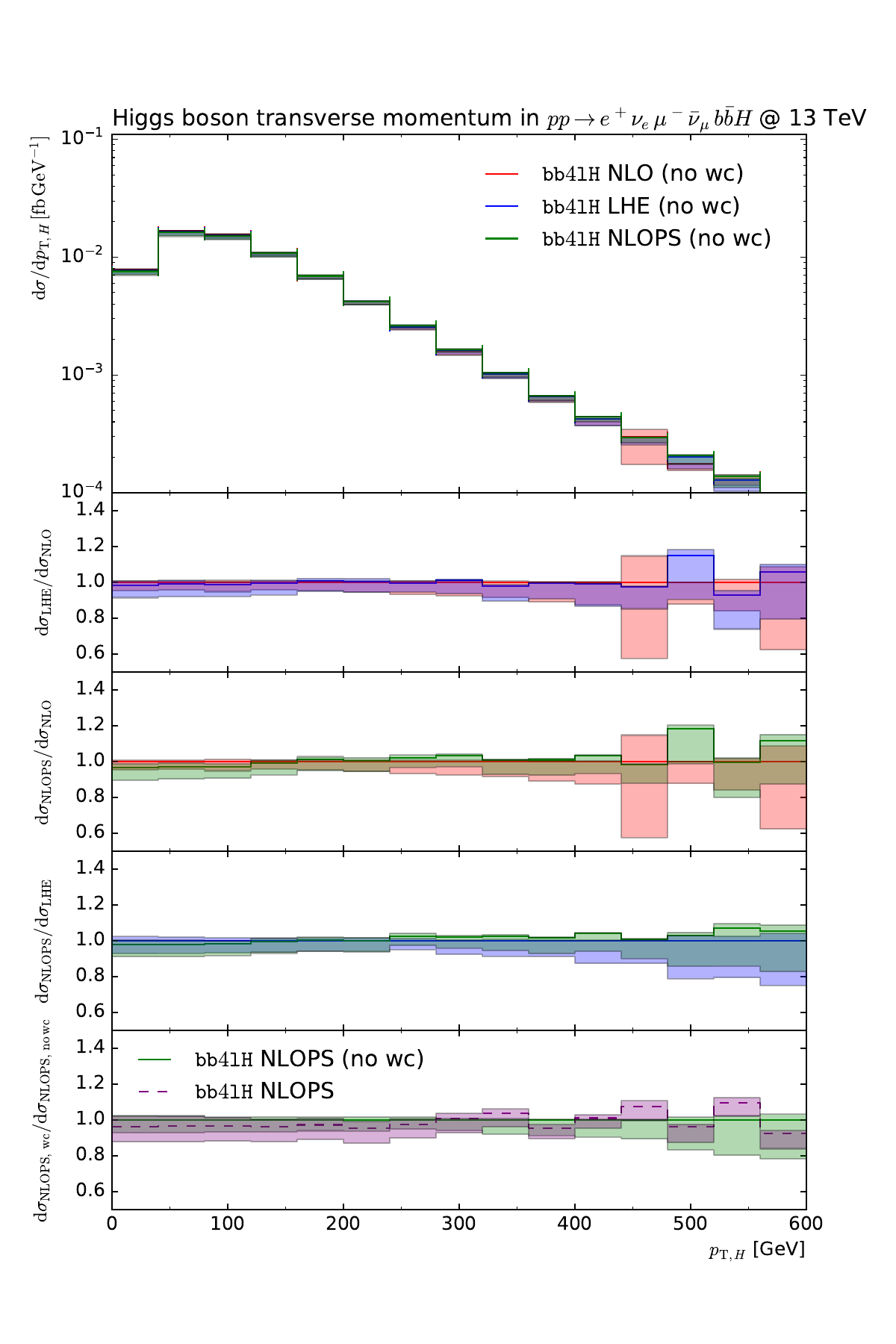}
    \caption{}
    \label{fig:dl_NLOPS_pTH}
  \end{subfigure}
  \hspace{0.01\linewidth}
  \begin{subfigure}{0.48\linewidth}
    \centering
    \includegraphics[width=\linewidth]{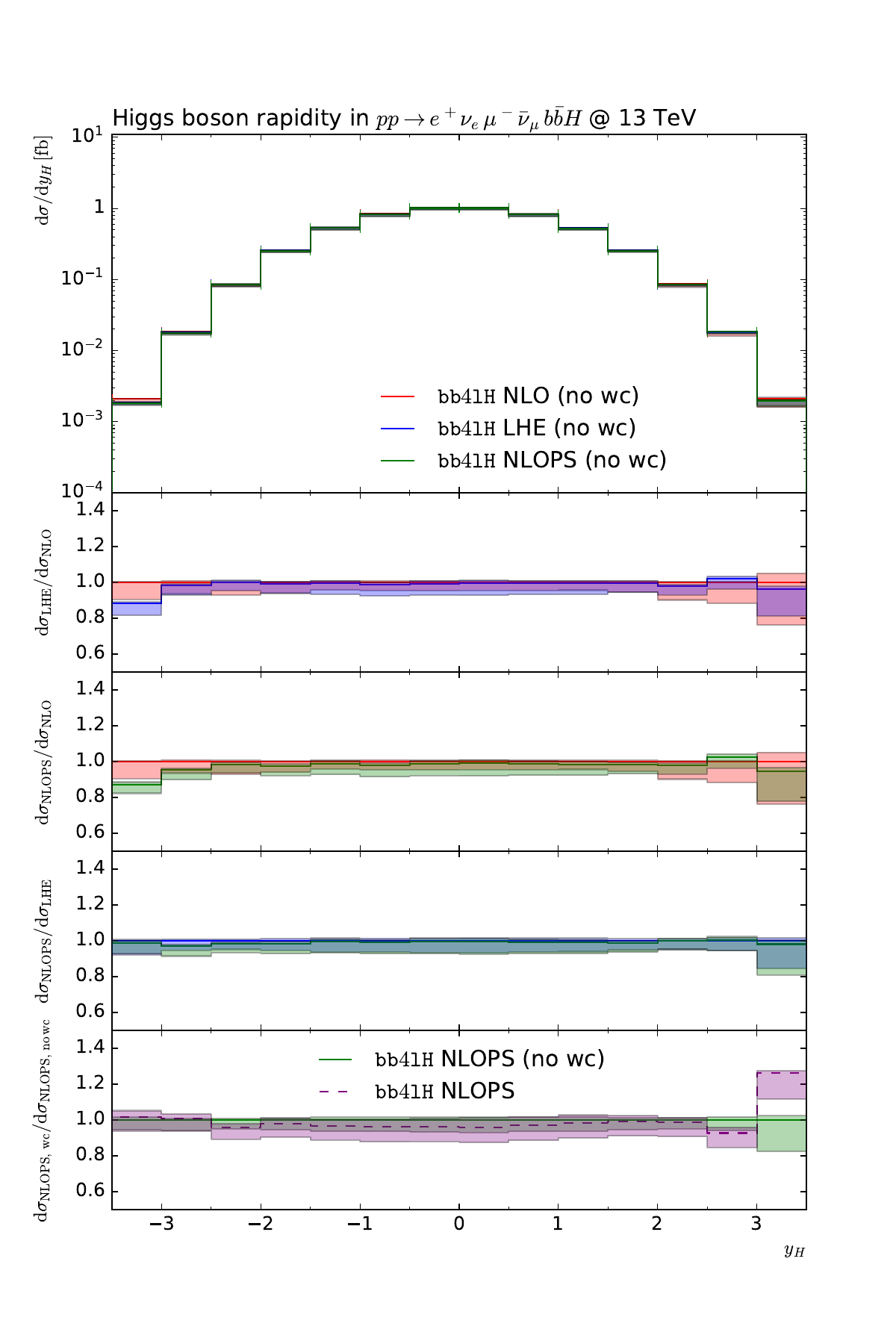}
    \caption{}
    \label{fig:dl_NLOPS_yH}
  \end{subfigure}
  \caption{NLO and NLOPS predictions of \bbfourlH{} for
    $pp\to e^+\nu_e\,\mu^-\bar\nu_\mu\,b\bar b\,H$ at $13\,\TeV$ in the
    dileptonic fiducial phase space defined in \refse{sec:setup_cuts}:
    distributions in the transverse momentum (\subref{fig:dl_NLOPS_pTH})
    and rapidity (\subref{fig:dl_NLOPS_yH}) of the Higgs boson.
    The upper frame shows fixed-order NLO predictions together with
    \bbfourlH{} predictions at the LHE and NLOPS level, all without the
    inverse-width expansion of \refse{sec:inverse-width} (labelled
    ``no wc'').
    The lower frames show the LHE/NLO, NLOPS/NLO and NLOPS/LHE ratios, and
    the ratio of NLOPS predictions with (dashed) and without the
    inverse-width expansion.
    The bands correspond to 7-point scale variations of the respective
    predictions.}
  \label{fig:dl_NLOPS_higgs}
\end{figure}

In \reffis{fig:dl_NLOPS_higgs}{fig:dl_NLOPS_top} we present distributions
in the transverse momentum and rapidity of the Higgs boson, in the
transverse momentum of the positron and in the missing transverse
momentum, in the invariant mass of the two $b$ jets and in the transverse
momentum of the leading $b$ jet, as well as in the invariant mass of the
reconstructed top quark and in the invariant mass of the positron--$b$-jet system.
The upper panels show absolute NLO, LHE and NLOPS predictions, while the
lower panels show the LHE/NLO, NLOPS/NLO and NLOPS/LHE ratios, as well as
the ratio of NLOPS predictions with and without the inverse-width
expansion.
The bands correspond to scale uncertainties, estimated through the 7-point variations
of \refeq{eq:setup_scalevar}.

Let us first focus on the distributions in the transverse momentum and
rapidity of the Higgs boson, shown in \reffi{fig:dl_NLOPS_higgs}.
Being treated as a stable and colour-neutral particle, the Higgs boson is
affected by QCD radiation only through the recoil of emissions from the
production subprocess and through the acceptance cuts on the other
final-state objects.
As a result, NLO, LHE and NLOPS predictions are in very good agreement.
In the bulk of the phase space, LHE and NLOPS predictions agree with
fixed-order NLO within $2\%$ and $4\%$, respectively.
The parton shower has a very small impact on both observables, and
NLOPS/LHE ratios deviate from one by at most $1$--$2\%$, apart from the
regions where Monte Carlo statistics is limited.
All observed differences are well within the scale uncertainties, which
amount to about $+2\%$ and $-7\%$ for LHE and NLOPS predictions and are
almost fully correlated between them, while the NLO scale band amounts to
$+1\%$ and $-5\%$.

\begin{figure}[tbp]
  \centering
  \begin{subfigure}{0.48\linewidth}
    \centering
    \includegraphics[width=\linewidth]{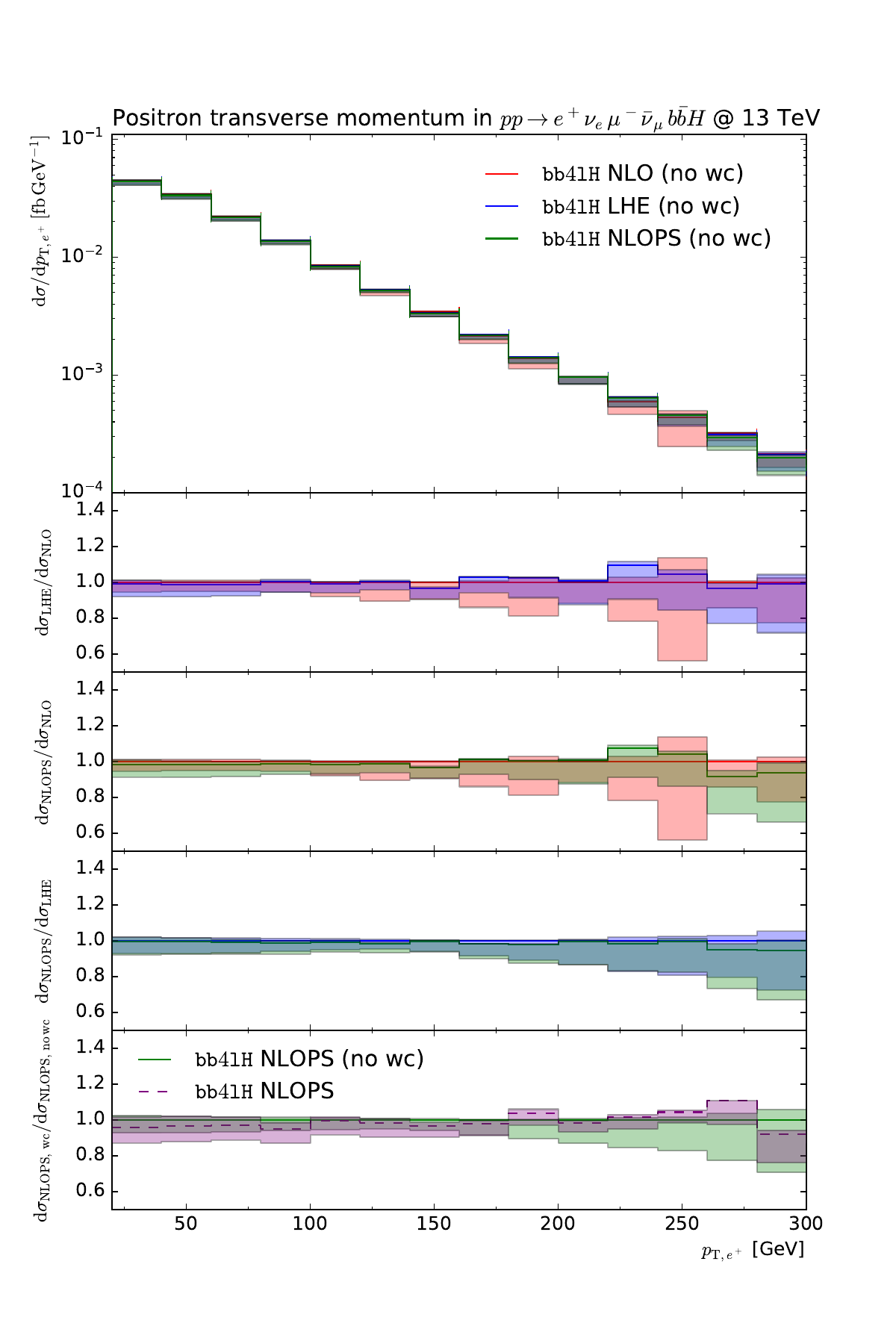}
    \caption{}
    \label{fig:dl_NLOPS_pTe}
  \end{subfigure}
  \hspace{0.01\linewidth}
  \begin{subfigure}{0.48\linewidth}
    \centering
    \includegraphics[width=\linewidth]{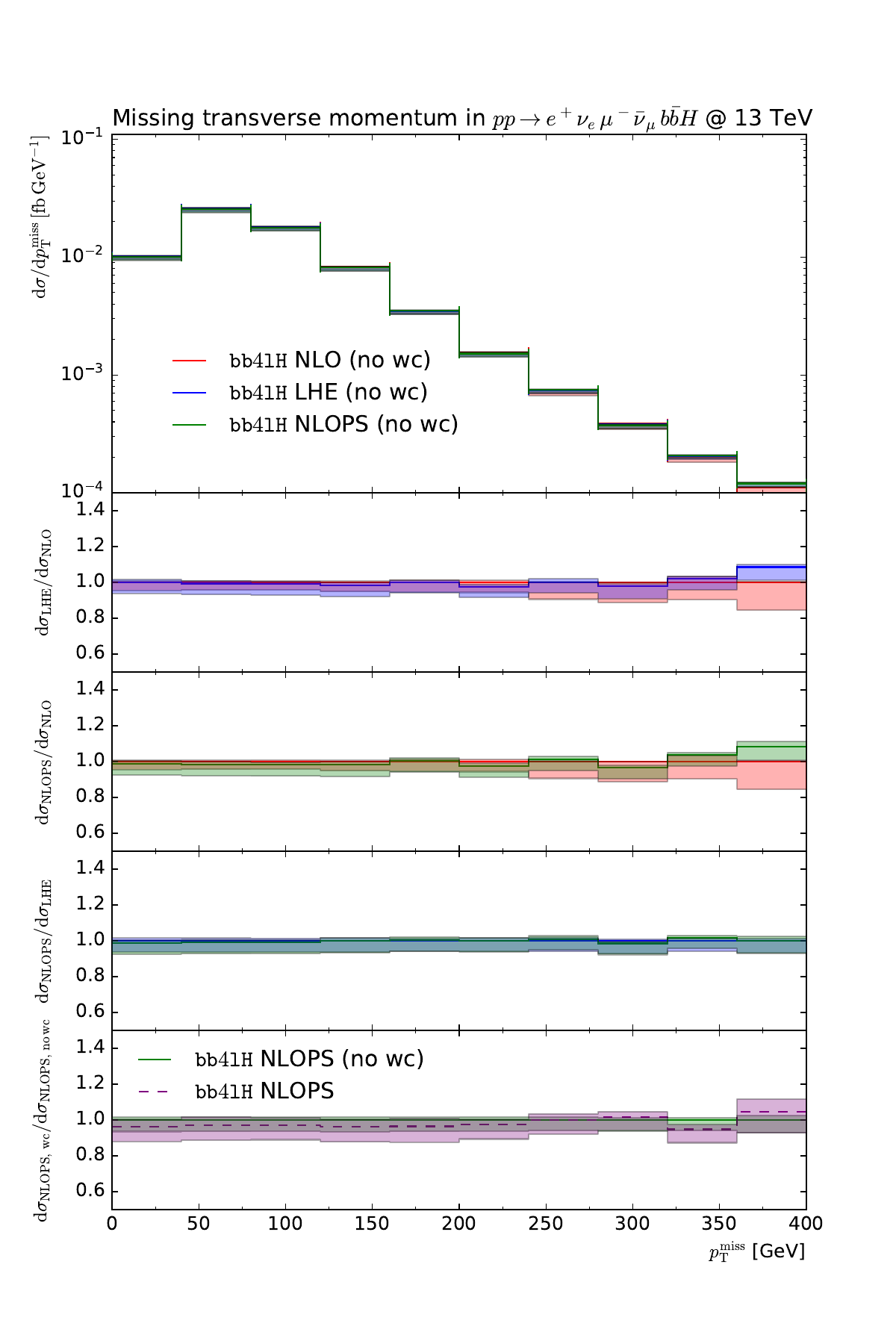}
    \caption{}
    \label{fig:dl_NLOPS_ptmiss}
  \end{subfigure}
  \caption{Distributions in the transverse momentum of the positron
    (\subref{fig:dl_NLOPS_pTe}) and in the missing transverse momentum
    (\subref{fig:dl_NLOPS_ptmiss}).
    Same acceptance cuts, predictions and ratios as in
    \reffi{fig:dl_NLOPS_higgs}.}
  \label{fig:dl_NLOPS_lep}
\end{figure}

The lowest frame of \reffi{fig:dl_NLOPS_higgs} shows the impact of the
inverse-width expansion of \refse{sec:inverse-width} on NLOPS predictions.
The expansion reduces the fiducial cross section by about $3\%$, in line with
the size of the spurious width effects estimated in
\refse{sec:inverse-width}.
At the same time, the scale uncertainty of the NLOPS prediction increases
from $+2\%$ and $-7\%$ to about $+5\%$ and $-9\%$.
A similar behaviour is observed for the leptonic and $b$-jet observables
in \reffis{fig:dl_NLOPS_lep}{fig:dl_NLOPS_bjets}.

A similarly stable behaviour is observed in the distributions in the
transverse momentum of the positron and in the missing transverse
momentum, shown in \reffi{fig:dl_NLOPS_lep}.
LHE/NLO and NLOPS/NLO ratios remain within a
few percent from one, while NLOPS/LHE ratios are consistent with one at
the level of $1$--$3\%$ in the entire plotted range.
This is expected, since QED radiation is switched off in our simulation, and leptonic
observables are sensitive to QCD radiation only through recoil effects
and through the acceptance cuts on $b$ jets.

\begin{figure}[tbp]
  \centering
  \begin{subfigure}{0.48\linewidth}
    \centering
    \includegraphics[width=\linewidth]{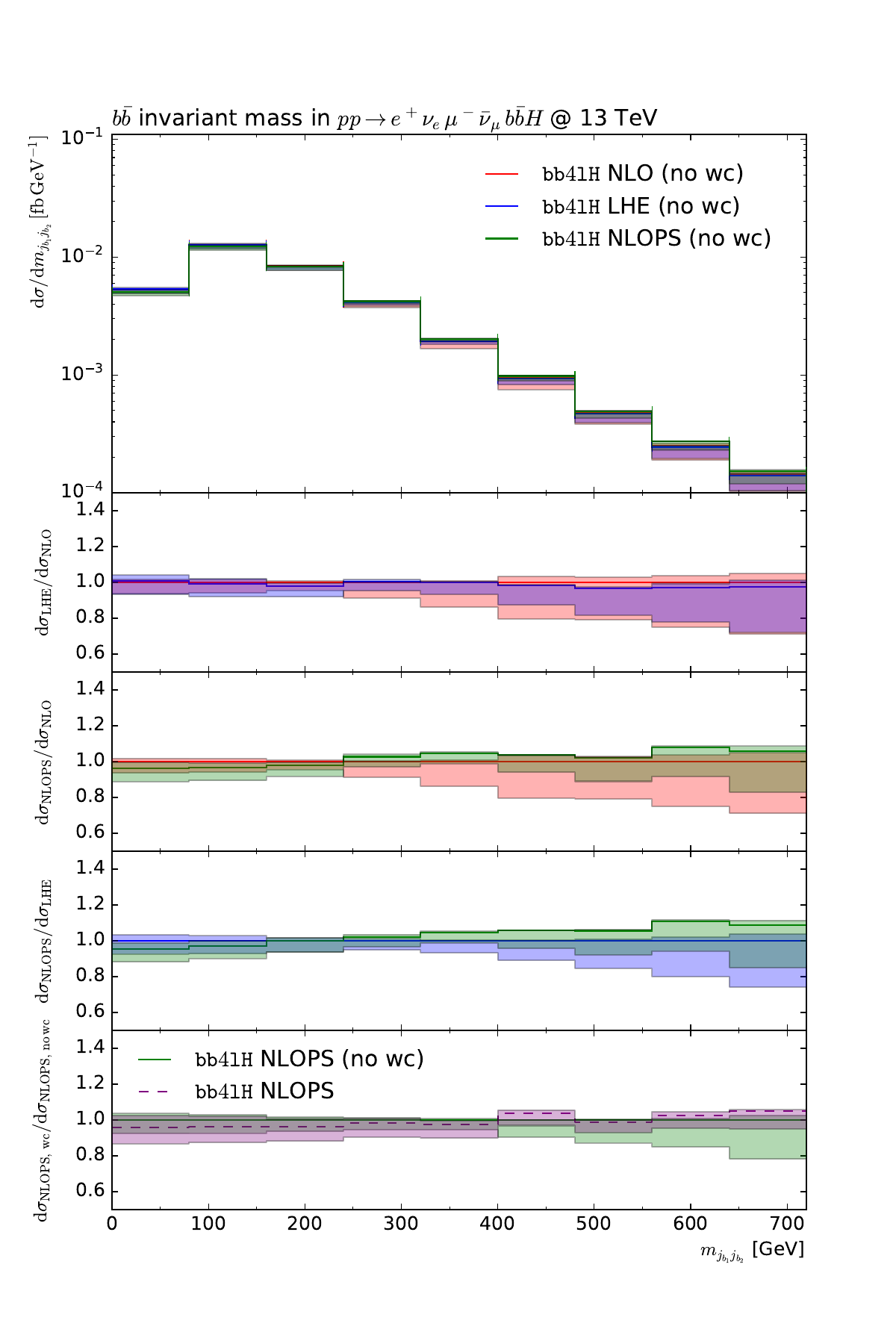}
    \caption{}
    \label{fig:dl_NLOPS_mbb}
  \end{subfigure}
  \hspace{0.01\linewidth}
  \begin{subfigure}{0.48\linewidth}
    \centering
    \includegraphics[width=\linewidth]{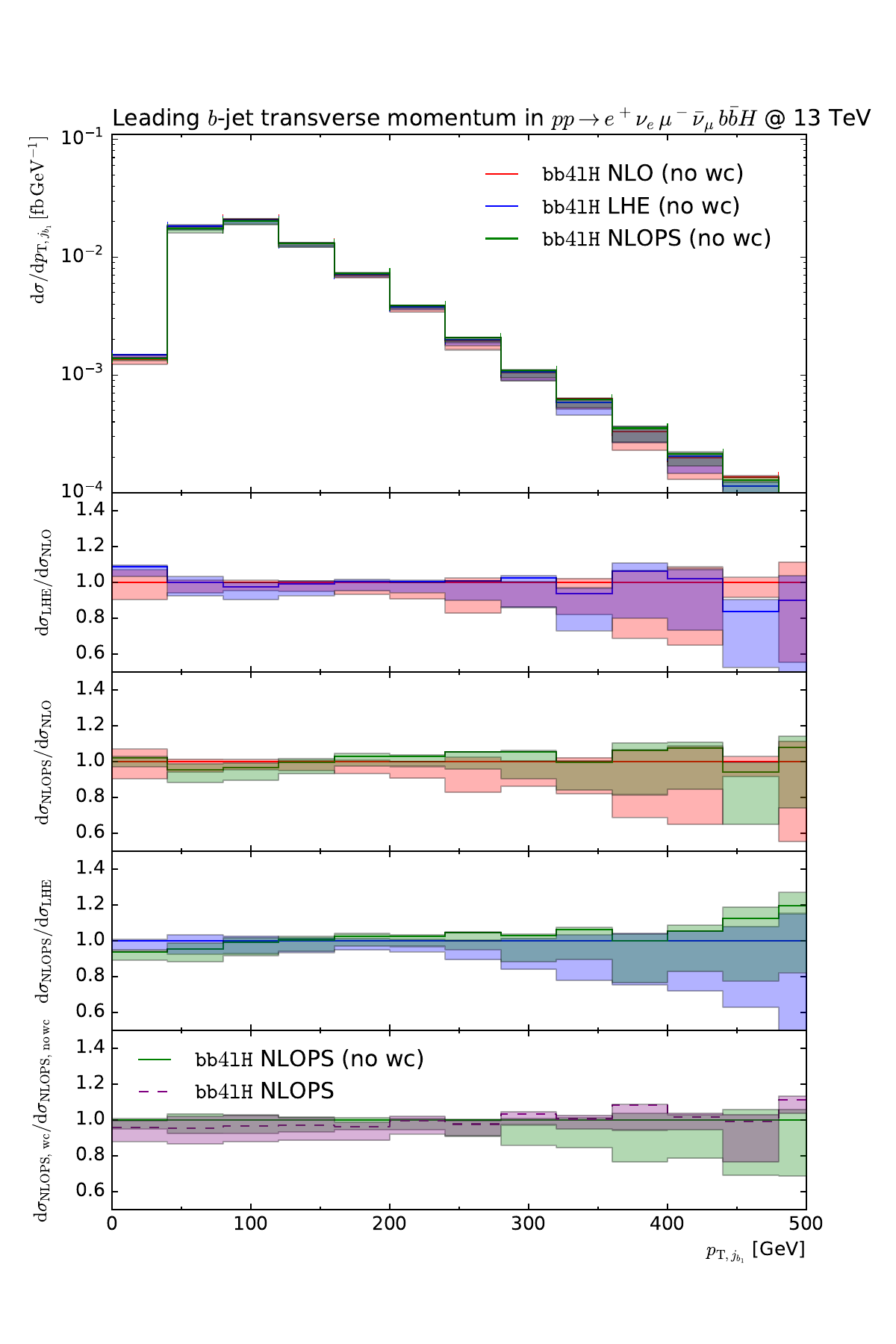}
    \caption{}
    \label{fig:dl_NLOPS_ptb1}
  \end{subfigure}
  \caption{Distributions in the invariant mass of the two $b$-jets
    (\subref{fig:dl_NLOPS_mbb}) and in the transverse momentum of the
    leading $b$-jet (\subref{fig:dl_NLOPS_ptb1}).
    Same acceptance cuts, predictions and ratios as in
    \reffi{fig:dl_NLOPS_higgs}.}
  \label{fig:dl_NLOPS_bjets}
\end{figure}

Also for observables that involve $b$ jets (\reffi{fig:dl_NLOPS_bjets})
the differences between fixed-order NLO and LHE predictions are rather
small.
In the distribution in the invariant mass of the two $b$ jets, LHE and
NLO predictions agree within $2$--$3\%$ in the entire plotted range, apart
from statistical fluctuations in the tail.
In the distribution in the transverse momentum of the leading $b$ jet,
the LHE prediction agrees with NLO within $2.5\%$ for
$40\,\GeV<p_{{\rm T},b_1}<200\,\GeV$, while it exceeds NLO by $9\%$
for $p_{{\rm T},b_1}<40\,\GeV$ and by up to $2\%$ for
$200\,\GeV<p_{{\rm T},b_1}<320\,\GeV$.
Thus, LHE/NLO ratios are rather flat and remain within $5\%$ from one,
apart from the first $p_{{\rm T},b_1}$ bin.
In contrast, NLOPS/NLO ratios feature shape distortions that range from
about $-4$--$5\%$ at small $m_{b\bar b}$ and $p_{{\rm T},b_1}$ to
$+5$--$8\%$ in the tails, and originate mostly from the parton shower.
More precisely, the parton shower reduces the cross section by about
$4\%$ and $6\%$ at small $m_{b\bar b}$ and $p_{{\rm T},b_1}$, respectively,
while in the tails it leads to positive corrections of up to $11\%$.
This hardening of the $b$-jet spectra is consistent with the
contamination of $b$ jets by soft QCD radiation from the production
subprocess, which is mostly generated by the parton shower, as discussed
in more detail below.

\begin{figure}[tbp]
  \centering
  \begin{subfigure}{0.48\linewidth}
    \centering
    \includegraphics[width=\linewidth]{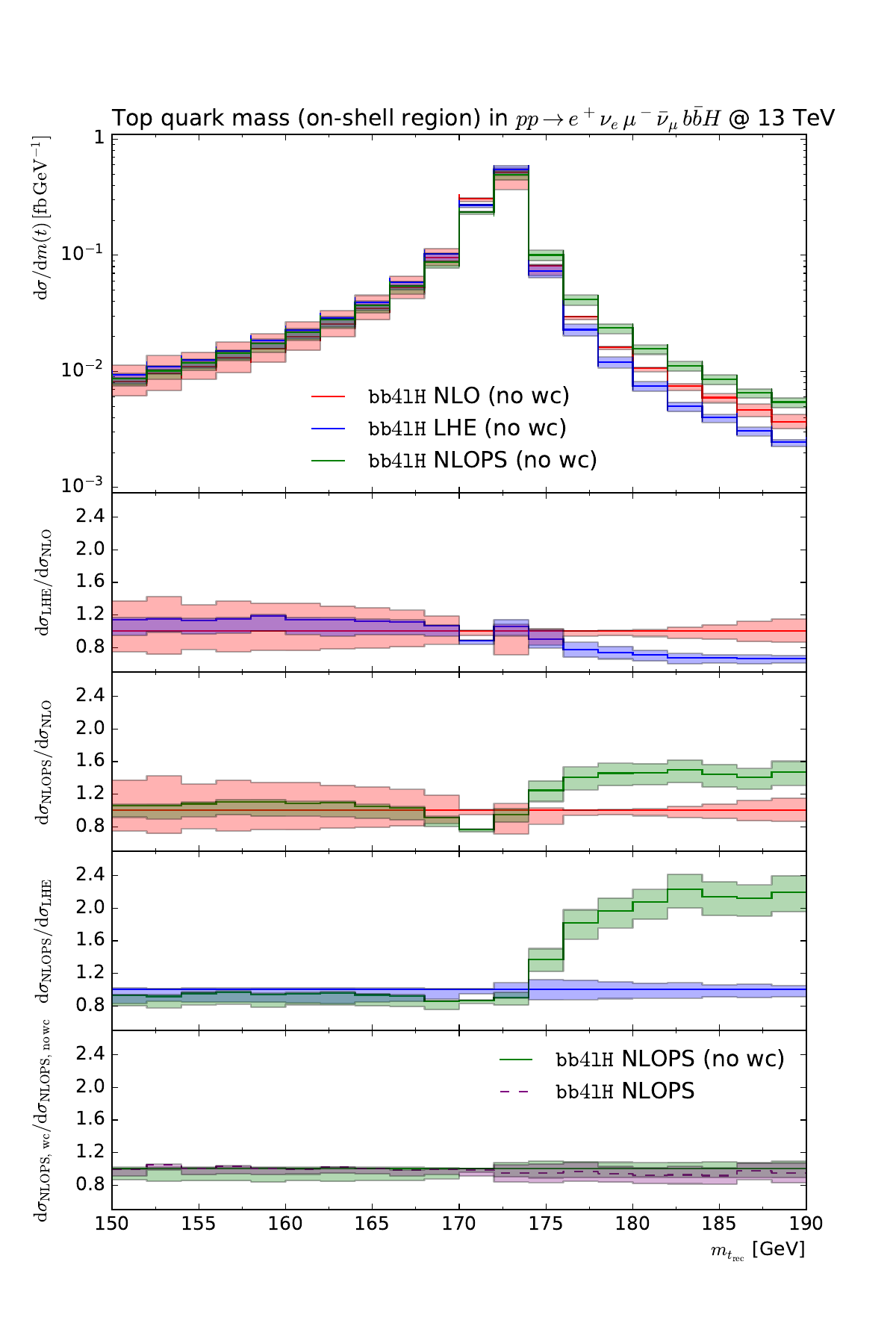}
    \caption{}
    \label{fig:dl_NLOPS_mt}
  \end{subfigure}
  \hspace{0.01\linewidth}
  \begin{subfigure}{0.48\linewidth}
    \centering
    \includegraphics[width=\linewidth]{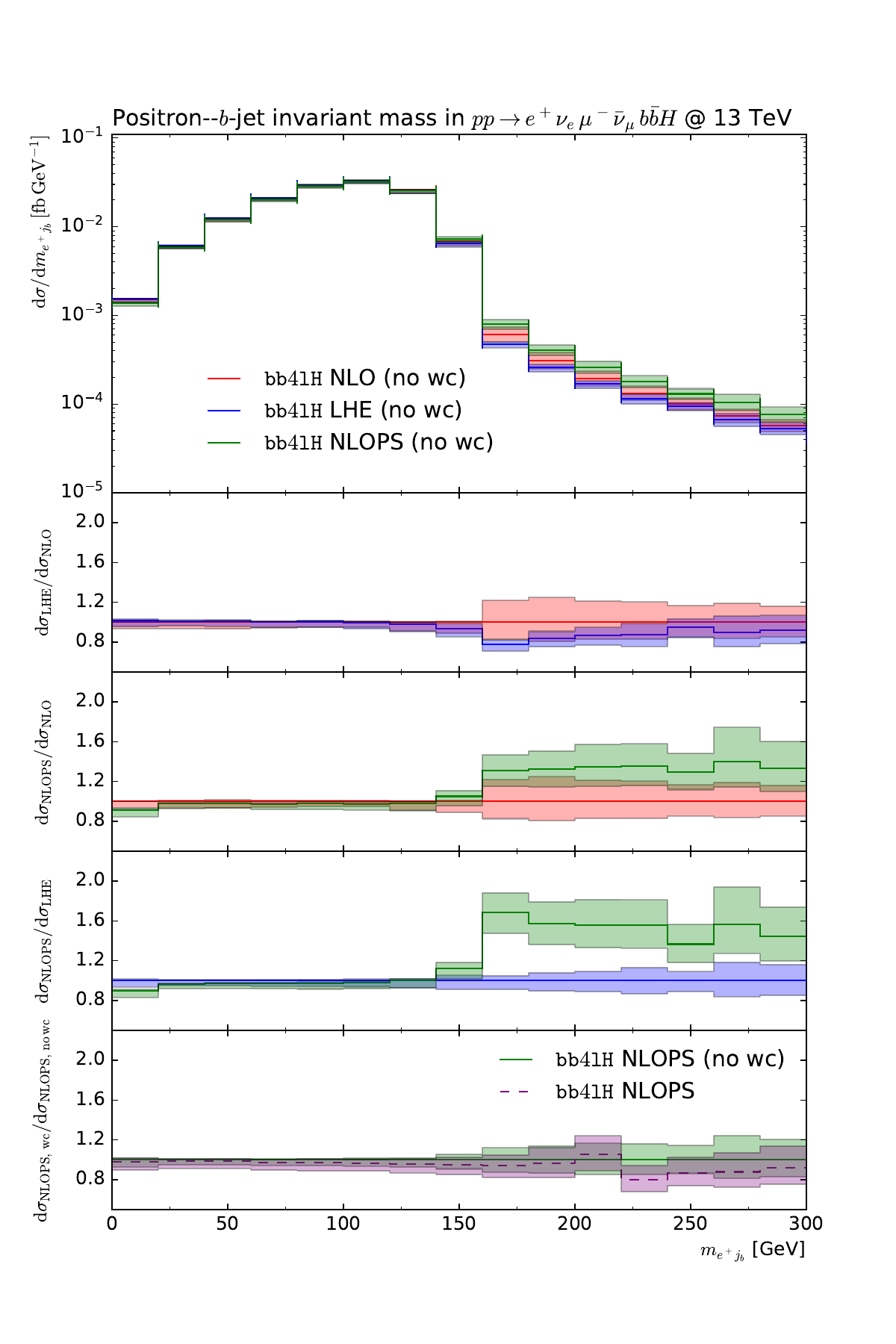}
    \caption{}
    \label{fig:dl_NLOPS_meb}
  \end{subfigure}
  \caption{Distributions in the invariant mass of the reconstructed top
    quark in the resonance region (\subref{fig:dl_NLOPS_mt}) and in the
    invariant mass of the positron and the $b$-jet
    (\subref{fig:dl_NLOPS_meb}).
    Same acceptance cuts, predictions and ratios as in
    \reffi{fig:dl_NLOPS_higgs}.}
  \label{fig:dl_NLOPS_top}
\end{figure}

In \reffi{fig:dl_NLOPS_mt} we show the distribution in the invariant mass
of the reconstructed top quark, $m_{t,{\rm rec}}$, in the vicinity of the
resonance, where the top quark is reconstructed from the positron, the
corresponding neutrino at Monte Carlo truth level and the associated $b$ jet.
At LO, the $b$ jet coincides with the $b$ quark from the top decay, and
the $m_{t,{\rm rec}}$ distribution follows the Breit--Wigner shape of the
top-quark propagator.
Beyond LO, QCD radiation distorts this shape in two ways.
Emissions from the top decay that are not clustered into the $b$ jet shift
events towards lower masses, while emissions that do not originate from
the top decay, mainly initial-state radiation and radiation off the top
quarks in the production subprocess, can be clustered into the $b$ jet and
shift events towards higher masses.
For a gluon with energy $E_g$ in the top-quark rest frame, the
corresponding shift in $m_{t,{\rm rec}}$ is of order $E_g$.
Thus, the regions below and above the peak are populated to a large extent
by relatively soft emissions, and fixed-order NLO predictions are
effectively only LO accurate in these regions.
Indeed, as we have checked by comparison with the LO prediction (not
shown), below the peak the NLO cross section exceeds LO by up to one order
of magnitude in the considered invariant mass range, while above the peak the NLO/LO ratio grows from $1.1$ at
$177\,\GeV$ to more than $1.4$ for $m_{t,{\rm rec}}\gtrsim 182\,\GeV$.

Below the peak, the LHE prediction exceeds NLO by $12$--$19\%$, which is
well within the large NLO scale uncertainty, ranging from $+26\%$ to
$+43\%$ and from $-19\%$ to $-28\%$.
This moderate enhancement can be attributed to the fact that \POWHEG{}
emissions from the top-quark decays involve $\alpha_s$ evaluated at their
transverse momentum, while at fixed order $\alpha_s$ is evaluated at the
hard scale $\mu_0=m_t+m_H/2$, and to the rather mild Sudakov suppression
of radiation in the top-quark decays.
In the peak bin, $172\,\GeV<m_{t,{\rm rec}}<174\,\GeV$, the LHE
prediction exceeds NLO by $5\%$, while in the two adjacent bins it lies
$9$--$12\%$ below NLO.
Above the peak, for $m_{t,{\rm rec}}>176\,\GeV$, the LHE prediction lies
$23$--$34\%$ below NLO and closely follows the LO Breit--Wigner shape.
This deficit is well outside the NLO scale band, which ranges from $+1\%$
and $-6\%$ at $177\,\GeV$ to $+15\%$ and $-13\%$ at $189\,\GeV$.
It reflects the fact that, in the production subprocess, the LHE level features only one (the hardest) emission, such that soft production-stage
emissions that could be clustered into the $b$-jet are strongly suppressed
by the corresponding Sudakov form factor, while at fixed order they are
described by unsuppressed $\mathcal{O}(\alpha_s)$ real emission.
Soft production-stage emissions are restored by the parton shower, which
generates multiple emissions with $\alpha_s$ evaluated at their transverse
momenta.
As a result, above the peak shower effects are huge, and the NLOPS/LHE
ratio grows from $1.4$ at $175\,\GeV$ to about $2$--$2.4$ for
$m_{t,{\rm rec}}\gtrsim 180\,\GeV$, such that the NLOPS prediction exceeds
NLO by $40$--$50\%$ for $m_{t,{\rm rec}}>176\,\GeV$.
Since the cross section above the peak is suppressed by one to two orders
of magnitude, even a small fraction of contaminated events has a large
relative impact.
In this region, the scale variations of the fixed-order NLO prediction,
which probe $\alpha_s$ only in the vicinity of the hard scale, clearly
underestimate the actual theoretical uncertainty.
In contrast, below and at the peak the parton shower has a rather
moderate impact, and NLOPS/LHE corrections amount to about $-5\%$ below
the peak and range from $-8\%$ to $-14\%$ in the bins around the peak.

Finally, in \reffi{fig:dl_NLOPS_meb} we present the distribution in the
invariant mass of the positron and the $b$ jet, i.e.\ the visible decay
products of the top quark.
At LO and for on-shell top quarks, this observable features a kinematic
edge at $m_{e^+j_b}^2=m_t^2-m_W^2$.
The region above the edge is populated only through QCD radiation,
off-shell and non-resonant effects, and single-resonant $\bar tW^+H$
contributions, where the positron does not stem from a top-quark
decay.
Below the edge, LHE and NLOPS predictions agree with NLO within $3\%$ and
$4\%$, respectively, apart from the first bin, where the NLOPS prediction
lies $9\%$ below NLO.
Above the edge, the LHE prediction lies $5$--$22\%$ below NLO, while shower
radiation enhances the cross section by a factor $1.4$--$1.7$, such that
NLOPS predictions exceed NLO by about $30$--$40\%$.
Similarly as for the region above the top-quark resonance, this behaviour
originates from the contamination of the $b$-jet through QCD radiation
that does not originate from the top decay, which is suppressed at the
LHE level and restored by the parton shower.
In this region fixed-order NLO predictions are to a large extent only LO
accurate, and their scale uncertainty of about $\pm(15$--$25)\%$
underestimates the actual theoretical uncertainty.

In summary, for Higgs-boson and leptonic observables, NLO, LHE and NLOPS
predictions agree at the few-percent level, while $b$-jet observables
feature shape differences of up to about $10\%$ that originate mostly from
the parton shower.
Large differences between fixed-order NLO and \POWHEG{} predictions arise
only in regions that are populated by soft QCD radiation, such as above
the peak of the reconstructed top-quark resonance and above the kinematic
edge of the $m_{e^+j_b}$ distribution, where fixed-order NLO predictions
are effectively only LO accurate.
Apart from regions that are populated only through QCD radiation, the
impact of the parton shower on top of LHE predictions is moderate.
This reflects the fact that in \bbfourlH{} the hardest emissions from
the production and from all top-quark decays are controlled by matrix
elements.


\subsection{Comparison to on-shell generators in the dileptonic channel}
\label{sec:results_dl_onoff}

In this section we compare predictions with the off-shell \bbfourlH{}
generator against the on-shell \ttH and \tWH generators, in the same
two--$b$-jet fiducial phase space and for the same observables as in
\refse{sec:results_dl_nlops}.
All generators are based on the same input parameters, PDFs, \POWHEG{}
and shower settings, as described in \refse{sec:setup}.
In order to quantify the relative importance of the $\ttbar H$ and $tWH$
production modes, we present pure \ttH predictions as well as the
combination \ttH{}+\tWH{}.
Moreover, both for \bbfourlH{} and \ttH{}+\tWH{} we compare predictions
at the LHE and NLOPS level, which illustrates the very different role of
the parton shower in the two approaches. For the \bbfourlH{}
predictions inverse width corrections are included throughout.

As discussed in \refse{Sec:Structure}, this comparison provides
quantitative information on various physics ingredients that are
included at NLOPS accuracy in \bbfourlH{}, while in the on-shell
generators they are either absent or only included in an approximate
way.
%
%
%
In particular, in the on-shell \tWH generator the second $b$ jet can only
arise from real emission or from initial-state parton-shower evolution,
such that $tWH$ contributions to the fiducial phase space with two
$b$ jets are at most LOPS accurate.

\begin{table}[tbp]
  \centering
  \begin{tabular}{ll|cc|cc}
\hline
level & generator & $\sigma_{\rm inc.}$ [fb] & $\sigma_{\rm inc.}/\sigma^{\rm NLOPS}_{\rm inc.,\,\bbfourlH{}}$ & $\sigma_{\rm fid.}$ [fb] & $\sigma_{\rm fid.}/\sigma^{\rm NLOPS}_{\rm fid.,\,\bbfourlH{}}$ \\
\hline
\multirow{4}{*}{LHE}   & \bbfourlH{}                 & 5.99(11)   & 1.00(3)    & 2.661(4)   & 1.007(2)   \\
                       & $\texttt{ttH}$              & 5.629(7)   & 0.939(17)  & 2.790(5)   & 1.055(2)   \\
                       & $\texttt{tWH}$              & 0.21088(13)& 0.0352(6)  & 0.03528(5) & 0.01335(3) \\
                       & $\texttt{ttH}+\texttt{tWH}$ & 5.840(7)   & 0.975(18)  & 2.825(5)   & 1.069(2)   \\
\hline
\multirow{4}{*}{NLOPS} & \bbfourlH{}                 & 5.99(11)   & 1.00(3)    & 2.643(4)   & 1.000(2)   \\
                       & $\texttt{ttH}$              & 5.629(7)   & 0.939(17)  & 2.451(4)   & 0.927(2)   \\
                       & $\texttt{tWH}$              & 0.21088(13)& 0.0352(6)  & 0.04347(6) & 0.01645(3) \\
                       & $\texttt{ttH}+\texttt{tWH}$ & 5.840(7)   & 0.975(18)  & 2.495(4)   & 0.944(2)   \\
\hline
  \end{tabular}
  \caption{Cross sections for
    $pp\to e^+\nu_e\,\mu^-\bar\nu_\mu\,b\bar b\,H$ at $13\,\TeV$ in the
    fully inclusive phase space ($\sigma_{\rm inc.}$) and in the
    dileptonic fiducial phase space ($\sigma_{\rm fid.}$) defined in
    \refse{sec:setup_cuts}.
    The rows report predictions of \bbfourlH{}, \ttH, \tWH and
    \ttH{}+\tWH{} at the LHE level and after showering with \Pythia{}
    (NLOPS), together with their ratios wrt the NLOPS prediction of
    \bbfourlH{}.
    The \tWH generator is used in the DR1 scheme.
    Numbers in parentheses indicate Monte Carlo uncertainties on the last
    digit(s).}
  \label{tab:xsec table}
\end{table}

Integrated cross sections at the LHE and NLOPS level are reported in
\refta{tab:xsec table}.
Let us first focus on the fully inclusive cross sections.
In this case, as expected from the unitarity of the \POWHEG{} matching,
LHE and NLOPS predictions coincide within Monte Carlo uncertainties.
As compared to \bbfourlH{}, the inclusive \ttH cross section features a
deficit of $6\%$, which is compensated by a $tWH$ contribution
of $3.5\%$, such that the combined \ttH{}+\tWH{} prediction lies
$2.7\%$ below \bbfourlH{}, which may point to $\ttbar H$--$tWH$ interference effects.

Moving to the fiducial phase space, we observe that the \bbfourlH{} cross
section is almost insensitive to the parton shower, i.e.\ NLOPS and LHE
predictions differ by less than $1\%$.
This is due to the fact that, in \bbfourlH{}, the hardest QCD emissions
from the production and from both top-quark decays are entirely
controlled by matrix elements.
This is not the case for the on-shell generators, where QCD radiation in
top-quark decays is entirely generated by the parton shower.
In the case of \ttH, the shower reduces the fiducial cross section by
$12\%$.
Similarly as for on-shell $\ttbar$ production~\cite{Jezo:2023rht}, this
can be attributed to negative shifts of the $b$-jet transverse momenta,
which result from QCD radiation that escapes the $b$ jets.
As a consequence, the fiducial acceptance of the on-shell generators
decreases from $48\%$ at the LHE level, where $b$ jets coincide with the
$b$ quarks from LO decays, to $43\%$ at the NLOPS level, while in
\bbfourlH{} it amounts to $44\%$ both at the LHE and NLOPS level.
In contrast, in the case of \tWH the parton shower increases the fiducial
cross section by $23\%$.
This is most likely due to the fact that the second $b$ jet of $tWH$
events can arise from initial-state $g\to b\bar b$ splittings that are
generated by the parton shower.
Nevertheless, due to the requirement of two $b$ jets, the fiducial
acceptance of \tWH amounts to only $21\%$, and its relative weight is
reduced from $3.6\%$ in the inclusive phase space to $1.7\%$ in the
fiducial phase space.

Comparing \bbfourlH{} to \ttH{}+\tWH{} in the fiducial phase space, we
find that the on-shell prediction exceeds \bbfourlH{} by $6.3\%$ at the
LHE level, while at the NLOPS level it lies $5.2\%$ below \bbfourlH{}.
Thus, the difference between on-shell and off-shell fiducial cross
sections is driven by the different modelling of QCD radiation in
top-quark decays, and changes sign when the parton shower is applied.
We also note that, at the NLOPS level, the deficit of the on-shell
prediction in the fiducial phase space is about twice as large as in the
inclusive phase space, and significantly larger than in the case of
semileptonic $\ttbar+tW$ production, where on-shell and off-shell
fiducial cross sections were found to agree at the percent
level~\cite{Jezo:2023rht}.

\begin{figure}[tp]
  \centering
  \begin{subfigure}{0.48\linewidth}
    \centering
    \includegraphics[width=\linewidth]{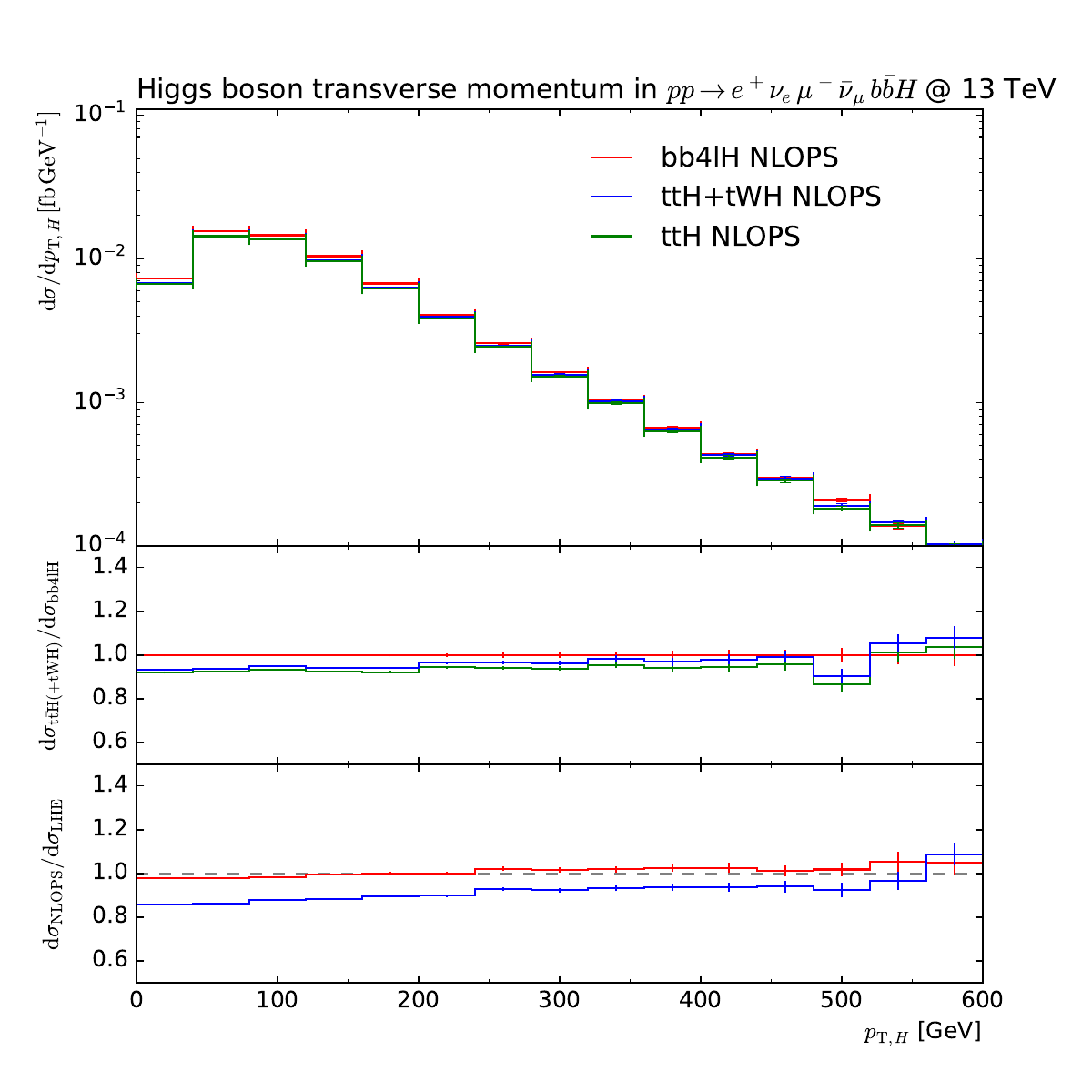}
    \caption{}
    \label{fig:dl_onoff_pTH}
  \end{subfigure}
  \hspace{0.01\linewidth}
  \begin{subfigure}{0.48\linewidth}
    \centering
    \includegraphics[width=\linewidth]{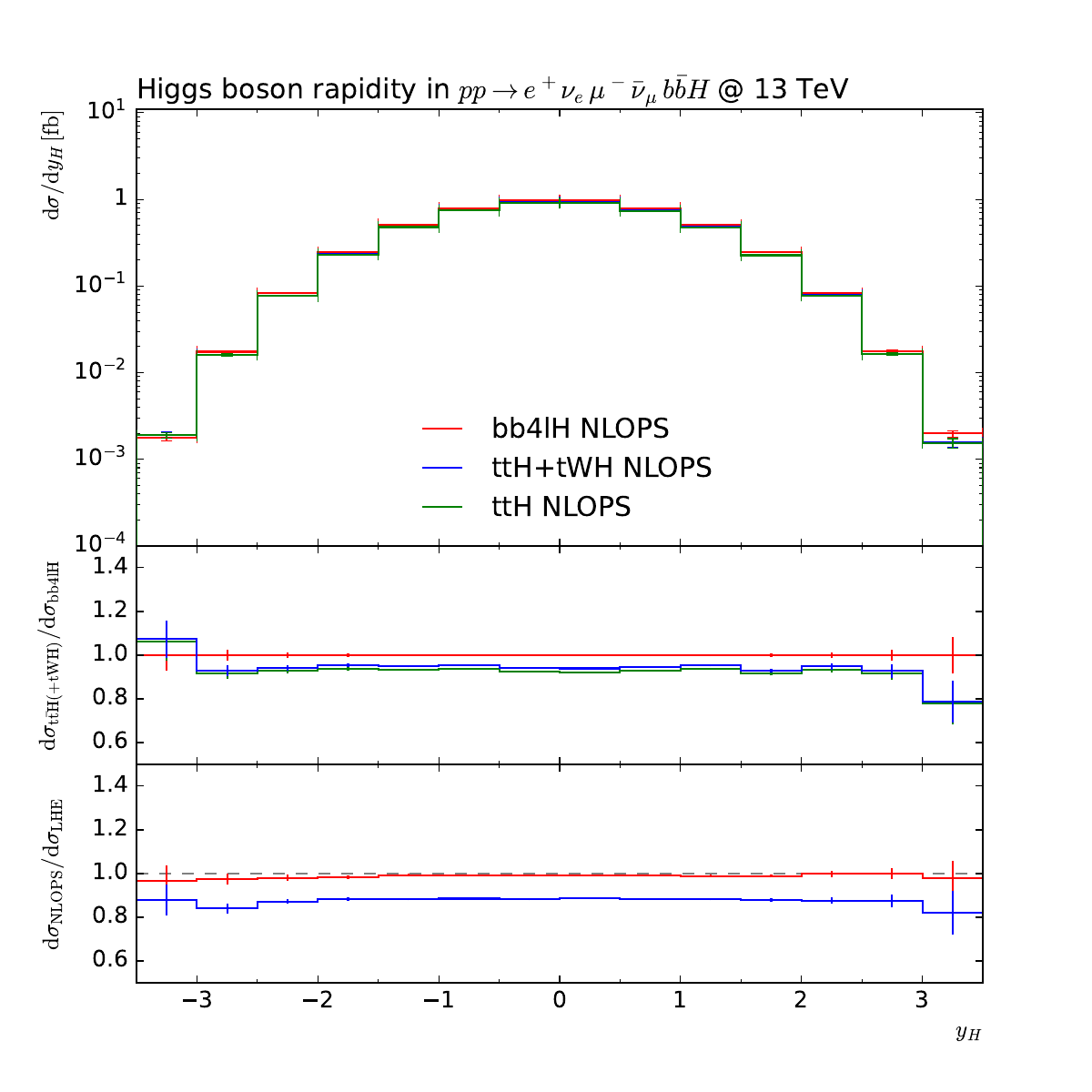}
    \caption{}
    \label{fig:dl_onoff_yH}
  \end{subfigure}
  \caption{Comparison of NLOPS predictions of the off-shell \bbfourlH{}
    generator and of the on-shell \ttH and \tWH generators for
    $pp\to e^+\nu_e\,\mu^-\bar\nu_\mu\,b\bar b\,H$ at $13\,\TeV$ in the
    dileptonic fiducial phase space defined in \refse{sec:setup_cuts}:
    distributions in the transverse momentum (\subref{fig:dl_onoff_pTH})
    and rapidity (\subref{fig:dl_onoff_yH}) of the Higgs boson.
    The upper frame shows NLOPS predictions of \bbfourlH{}, \ttH{}+\tWH{}
    and pure \ttH.
    The middle frame shows ratios wrt \bbfourlH{} for the case of
    \ttH{}+\tWH{} and \ttH, while the lower frame shows the NLOPS/LHE
    ratios of \bbfourlH{} and \ttH{}+\tWH{}.
    The \tWH generator is used in the DR1 scheme.
    Error bars indicate Monte Carlo uncertainties.}
  \label{fig:dl_onoff_higgs}
\end{figure}

In \reffis{fig:dl_onoff_higgs}{fig:dl_onoff_top} we turn to
differential distributions.
The upper frames show NLOPS predictions of \bbfourlH{},
\ttH{}+\tWH{} and pure \ttH, the middle frames show ratios wrt
\bbfourlH{}, and the lower frames show the NLOPS/LHE ratios of
\bbfourlH{} and \ttH{}+\tWH{}.

\begin{figure}[tbp]
  \centering
  \begin{subfigure}{0.48\linewidth}
    \centering
    \includegraphics[width=\linewidth]{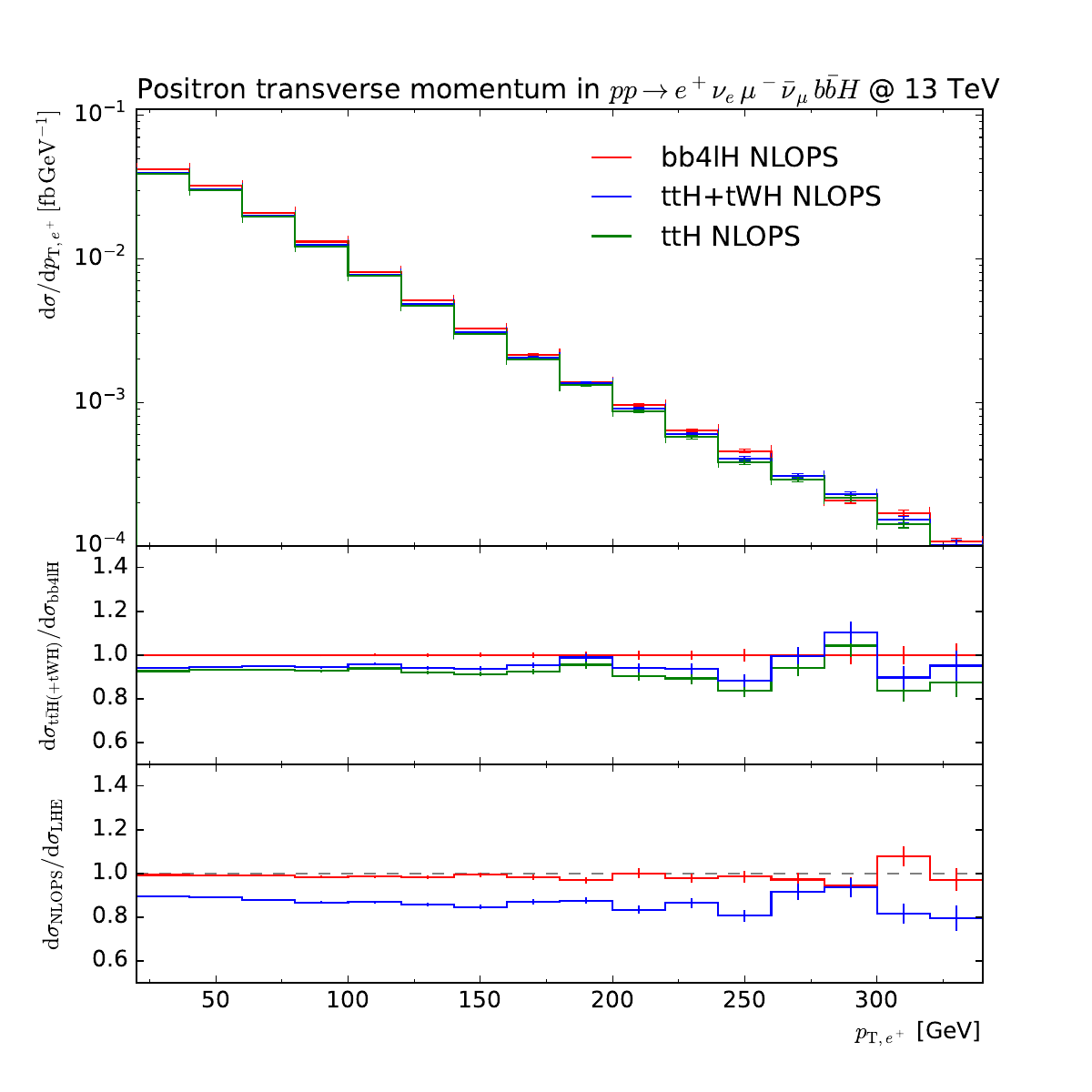}
    \caption{}
    \label{fig:dl_onoff_pTe}
  \end{subfigure}
  \hspace{0.01\linewidth}
  \begin{subfigure}{0.48\linewidth}
    \centering
    \includegraphics[width=\linewidth]{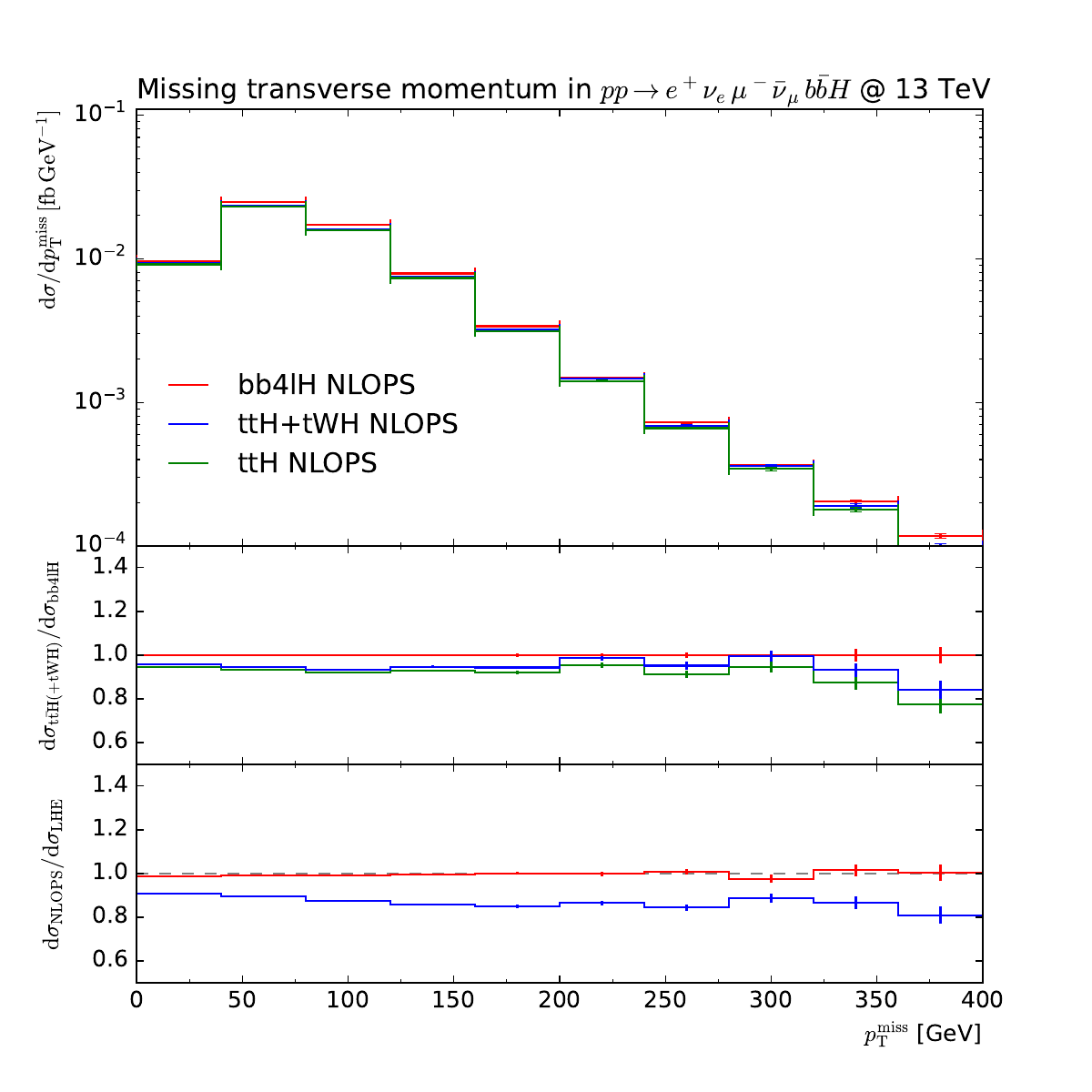}
    \caption{}
    \label{fig:dl_onoff_ptmiss}
  \end{subfigure}
  \caption{Distributions in the transverse momentum of the positron
    (\subref{fig:dl_onoff_pTe}) and in the missing transverse momentum
    (\subref{fig:dl_onoff_ptmiss}).
    Same acceptance cuts, predictions and ratios as in
    \reffi{fig:dl_onoff_higgs}.}
  \label{fig:dl_onoff_lep}
\end{figure}

Let us first consider the distributions in the transverse
momentum and rapidity  of the Higgs boson, shown in \reffi{fig:dl_onoff_higgs}.
In the rapidity distribution, the ratio of the \ttH{}+\tWH{} and
\bbfourlH{} predictions is nearly constant, and in the central region the
on-shell prediction lies $4$--$7\%$ below \bbfourlH{}, consistently with
the fiducial cross sections of \refta{tab:xsec table}.
The $tWH$ contribution amounts to $1$--$2\%$ in the entire rapidity
range.
Also the NLOPS/LHE ratios are nearly independent of $y_H$ and behave as
in the case of the fiducial cross section, with shower effects of about
$-12\%$ in the on-shell generators and of only a few percent in
\bbfourlH{}.

In the $p_{{\rm T},H}$ distribution the on-shell deficit amounts to
$5$--$7\%$ up to $200\,\GeV$, while it decreases to a few percent above,
and in the tail the two predictions are consistent within statistical
uncertainties.
This trend is mirrored by the NLOPS/LHE ratio of the on-shell generators,
which increases from $0.86$ at small $p_{{\rm T},H}$ to $0.92$--$0.97$
above $240\,\GeV$.
This suggests that the deficit is related to the acceptance losses
induced by shower radiation in top-quark decays, which are less
pronounced in the presence of boosted top quarks recoiling against a
hard Higgs boson.

\begin{figure}[tbp]
  \centering
  \begin{subfigure}{0.48\linewidth}
    \centering
    \includegraphics[width=\linewidth]{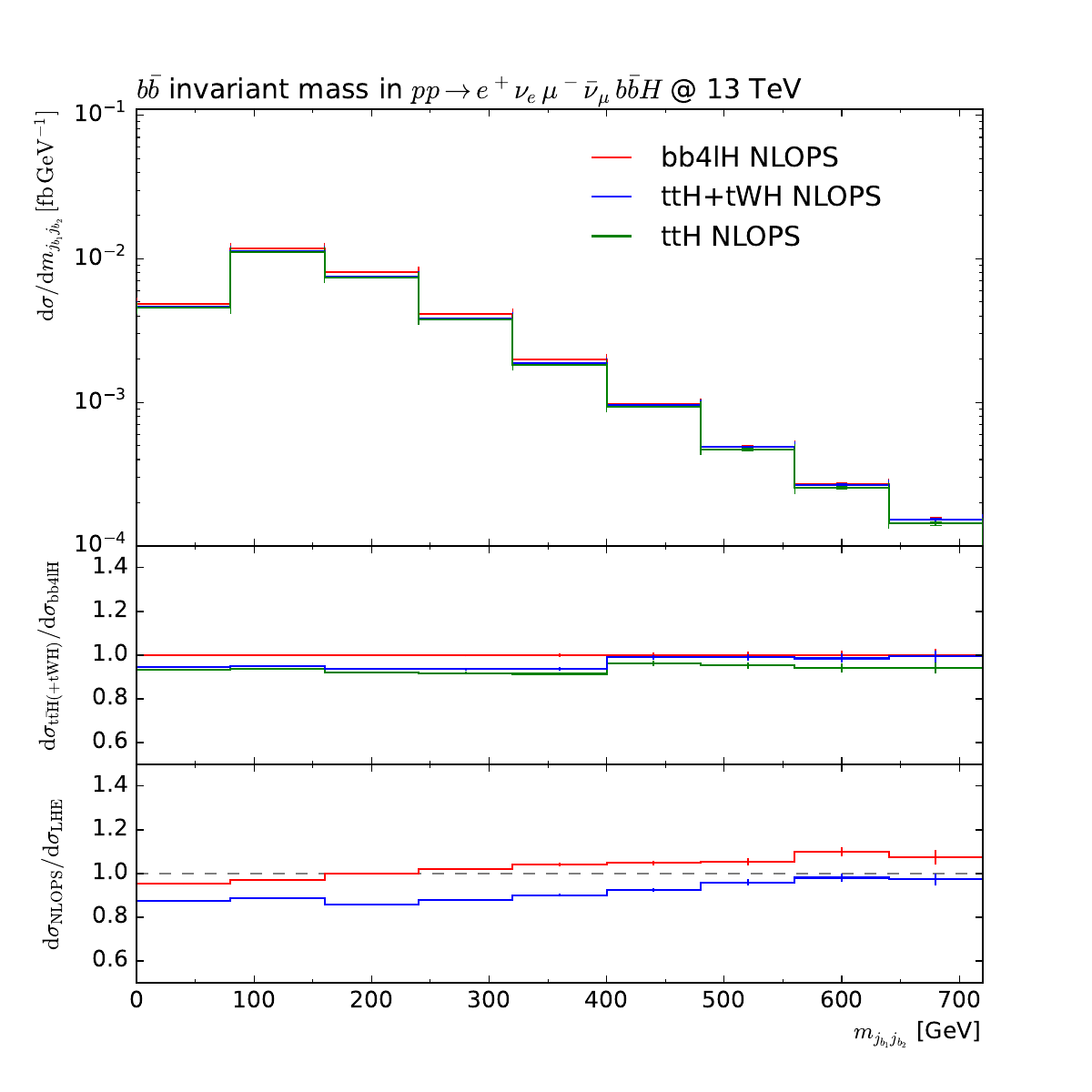}
    \caption{}
    \label{fig:dl_onoff_mbb}
  \end{subfigure}
  \hspace{0.01\linewidth}
  \begin{subfigure}{0.48\linewidth}
    \centering
    \includegraphics[width=\linewidth]{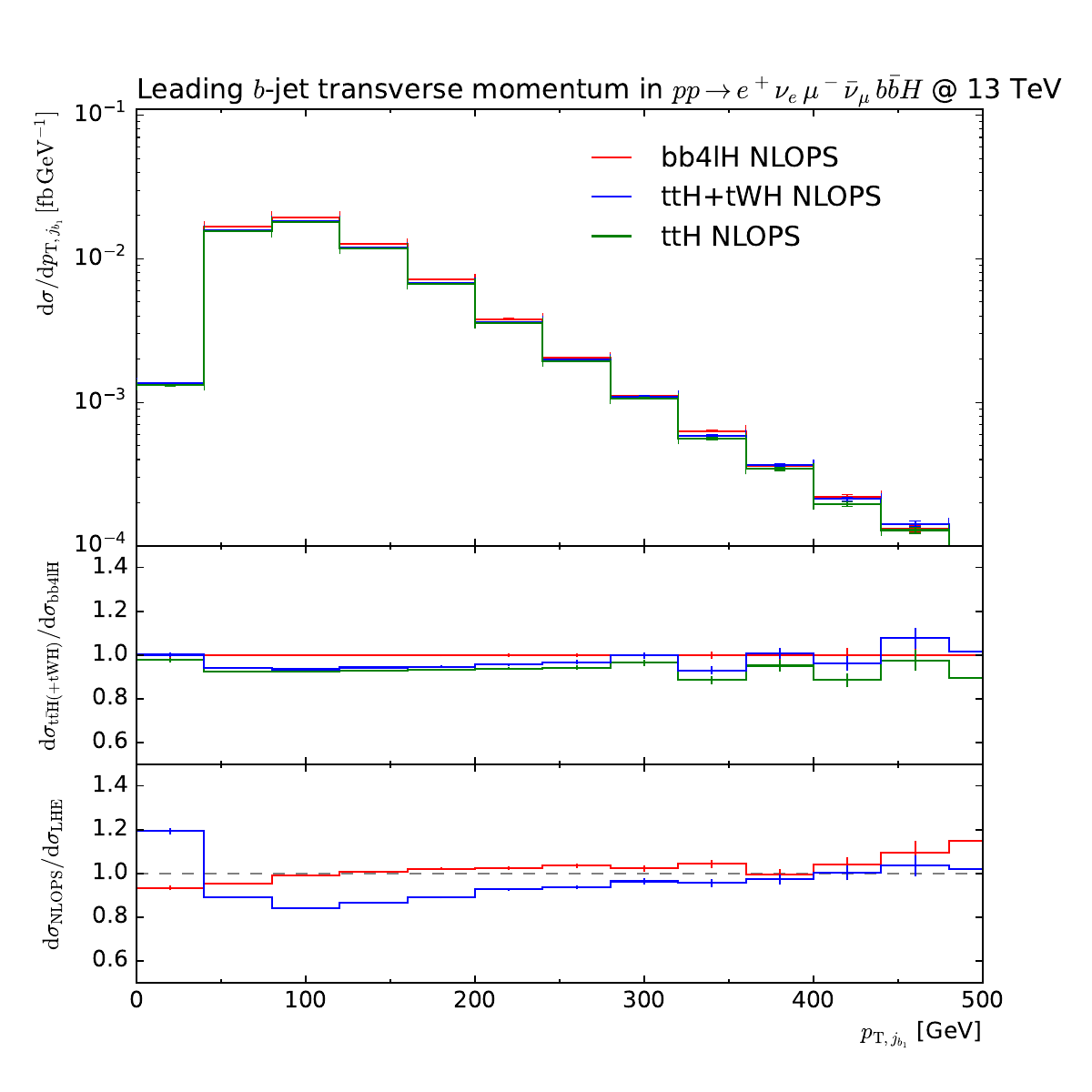}
    \caption{}
    \label{fig:dl_onoff_ptb1}
  \end{subfigure}
  \caption{Distributions in the invariant mass of the two $b$-jets
    (\subref{fig:dl_onoff_mbb}) and in the transverse momentum of the
    leading $b$-jet (\subref{fig:dl_onoff_ptb1}).
    Same acceptance cuts, predictions and ratios as in
    \reffi{fig:dl_onoff_higgs}.}
  \label{fig:dl_onoff_bjets}
\end{figure}

Also in the distributions in the positron transverse momentum and in the
missing transverse momentum (\reffi{fig:dl_onoff_lep}), the ratios of
on-shell and off-shell predictions are nearly constant up to about
$200\,\GeV$, and their values are consistent with the fiducial cross
section.
At higher transverse momenta, the relative weight of $tWH$ grows to about
$4$--$8\%$.
Here we note that $tWH$ predictions in the DR1 scheme do not include
$\ttbar H$--$tWH$ interference effects, which in the case of $\ttbar+tW$
production were found to be negative and comparable in size to the $tW$
contribution in the tail of the lepton-$p_{\rm T}$
distribution~\cite{Jezo:2023rht}.
The NLOPS/LHE ratios behave in a similar way as for the Higgs-boson
observables: in the bulk of the phase space shower effects amount to
$-10\%$ to $-16\%$ in the on-shell generators, and to only $1$--$4\%$ in
\bbfourlH{}.

For the distributions in the invariant mass of the two $b$-jets and in
the transverse momentum of the leading $b$ jet
(\reffi{fig:dl_onoff_bjets}), the on-shell prediction lies $3$--$7\%$
below \bbfourlH{} in the bulk of the phase space, while the two
predictions agree within a few percent for
$m_{b\bar b}\gtrsim 400\,\GeV$ and $p_{{\rm T},j_{b_1}}\gtrsim 240\,\GeV$.
In the $p_{{\rm T},j_{b_1}}$ distribution the ratio features a non-trivial
shape.
In the first bin, $p_{{\rm T},j_{b_1}}<40\,\GeV$, on-shell and off-shell
predictions agree within $1\%$, while for
$40\,\GeV<p_{{\rm T},j_{b_1}}<160\,\GeV$ the on-shell prediction lies about
$6\%$ below \bbfourlH{}.
This pattern is closely related to the behaviour of shower effects in the
on-shell generators, which shift $b$ jets from intermediate to small
transverse momenta: the NLOPS/LHE ratio amounts to $+19\%$ in the first
bin and to $-11\%$ to $-16\%$ for $40\,\GeV<p_{{\rm T},j_{b_1}}<160\,\GeV$.
In \bbfourlH{} shower effects vary only between $-7\%$ and $+6\%$ up to
$p_{{\rm T},j_{b_1}}\simeq 360\,\GeV$.
A similar behaviour is observed in the $m_{b\bar b}$ distribution, where
shower effects in the on-shell generators amount to $-10\%$ to $-14\%$ up
to $400\,\GeV$, while in \bbfourlH{} they remain within $10\%$ in the
entire plotted range.
In the light of the large magnitude of shower effects in the on-shell
generators, the agreement with \bbfourlH{} at the level of $5\%$ in the
normalisation, and of a few percent in the shapes, is quite remarkable.
This indicates that \Pythia{}, which applies matrix-element corrections
in top-quark decays, provides a fairly good description of QCD radiation
off $b$ quarks for these observables.

\begin{figure}[tbp]
  \centering
  \begin{subfigure}{0.48\linewidth}
    \centering
    \includegraphics[width=\linewidth]{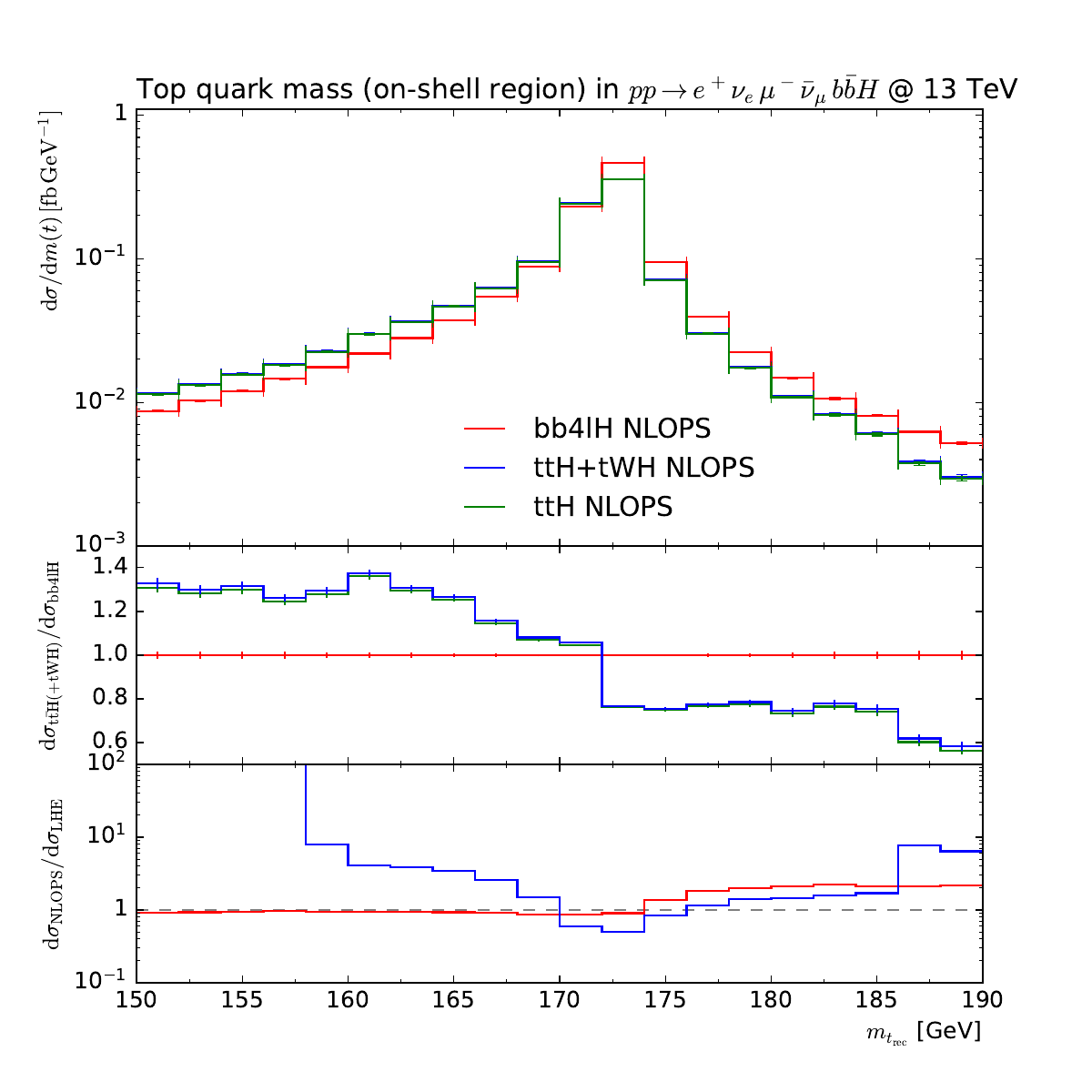}
    \caption{}
    \label{fig:dl_onoff_mt}
  \end{subfigure}
  \hspace{0.01\linewidth}
  \begin{subfigure}{0.48\linewidth}
    \centering
    \includegraphics[width=\linewidth]{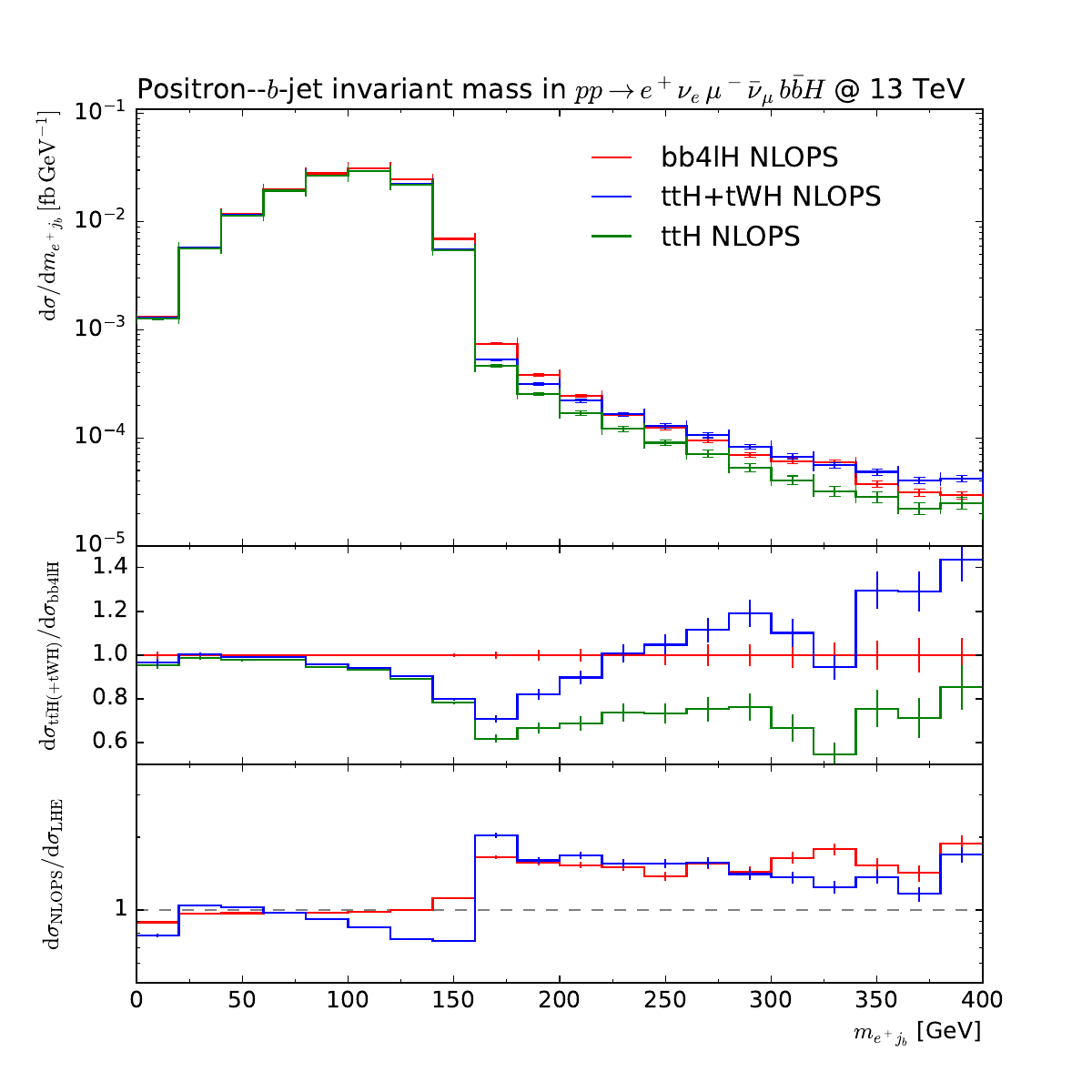}
    \caption{}
    \label{fig:dl_onoff_meb}
  \end{subfigure}
  \caption{Distributions in the invariant mass of the reconstructed top
    quark in the resonance region (\subref{fig:dl_onoff_mt}) and in the
    invariant mass of the positron and the $b$-jet
    (\subref{fig:dl_onoff_meb}).
    Same acceptance cuts, predictions and ratios as in
    \reffi{fig:dl_onoff_higgs}.}
  \label{fig:dl_onoff_top}
\end{figure}

Much larger differences between on-shell and off-shell generators show up
in the reconstructed top-quark line shape (\reffi{fig:dl_onoff_mt}).
Below the resonance, for $m_{t,{\rm rec}}<166\,\GeV$, the on-shell
prediction exceeds \bbfourlH{} by $26$--$38\%$, and the excess decreases
to $6$--$16\%$ in the bins immediately below the peak.
In contrast, in the peak bin the on-shell prediction lies about $23\%$
below \bbfourlH{}, and above the peak the deficit amounts to
$20$--$25\%$.
Since pure \ttH and \ttH{}+\tWH{} predictions are almost identical, these
differences are entirely due to the description of $\ttbar H$ production
and decay.
The NLOPS/LHE ratios reveal that the line shape of the on-shell
generators is completely controlled by the parton shower.
In fact, at the LHE level, where top-quark decays are described at LO,
the on-shell line shape essentially corresponds to the Breit--Wigner
distribution of top-quark virtualities.
Parton showering depletes the peak region by about a factor two and
shifts events towards the tails of the resonance, where the NLOPS/LHE
ratio reaches factors of $3$ to $4$ below the peak.
The sharp edges of the on-shell predictions at
$m_{t,{\rm rec}}\simeq 159\,\GeV$ and $186\,\GeV$ are a technical artefact
of the decay procedure of \refse{Sec:Top decay}, where the Breit--Wigner
smearing~\eqref{eq:top-BR} of the top-quark virtualities is restricted to
$|M-m_t|<10\,\Gamma_t$.
Beyond these limits, the on-shell line shape is populated only through
QCD radiation that escapes from, or is clustered into, the $b$ jet.
As a result, at the LHE level, where top-quark decays are described at LO,
the region below $159\,\GeV$ is essentially empty, while above
$186\,\GeV$ the NLOPS/LHE ratio jumps to $6$--$8$ and the deficit of the
on-shell prediction with respect to \bbfourlH{} increases to about
$40\%$.
In \bbfourlH{}, where the line shape at the LHE level is NLO accurate and
includes exact off-shell effects, shower effects amount to $-4\%$ to
$-14\%$ below and at the peak, while above the peak they reach about a
factor two as a result of $b$-jet contamination, as discussed in
\refse{sec:results_dl_nlops}.
The differences between the two approaches are thus of the same order as
the shower effects in the on-shell generators, and can be attributed to
the different modelling of QCD radiation in top-quark decays.
Given the relevance of the reconstructed top-quark line shape for the
kinematic reconstruction of $\ttbar H$ events, differences of this size
underline the importance of a consistent NLOPS description of top-quark
decays and off-shell effects, as provided by \bbfourlH{}.

Finally, the distribution in $m_{e^+j_b}$, shown in
\reffi{fig:dl_onoff_meb}, features an interesting interplay of shower and
$tWH$ effects.
For $m_{e^+j_b}\lesssim 80\,\GeV$, on-shell and off-shell predictions
agree within $3\%$, while approaching the kinematic edge the on-shell
prediction falls $4$--$20\%$ below \bbfourlH{}.
This behaviour is consistent with the shower effects in the on-shell
generators, which reduce the cross section by up to $25\%$ in the
vicinity of the edge, as a result of QCD radiation that escapes the
$b$-jet, while in \bbfourlH{} shower effects remain between $-4\%$ and
$+13\%$ for $m_{e^+j_b}>20\,\GeV$.
Above the edge, shower effects are large in both approaches, and enhance
the cross section by a factor $1.4$--$1.8$ in \bbfourlH{} and
$1.6$--$2.0$ in the on-shell generators, consistently with the
$b$-jet contamination effects discussed in \refse{sec:results_dl_nlops}.
At the same time, the relative weight of $tWH$ production grows very
rapidly.
While pure \ttH predictions lie $25$--$38\%$ below \bbfourlH{} between
$160$ and $280\,\GeV$, the $tWH$ contribution amounts to about $13\%$ of
the \ttH{}+\tWH{} cross section right above the edge, to $27$--$33\%$
between $220$ and $280\,\GeV$, and to $36$--$45\%$ for
$m_{e^+j_b}\gtrsim 280\,\GeV$.
This is due to $\bar tW^+H$ production, where the positron originates
from the $W^+$ boson that is produced in association with the anti-top
quark, such that $m_{e^+j_b}$ is not bounded by the kinematic edge.
As a result, \ttH{}+\tWH{} lies $10$--$29\%$ below \bbfourlH{} right
above the edge, while it exceeds \bbfourlH{} by $1$--$12\%$ between
$220$ and $280\,\GeV$ and by up to $44\%$ at higher invariant masses.
Since the DR1 scheme does not include $\ttbar H$--$tWH$ interference
effects, which in the analogous case of $\ttbar+tW$ production were
found to be negative and to reach $20\%$ in the tail of the
$m_{e^+j_b}$ distribution~\cite{Jezo:2023rht}, this excess may point to
sizeable destructive interference effects.
The region above the edge is thus particularly sensitive to the
separation of $\ttbar H$ and $tWH$ production, which can be described in
a consistent way only in the off-shell approach.

In summary, in spite of the large shower effects in the on-shell
generators, the combination of on-shell \ttH and \tWH generators
reproduces \bbfourlH{} at the level of $3\%$ for the inclusive and $5\%$
for the fiducial cross section, and of a few percent in the shapes of
Higgs-boson, leptonic and $b$-jet observables.
In contrast, differences of $20$--$40\%$ show up in the reconstructed
top-quark line shape and in the region above the kinematic edge of the
$m_{e^+j_b}$ distribution, where the description of top-quark decays,
off-shell effects and the separation of $\ttbar H$ and $tWH$ production
play a crucial role.

\subsection{Comparison to on-shell generators in the semileptonic channel}
\label{sec:results_sl_onoff}

\begin{figure}[tbp]
  \centering
  \begin{subfigure}{0.48\linewidth}
    \centering
    \includegraphics[width=\linewidth]{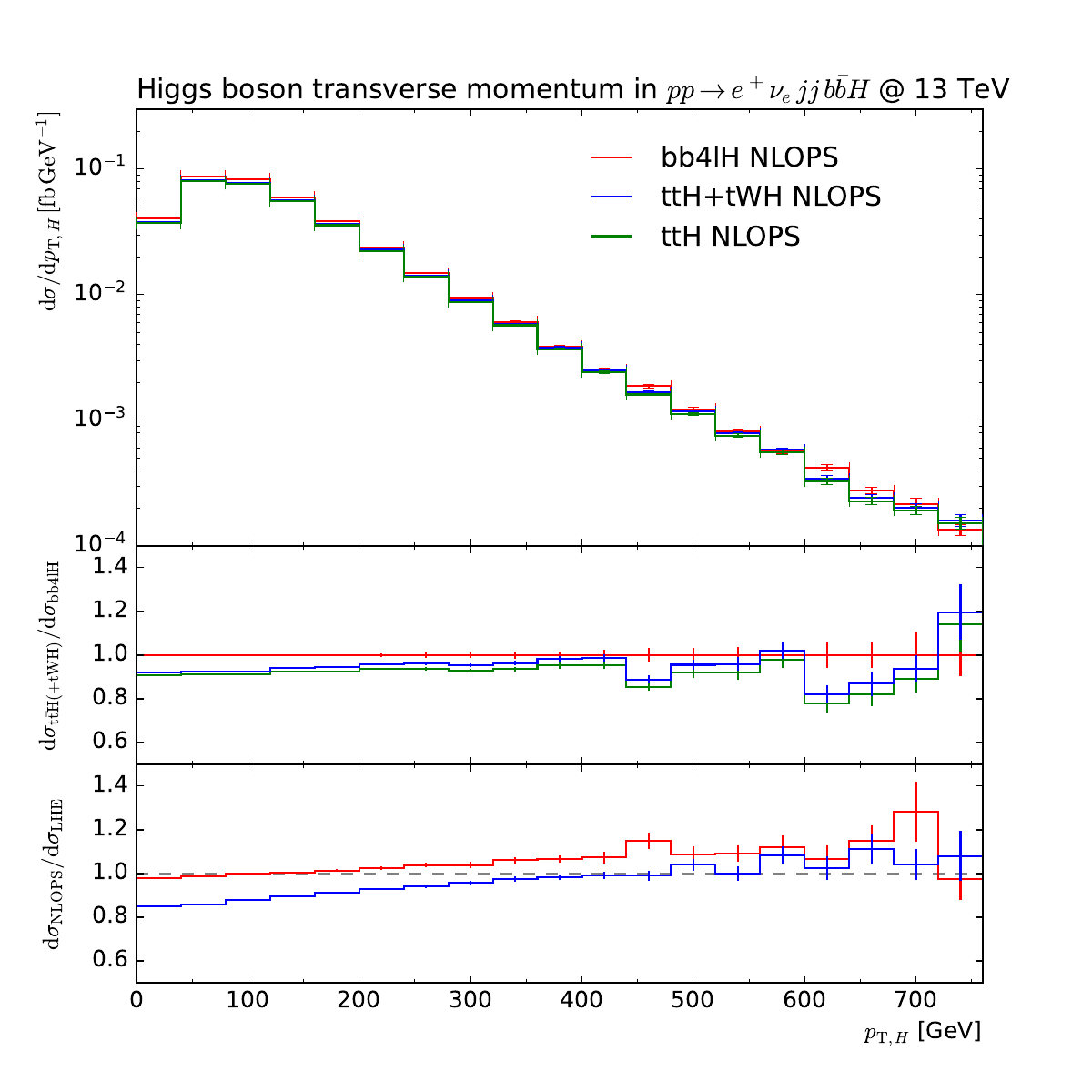}
    \caption{}
    \label{fig:sl_onoff_pTH}
  \end{subfigure}
  \hspace{0.01\linewidth}
  \begin{subfigure}{0.48\linewidth}
    \centering
    \includegraphics[width=\linewidth]{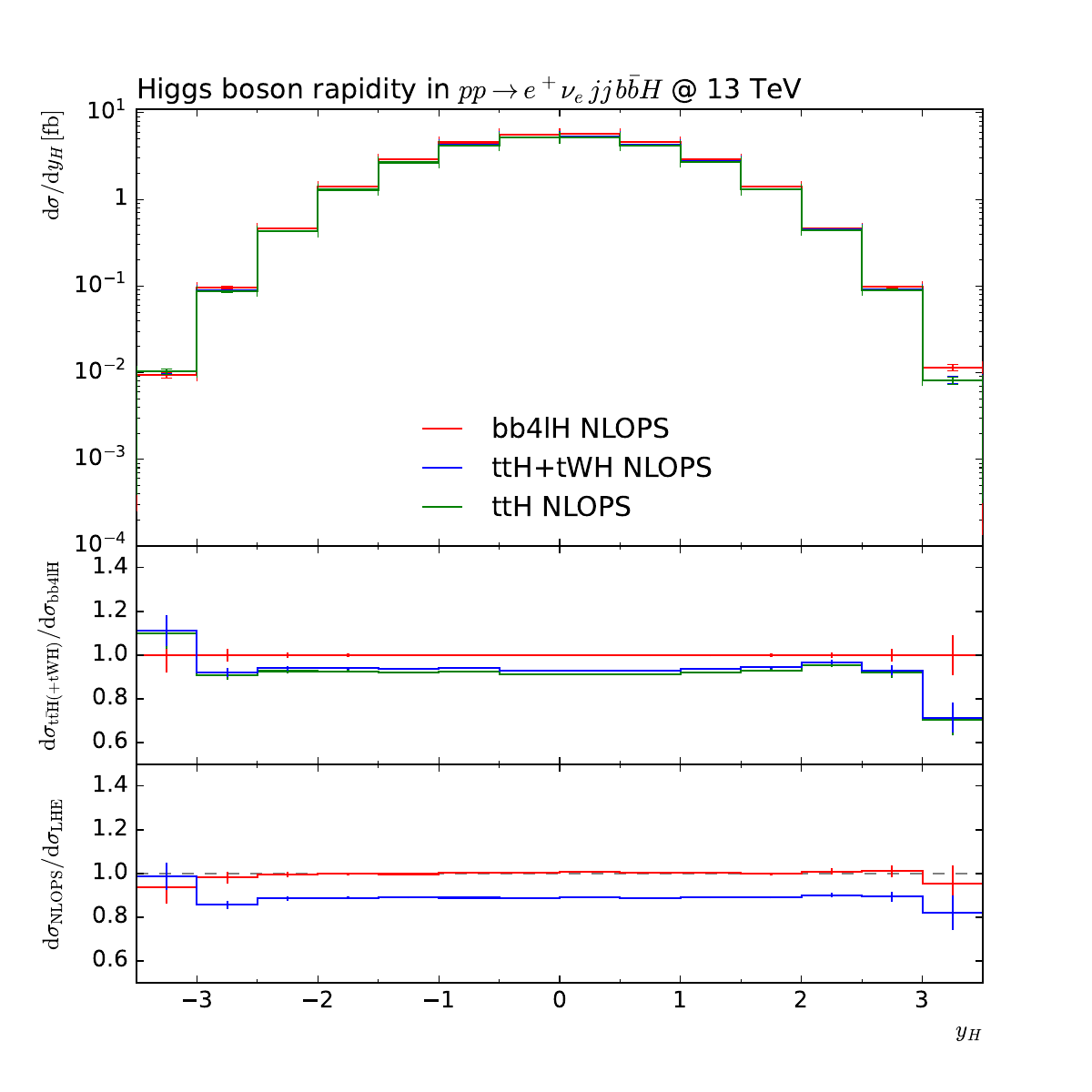}
    \caption{}
    \label{fig:sl_onoff_yH}
  \end{subfigure}
  \caption{Comparison of NLOPS predictions of the off-shell \bbfourlH{}
    generator and of the on-shell \ttH and \tWH generators for
    $pp\to e^+\nu_e\,jj\,b\bar b\,H$ at $13\,\TeV$ in the
    semileptonic fiducial phase space defined in \refse{sec:setup_cuts}:
    distributions in the transverse momentum (\subref{fig:sl_onoff_pTH})
    and rapidity (\subref{fig:sl_onoff_yH}) of the Higgs boson.
    Same predictions and ratios as in \reffi{fig:dl_onoff_higgs}.}
  \label{fig:sl_onoff_higgs}
\end{figure}

In this section we extend the comparison of on-shell and off-shell
generators to the semileptonic channel.
To this end we consider the process
$pp\to e^+\nu_e\,jj\,b\bar b\,H$ at $\sqrt{s}=13\,\TeV$, where the
top quark decays leptonically, while the anti-top quark decays
hadronically via $\bar t\to\bar b\,W^-\to\bar b\,jj$, and we apply
the semileptonic fiducial cuts of \refse{sec:setup_cuts}, i.e.\ we
require one positron, at least two $b$ jets and at least two light jets.
In \bbfourlH{} semileptonic events are obtained from dileptonic events as
described in \refse{sec:semileptonic}, such that the hardest emission
from the hadronic $W$ decay is generated by \POWHEG{}, and $\ttbar H$-type
events receive up to four \POWHEG{} emissions.
In contrast, in the on-shell generators the hadronic $W$ decay is
included at LO, and QCD radiation off the light quarks is entirely
generated by the parton shower.
Thus, as compared to the dileptonic channel, the semileptonic channel
provides additional sensitivity to the modelling of QCD radiation in
hadronic $W$ decays, which plays an important role for the
reconstruction of $\ttbar H$ events.

In \reffis{fig:sl_onoff_higgs}{fig:sl_onoff_jets} we present
distributions in the transverse momentum and rapidity of the Higgs boson,
in the transverse momentum of the system of the two hardest light jets,
$j_1j_2$, which approximates the hadronically decaying $W^-$ boson, and
in the transverse momentum of the leading light jet.
Predictions and ratios are the same as in \refse{sec:results_dl_onoff}.
For brevity, we restrict ourselves to this limited set of observables.
Other observables can be studied in the same way, keeping in mind the
range of validity of the semileptonic approach discussed in
\refse{sec:semileptonic}.
In particular, a realistic experimental analysis will involve the
reconstruction of the hadronically decaying $W$ boson, including the
grooming of QCD contamination.

The distributions in the transverse momentum and rapidity of the Higgs
boson, shown in \reffi{fig:sl_onoff_higgs}, behave in a similar way as in
the dileptonic channel.
In the rapidity distribution the ratio of the \ttH{}+\tWH{} and
\bbfourlH{} predictions is nearly constant, and in the central region the
on-shell prediction lies $6$--$7\%$ below \bbfourlH{}, as compared to a
deficit of about $5\%$ in the dileptonic fiducial cross section
(see \refta{tab:xsec table}).
The relative weight of $tWH$ production amounts to $1$--$2\%$, as in the
dileptonic channel.
Also the NLOPS/LHE ratios are remarkably similar to the dileptonic case:
parton showering reduces the on-shell prediction by about $11\%$, while
\bbfourlH{} is almost insensitive to the shower.
Since in the semileptonic channel the cuts on the light jets are
sensitive to QCD radiation in the hadronic $W$ decay, one might expect
more pronounced shower effects in the on-shell generators.
The fact that this is not the case suggests that acceptance losses due to
radiation off the light quarks are, to a large extent, compensated by
additional light jets from shower radiation, which can fulfil the
requirement of two light jets.

In the $p_{{\rm T},H}$ distribution the on-shell deficit decreases from
$7$--$8\%$ at small transverse momenta to $4$--$6\%$ between $120$ and
$360\,\GeV$, and in the tail the two predictions agree within
statistical uncertainties.
Here shower effects feature a more pronounced kinematic dependence than
in the dileptonic channel.
In the case of \bbfourlH{}, the NLOPS/LHE ratio grows from $0.98$ at
small $p_{{\rm T},H}$ to $1.06$--$1.08$ at $360$--$440\,\GeV$, while in
the on-shell generators it increases from $0.85$ to $0.98$--$0.99$ in the
same range.
Since this trend is similar in the two approaches, and is less
pronounced in the dileptonic channel (\reffi{fig:dl_onoff_pTH}), it is
most likely related to the light-jet requirement, which, in the presence
of a hard Higgs boson, can be fulfilled more easily through additional
jets from initial-state shower radiation.
This interpretation is consistent with the fact that initial-state
radiation is handled in a similar way in the different generators.

\begin{figure}[tbp]
  \centering
  \begin{subfigure}{0.48\linewidth}
    \centering
    \includegraphics[width=\linewidth]{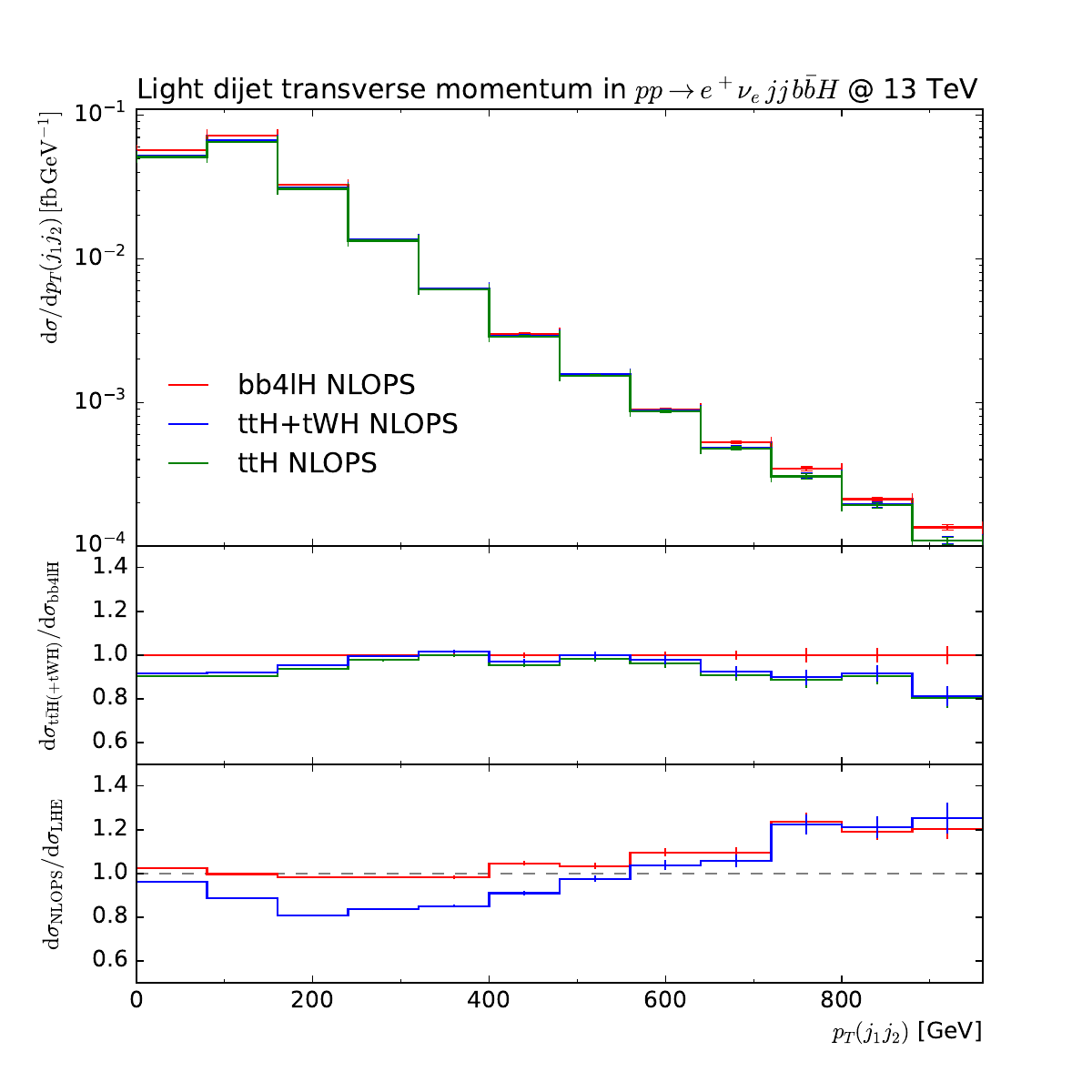}
    \caption{}
    \label{fig:sl_onoff_ptjj}
  \end{subfigure}
  \hspace{0.01\linewidth}
  \begin{subfigure}{0.48\linewidth}
    \centering
    \includegraphics[width=\linewidth]{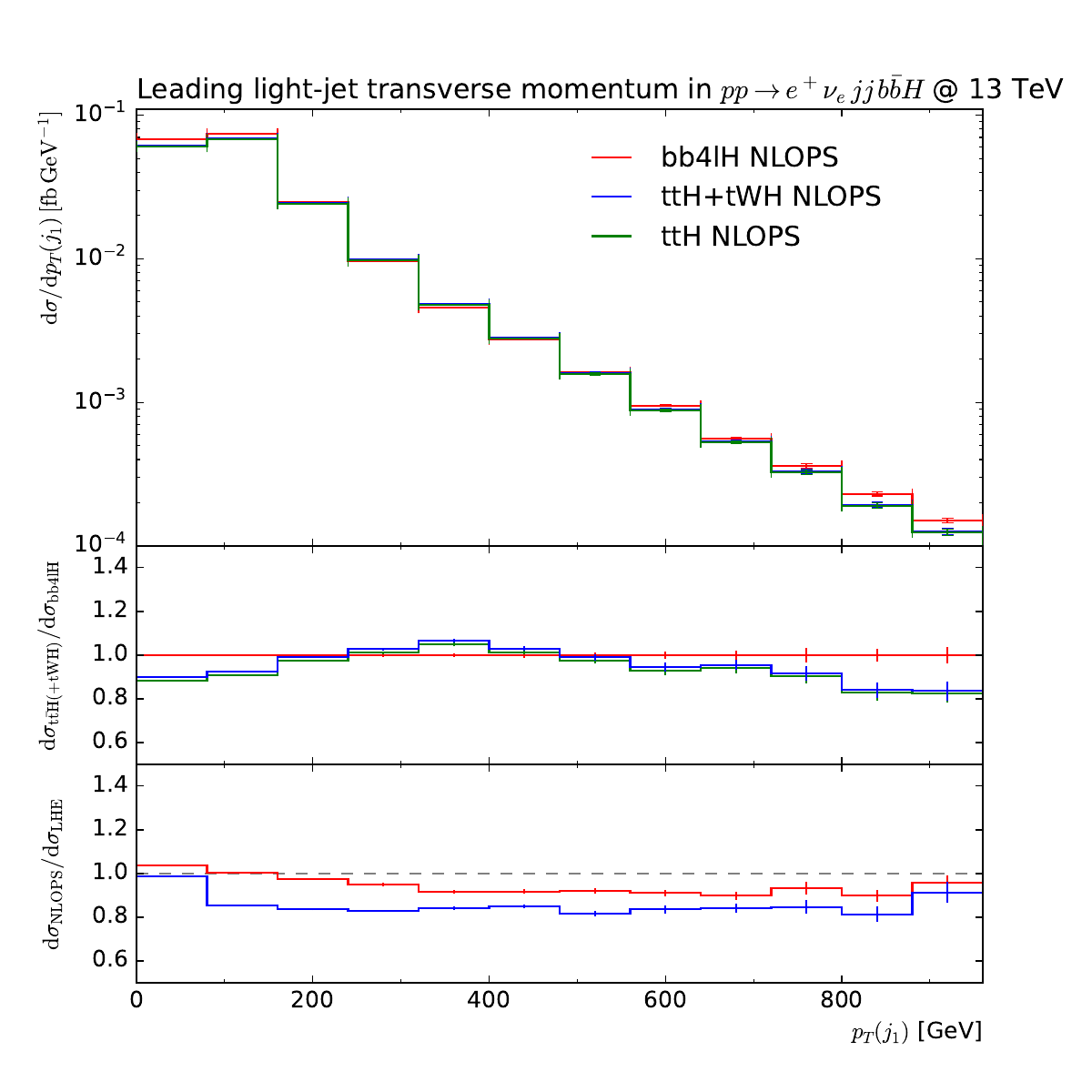}
    \caption{}
    \label{fig:sl_onoff_ptj1}
  \end{subfigure}
  \caption{Distributions in the transverse momentum of the system of the
    two hardest light jets (\subref{fig:sl_onoff_ptjj}) and of the
    leading light jet (\subref{fig:sl_onoff_ptj1}).
    Same acceptance cuts, predictions and ratios as in
    \reffi{fig:sl_onoff_higgs}.}
  \label{fig:sl_onoff_jets}
\end{figure}

Observables that involve light jets are directly sensitive to QCD
radiation in the hadronic $W$ decay.
In \reffi{fig:sl_onoff_jets} we show the distributions in the transverse
momentum of the light-dijet system, $p_{{\rm T},j_1j_2}$, and of the
leading light jet, $p_{{\rm T},j_1}$.
In the on-shell generators these observables are potentially subject to large shower
effects.
For $p_{{\rm T},j_1j_2}$ the NLOPS/LHE ratio amounts to $0.80$--$0.85$
between $160$ and $400\,\GeV$, and for $p_{{\rm T},j_1}$ it is around
$0.80$--$0.85$ above $80\,\GeV$, while in the first $p_{{\rm T},j_1}$
bin shower effects are at the percent level.
This behaviour is due to QCD radiation off the light quarks that escapes
the jets and shifts the light-jet spectrum towards lower transverse
momenta, similarly as observed for $b$ jets in the dileptonic channel
(\reffi{fig:dl_onoff_ptb1}).
In \bbfourlH{}, where the hardest emission from the hadronic $W$ decay is
already included at the LHE level, shower effects are much smaller.
For $p_{{\rm T},j_1j_2}$ they remain within $5\%$ up to about
$560\,\GeV$, while at higher transverse momenta they reach up to $25\%$,
similarly as in the on-shell generators, and for $p_{{\rm T},j_1}$ they
vary from $+4\%$ in the first bin to about $-10\%$ in the tail.

Comparing on-shell and off-shell predictions, we find that the deficit
of the on-shell prediction, which amounts to $7$--$10\%$ in the region of
small and moderate transverse momenta, disappears at higher $p_{\rm T}$.
In the $p_{{\rm T},j_1j_2}$ distribution \ttH{}+\tWH{} and \bbfourlH{}
agree within $3\%$ between $240$ and $640\,\GeV$, while in the tail the
on-shell prediction lies $8$--$20\%$ below \bbfourlH{}.
In the $p_{{\rm T},j_1}$ distribution the on-shell prediction lies
$8$--$10\%$ below \bbfourlH{} for $p_{{\rm T},j_1}<160\,\GeV$, while it
exceeds \bbfourlH{} by $3$--$7\%$ between $240$ and $480\,\GeV$, and in
the tail it falls $5$--$15\%$ below \bbfourlH{}.
The resulting shape difference of $15$--$20\%$ is of the same order as
the shower effects in the on-shell generators.
This indicates that it originates from the different modelling of QCD
radiation in hadronic $W$ decays, which is NLOPS accurate in \bbfourlH{},
while in the on-shell generators it is entirely controlled by \Pythia{}.
The relative weight of $tWH$ production is again at the level of
$1$--$2\%$ in both distributions.

In summary, in the semileptonic channel we find a similar level of
agreement between on-shell and off-shell generators as in the dileptonic
channel for Higgs-boson observables, with an overall deficit of the
on-shell prediction of $6$--$7\%$.
Light-jet observables, which probe QCD radiation in hadronic $W$ decays,
feature shape differences of up to about $20\%$, which are of the same
order as the sizeable shower effects in the on-shell generators.
In contrast, in \bbfourlH{}, where the hardest emissions from all
top-quark and $W$-boson decays are controlled by matrix elements, shower
effects are typically at the few-percent level and do not exceed about
$10\%$, apart from the tail of the $p_{{\rm T},j_1j_2}$ distribution.

\FloatBarrier
\section{Summary and conclusions}
\label{se:conclusions}

We have presented a new NLOPS generator for off-shell
$\ttbar H+tWH$ production with dileptonic and semileptonic decays, based
on exact NLO QCD matrix elements for
$pp\to b\bar b\,e^+\nu_e\,\mu^-\bar\nu_\mu\,H$ in the 4FNS with massive $b$ quarks.
This new generator is dubbed \bbfourlH{} and provides, for the first time, a unified NLOPS description of $\ttbar H$ and $tWH$ production, including interference effects and NLO
accurate top-quark decays.
The parton-shower matching is based on the \POWHEGRES{} method with matrix-element--based
resonance histories and an inverse-width expansion that avoids spurious
width effects~\cite{Jezo:2023rht}.
In addition, we have presented new on-shell \POWHEG{} generators for
$\ttbar H$ and $tWH$ production, the latter being the first of its kind,
which enable a tuned comparison against \bbfourlH{} within the same
framework.
All generators have been validated against fixed-order results, and
\bbfourlH{} also reproduces the NWA in the zero-width limit both at LO and NLO.

For Higgs-boson and leptonic observables, NLO, LHE and NLOPS predictions
of \bbfourlH{} agree at the few-percent level, and for $b$-jet observables
within about $10\%$, while large differences arise above the peak of the
reconstructed top-quark line shape, which is populated by soft QCD
radiation clustered into the $b$-jet, such that fixed-order NLO
predictions are effectively only LO accurate there.
Since the hardest emissions from all subprocesses are controlled by
matrix elements, shower effects in \bbfourlH{} are moderate, apart from
regions that are populated only by soft QCD radiation.

In the comparison against the on-shell generators, inclusive cross
sections agree within $2\%$, while in the fiducial phase space the
on-shell predictions lie about $5\%$ ($6$--$7\%$) below \bbfourlH{} in the
dileptonic (semileptonic) channel.
This deficit is driven by shower radiation in top-quark decays, which
reduces the on-shell fiducial cross section by about $11\%$.
Apart from that, Higgs-boson, leptonic and $b$-jet distributions agree at
the few-percent level.
Much larger differences, of $20$--$40\%$, show up in the reconstructed
top-quark line shape and above the kinematic edge of the
positron--$b$-jet invariant mass, where $tWH$ contributes up to
$30$--$45\%$ and the missing $\ttbar H$--$tWH$ interference in the DR1
scheme becomes important.
In the semileptonic channel, light-jet observables feature shape
differences of $15$--$20\%$.

These findings indicate that a consistent off-shell description is
particularly relevant for observables sensitive to the top-quark
resonance and to $tWH$ production.
Since the Higgs boson is decayed by the parton shower, the same events
can be used for different Higgs-decay channels, and the approach can be
extended to fully hadronic top-quark decays.
Moreover, \bbfourlH{} supports modified top-Yukawa couplings, which makes
it possible to study scenarios where $tWH$ production and its
interference with $\ttbar H$ production are strongly enhanced. We leave an investigation of such effects, including a detailed investigation of the $tWH$ phase-space to future studies.

The new \bbfourlH{} and \tWH generators are available upon request and will be added to the official \RES repository.

\section*{Acknowledgments}
J.~L.~was supported by the Science and Technology Research Council (STFC) under the Consolidated Grants ST/T00102X/1 and ST/C004242/1 (UKRI4242).
We would like to thank S.~F.~Ravasio, F.~Buccioni, and S.~Pozzorini for valuable discussions.
We acknowledge the use of the DiRAC Cumulus HPC facility under Grant No. PPSP351 and the HPC cluster PALMA II of the University of Münster, subsidized by the DFG Grant No. INST 211/667-1.

\FloatBarrier
\bibliography{paper}
 
\end{document}